\documentclass[useAMS,usenatbib]{mn2e}  
\usepackage {graphicx,subfig,aas_macros,float,amsmath}

\newcommand{\equ}[1]{eq.~(\ref{eq:#1})}
\newcommand{\equs}[1]{eqs.~(\ref{eq:#1})}
\newcommand{\equref}[1]{(\ref{eq:#1})}

\newcommand{\Equ}[1]{Eq.~(\ref{eq:#1})}
\newcommand{\Equs}[1]{Eqs.~(\ref{eq:#1})}
\newcommand{\equnp}[1]{eq.~\ref{eq:#1}}
\newcommand{\equsnp}[1]{eqs.~\ref{eq:#1}}
\newcommand{\se}[1]{\S\ref{sec:#1}}
\newcommand{\fig}[1]{Fig.~\ref{fig:#1}}

\newcommand{\Fig}[1]{Figure~\ref{fig:#1}}

\newcommand{\tab}[1]{Table~\ref{tab:#1}}
\newcommand{\be}{\begin{equation}}
\newcommand{\ee}{\end{equation}}
\newcommand{\bea}{\begin{eqnarray}}
\newcommand{\eea}{\end{eqnarray}}

\newcommand{\no}{\noindent}

\newcommand{\msun}{{\rm M}_\odot}

\newcommand{\ifm}[1]{\relax\ifmmode#1\else$\mathsurround=0pt #1$\fi}
\newcommand{\kms}{\ifmmode\,{\rm km}\,{\rm s}^{-1}\else km$\,$s$^{-1}$\fi}

\newcommand{\kpc}{\,{\rm kpc}}
\newcommand{\pc}{\,{\rm pc}}

\newcommand{\K}{\,{\rm K}}

\newcommand{\ltsima}{$\; \buildrel < \over \sim \;$}
\newcommand{\lsim}{\lower.5ex\hbox{\ltsima}}
\newcommand{\gtsima}{$\; \buildrel > \over \sim \;$}
\newcommand{\gsim}{\lower.5ex\hbox{\gtsima}}

\def\la{\langle}

\def\sy{\,M_\odot\, {\rm yr}^{-1}}

\def\cmc{\,{\rm cm}^{-3}}

\def\M*{M_{\rm *}}
\def\Mv{M_{\rm v}}
\def\Rv{R_{\rm v}}
\def\Vv{V_{\rm v}}
\def\tv{t_{\rm v}}
\def\tg{t_{\rm growth}}
\def\Tv{T_{\rm v}}
\def\Ts{T_{\rm s}}
\def\Rs{R_{\rm s}}
\def\rhob{\rho_{\rm b}}
\def\rhos{\rho_{\rm s}}

\def\cb{c_{\rm b}}
\def\cs{c_{\rm s}}
\def\cbs{c_{\rm b,s}}
\def\vb{v_{\rm b}}
\def\vs{v_{\rm s}}
\def\vbs{v_{\rm b,s}}
\def\qb{q_{\rm b}}
\def\qs{q_{\rm s}}
\def\qbs{q_{\rm b,s}}
\def\tkh{t_{\rm KH}}
\def\tsc{t_{\rm sc}}
\def\Mb{M_{\rm b}}
\def\Ms{M_{\rm s}}
\def\Mbs{M_{\rm b,s}}
\def\Or{\omega_{_{\rm R}}}
\def\Oi{\omega_{_{\rm I}}}
\def\Pr{\varpi_{_{\rm R}}}
\def\Pi{\varpi_{_{\rm I}}}

\usepackage{color}

\begin{document} 

\large 

\title[Kelvin Helmholtz Instability in Cold Flows]
{Instability of Supersonic Cold Streams Feeding Galaxies I: Linear Kelvin-Helmholtz Instability with Body Modes}

\author[Mandelker et al.] 
{\parbox[t]{\textwidth} 
{ 
Nir Mandelker$^1$\thanks{E-mail: nir.mandelker@mail.huji.ac.il },
Dan Padnos$^1$,
Avishai Dekel$^1$,
Yuval Birnboim$^1$, 
Andreas Burkert$^{2,3}$, 
Mark R. Krumholz$^{4,5}$,
Elad Steinberg$^1$
} 
\\ \\  
$^1$Centre for Astrophysics and Planetary Science, Racah Institute of Physics, The Hebrew University, Jerusalem 91904, Israel\\ 
$^2$Unitversit${\ddot{a}}$ts-Sternwarte M${\ddot{u}}$nchen, Scheinerstr. 1, D-81679 Munich, Germany\\
$^3$Max-Planck Institute for Extraterrestrial Physics, Giessenbachstr. 1, D-85748 Garching, Germany\\
$^4$Department of Astronomy and Astrophysics, University of California, Santa Cruz, CA 95064, USA\\
$^5$Research School of Astronomy \& Astrophysics, Australian National University, Cotter Road, Weston, ACT 2611, Australia} 
\date{} 
 
\pagerange{\pageref{firstpage}--\pageref{lastpage}} \pubyear{0000} 
 
\maketitle 
 
\label{firstpage} 
 
\begin{abstract} 
Massive galaxies at high redshift are predicted to be fed from the cosmic web by narrow, dense streams of cold gas that penetrate through the hot medium encompassed by a stable shock near the virial radius of the dark-matter halo. Our long-term goal is to explore the heating and dissipation rate of the streams and their fragmentation and possible breakup, in order to understand how galaxies are fed, and how this affects their star-formation rate and morphology. We present here the first step, where we analyze the linear Kelvin-Helmholtz instability (KHI) of a cold, dense slab or cylinder in 3D flowing supersonically through a hot, dilute medium. The current analysis is limited to the adiabatic case with no gravity. By analytically solving the linear dispersion relation, we find a transition from a dominance of the familiar rapidly growing \textit{surface} modes in the subsonic regime to more slowly growing \textit{body} modes in the supersonic regime. The system is parametrized by three parameters: the density contrast between stream and medium, the Mach number of stream velocity with respect to the medium, and the stream width with respect to the halo virial radius. A realistic choice for these parameters places the streams near the mode transition, with the KHI exponential-growth time in the range 0.01-10 virial crossing times for a perturbation wavelength comparable to the stream width. We confirm our analytic predictions with idealized hydrodynamical simulations. Our linear estimates thus indicate that KHI may be effective in the evolution of streams before they reach the galaxy. More definite conclusions await the extension of the analysis to the nonlinear regime and the inclusion of cooling, thermal conduction, the halo potential well, self-gravity and magnetic fields. 
\end{abstract} 
 

\section{Introduction} 
\label{sec:intro}
\smallskip
According to the standard $\Lambda {\rm CDM}$ model of cosmology, the most 
massive haloes at any epoch lie at the nodes of the cosmic web, and are 
penetrated by cosmic filaments of dark matter \citep[e.g.][]{Bond96,Springel05,
Dekel09}. These represent high-sigma peaks in the density fluctuation field, 
much more massive than the Press-Schechter mass, $\M*$, of typical haloes at 
that time \citep{Press74}. At redshift $z=1-4$, when star-formation is at its 
peak and most of the mass is assembled into galaxies \citep{Madau98,hopkins06,
Madau14}, such haloes have virial masses of $\Mv\sim 10^{12}\msun$ and above, 
larger than the critical mass for shock heating $M_{\rm shock}\lsim 10^{12}\msun$ 
\citep{bd03,db06}. They thus contain hot gas at the virial temperature, $\Tv \sim 10^6\K$. 
However, for such high-sigma peaks the filaments that feed the halo are significantly 
narrower than the virial radius, and the gas residing in them is much denser 
than the halo gas. Therefore, the radiative cooling time of the stream gas is 
shorter than the local compression time, preventing the formation of a stable 
virial shock within the streams. The streams are thus expected to remain cold, 
with temperatures of $\Ts\gsim 10^4\K$, allowing them to penetrate efficiently 
through the hot halo circumgalactic medium (CGM) onto the central galaxy 
\citep{db06}. 

\smallskip
The above theoretical picture is supported by cosmological simulations 
\citep{Keres05,Ocvirk08,Dekel09,CDB,FG11,vdv11}. In these simulations, cold streams 
with widths of a few to ten percent of the virial radius penetrate deep into the 
halo. This gas supply allows the high star-formation rates (SFRs) of 
$\sim 20-200\sy$ observed in massive star-forming galaxies (SFGs) with baryonic masses 
of $\sim 10^{11}\msun$ at $z\sim 2$ \citep{Genzel06,Forster06,Elmegreen07,Genzel08,
Stark08}. These high SFR values are only a factor of $\lsim 2$ lower than the accretion 
rate at the virial radius, implying that at least half the gas mass flux brought into 
the halo by the streams must reach the central galaxy \citep{Dekel09}, irrespective of 
what happens to the stream structure and thermal properties along the way. The streams 
also play a key role in the buildup of angular momentum in disk galaxies \citep{Pichon11,Kimm11,Stewart11,Stewart13,Codis12,Danovich12,Danovich15}. 

\smallskip
Cosmological simulations indicate that the streams maintain roughly constant 
inflow velocities as they travel from the outer halo to the central galaxy 
\citep{Dekel09,Goerdt15a}. The constant velocity, as opposed to the expected 
gravitational acceleration, indicates energy loss into radiation which may be 
observed as Lyman-$\alpha$ cooling emission \citep{Dijkstra09,Goerdt10,FG10}, 
though the dissipation process has not been explored yet. Based on radiative 
transfer models, the total luminosity and the spatial structure of the emitted 
radiation appear similar to Lyman-$\alpha$ ``blobs" observed at $z>2$ 
\citep{Steidel00,Matsuda06,Matsuda11}. The models find that roughly half the 
radiation comes from the dissipation of gravitational energy while the other 
half is due to heating from the UV background (\citealp{Goerdt10}, though see 
also \citealp{FG10} who found somewhat lower luminosities in their simulations). 
Radiative transfer models also show that a central quasar can power the emission 
by supplying seed photons which scatter inelastically within the filaments, 
producing Lyman-$\alpha$ cooling emission that extends to several hundred $\kpc$ 
and appears similar to observed structures \citep{Cantalupo14}. Recent observations 
using the MUSE integral-field instrument suggest that such extended Lyman-$\alpha$ 
emitting nebulae are ubiquitous around the brightest quasars at $z\sim 3.5$ \citep{Borisova16}. 
In addition to emission, the cold streams consisting of mostly neutral Hydrogen should 
also be visible in absorption, and can account for observed Lyman-limit systems (LLSs) 
and damped Lyman-$\alpha$ systems (DLAs) \citep{Fumagalli11,Goerdt12,vdv12}. Observations 
using absorption features along quasar sight-lines to probe the CGM of massive SFGs at 
$z\sim 1-2$ reveal low-metallicity, co-planar, co-rotating accreting material \citep{Bouche13,
Bouche16,Prochaska14}, providing further observational support for the cold-stream paradigm. 
Strong Lyman-$\alpha$ absorption has also been detected in the CGM of massive sub-millimeter 
galaxies (SMGs) at $z\sim 2$ \citep{Fu16}. 

\smallskip
Despite the growing evidence from simulations and observations that cold 
streams are a fundamental part of galaxy formation at high redshift, several 
important questions remain regarding their evolution. How much of the stream 
energy is dissipated as they travel through the CGM? What are the implications 
on the emitted radiation and the mass inflow rate onto the galaxy? How do the 
streams join the galaxy, in terms of coherency versus fragmentation/clumpiness, 
temperature, and velocity? How does this affect the growth of angular momentum 
and the SFR in the disk? 

\smallskip
While there is some preliminary observational evidence for the fragmentation 
of cold streams \citep{Cantalupo14}, most attempts to address these questions 
have used cosmological simulations. Unfortunately, in current cosmological 
simulations the resolution within the streams is never better than a few hundred 
$\pc$, and is often on the order of a $\kpc$, comparable to the stream width 
itself. They thus cannot resolve the detailed physical processes associated 
with stream instabilities necessary to properly address these questions. 
Grid-based adaptive-mesh-refinement (AMR) codes show streams that remain cold 
and coherent outside of $\sim 0.3\Rv$, inside of which a messy interaction region 
is seen where the streams collide, fragment, and experience strong torques before 
settling onto the disc \citep{CDB,Danovich15}. These simulations exhibit high gas 
mass accretion rates onto the central galaxy, roughly half the virial accretion rate 
\citep{Dekel13}. Simulations using the moving mesh code AREPO \citep{Springel10,Vogelsberger12} 
suggest that the streams heat up at $\sim 0.25-0.5\Rv$, with most of the accreted 
gas heating to roughly the virial temperature before falling onto the galaxy 
\citep{Nelson13}. Nevertheless, the mass inflow rate onto the central galaxy 
ends up very similar to the virial accretion rate, likely because the dense 
stream gas in the inner halo rapidly cools after heating. It is unclear whether 
this gas is ever in hydrostatic equilibrium within the halo. The same study argued 
that previous reports of streams remaining cold and coherent in Smooth-Particle-Hydrodynamic 
(SPH) simulations were due to numerical inaccuracies associated with the standard 
formulation of SPH. 

\smallskip
Since current cosmological simulations are far from being able to properly 
resolve the streams, a more fundamental analytical and numerical approach is 
warranted. The physical problem of the evolution of a supersonic, cold, dense, 
gas stream in a hot, dilute medium has not been addressed thus far in the 
literature, even at its simplest hydrodynamic level of Kelvin-Helmholtz instability 
(KHI)\footnote{The problem of a hot jet travelling supersonically in a cold medium 
has been studied, see references in \se{linear}.}. This is the first in a series 
of papers where we study this issue, in the context of cold streams feeding massive 
galaxies at high redshift, using analytic models of increasing complexity together 
with idealized simulations, and concluding with full-scale cosmological simulations 
with tailored mesh refinement in the streams. 

\smallskip
In this paper we take the first step and address KHI under fully compressible 
conditions\footnote{We use ``compressible" to refer to flows with arbitrary 
Mach number, supersonic ($M>1$), transonic ($M\sim 1$), or subsonic ($M<1$). 
We use ``incompressible" to refer to the limit where $M\rightarrow 0$, equivalent 
to taking the sound speed $c \rightarrow \infty$. Since pressure and density 
are related through the sound speed, $dP\propto c^2 d\rho$, the density is 
effectively constant in the incompressible limit.}. We derive the dispersion 
relation for the growth of linear instabilities in a confined planar-slab and 
cylinder, and ask whether such instabilities grow to non-linear amplitudes in 
a virial crossing time. In a forthcoming paper (Padnos et al., in preparation), 
we will address in detail, analytically and using idealized simulations, the 
non-linear evolution of these instabilities. In future work we will add one-by-one 
thermal conduction, cooling, the external potential of the host halo, self gravity 
and magnetic fields. In the final phase, we will study cosmological simulations 
with forced mesh refinement in the streams, to explore the effect of stream 
instability on galaxy formation at the halo centre and the effects of feedback 
on the streams. 

\smallskip
This paper is organized as follows: In \se{linear} we 
summarize the derivation of the linear dispersion relation 
for compressible KHI in different idealized geometries. Mathematical 
details of the derivations are provided in several appendices, 
which may be of interest to the mathematically inclined reader. 
In \se{sim} we use numerical simulations to test the analytic 
predictions of the preceding section. In \se{tkh} we apply the 
analytic formalism to the case of cosmic cold streams and estimate 
the number of e-foldings of growth experienced by initially small 
perturbations within a virial crossing time. In \se{disc}, we 
speculate as to the effects of additional physics not included 
in our analysis, presenting an outline for future work. We discuss 
our results and summarise our conclusions in \se{conc}.

\section{Compressible KHI} 
\label{sec:linear} 

\smallskip
In this section we derive the dispersion relations for 
compressible KHI in several different geometries. For 
simplicity and analytic tractability, we begin by deriving 
the relation in planar geometry, first discussing a ``two-zone 
instability", or a \textit{sheet}, where two semi-infinite 
fluids are separated by a single planar interface, and then a 
``three-zone instability", or a \textit{slab}, where one fluid 
is confined to a planar slab of finite thickness and surrounded 
by a second (background) fluid from both sides. KHI in a slab is 
qualitatively different than in a sheet due to the appearance 
of \textit{body modes}, unstable perturbations caused by waves 
reverberating back and forth between the slab boundaries, that 
dominate the instability at high Mach numbers (\se{reflected}). 
We then derive the dispersion relation for a cylindrical stream, 
and show that the behaviour of linear perturbations with wavelengths 
comparable to or smaller than the stream radius is effectively 
identical to that of perturbations in a slab.

\smallskip
Several previous studies have addressed linear stability 
of astrophysical jets to KHI in both planar and cylindrical 
geometries, both analytically and numerically \citep[e.g.][]{Ferrari78,
Birkinshaw84,Birkinshaw90,Payne_Cohn85,Hardee87,Hardee_Norman88a,
Hardee_Norman88b,Bodo94,Perucho04}. These studies focus primarily 
on hot, dilute jets travelling in cold, dense media. As we will see 
below, the main difference between such a scenario and that studied 
here, of cold streams in a hot medium, is the ratio of the stream 
sound crossing time to the KHI exponential-growth time. While this 
can be important for the overall stability of the stream (see \se{tkh}), 
it does not fundamentally alter the linear dispersion relation, and 
our derivation is similar to those presented in \citet{Payne_Cohn85} 
(for the cylinder) and \citet{Hardee_Norman88a} (for the slab). However, 
there are certain differences in our approaches and conclusions which 
we highlight in the text, and we find our analysis to be more complete, 
addressing a sheet, a slab, and a cylinder in a self-contained and consistent 
way.

\subsection{General KHI in Planar Coordinates}
\label{sec:general}

\smallskip
We begin with the basic equations of hydrodynamics, which 
represent conservation of mass (the continuity equation), 
momentum (the Euler equation) and energy: 
\be 
\label{eq:continuity}
\frac{\partial \rho}{\partial t} + \left({\vec {v}}\cdot{\vec {\nabla}}\right)\rho + \rho{\vec {\nabla}}\cdot{\vec {v}} = 0,
\ee
\be 
\label{eq:Euler}
\rho\left[\frac{\partial {\vec {v}}}{\partial t} + \left({\vec {v}}\cdot{\vec {\nabla}}\right){\vec {v}}\right] + {\vec {\nabla}}P = 0,
\ee
\be 
\label{eq:energy}
\frac{\partial P}{\partial t} + \left({\vec {v}}\cdot{\vec {\nabla}}\right)P - c^2\left[\frac{\partial \rho}{\partial t} + \left({\vec {v}}\cdot{\vec {\nabla}}\right)\rho\right] = 0.
\ee
{\no}Above, $\rho$ is the fluid's density, $\vec{v}$ its velocity and $P$ 
the pressure in the fluid. We assume an ideal equation of state, so 
$c=(\gamma P/\rho)^{1/2}$ is the sound speed, where $\gamma$ is the adiabatic 
index of the fluid, $\gamma = ( \partial {\rm ln P} / \partial {\rm ln \rho} )_{\rm s}$. 

\smallskip
We consider a time-independent flow in the ${\hat {z}}$ direction, 
where the flow velocity and the fluid density are arbitrary functions 
of $x$ (in Cartesian coordinates): 
\be 
\label{eq:equilibrium}
\rho_0(x,y,z) = \rho_0(x),\:\:\: {\vec {v}}_0(x,y,z) = v_0(x){\hat {z}}.
\ee
{\no}In this case, with no external forces, \equ{Euler} dictates that the 
pressure is uniform, $P_0(x,y,z) = P_0$. On top of this equilibrium 
flow, we impose small perturbations in the fluid variables, 
$\rho = \rho_0 + \rho_1$, $\vec{v} = v_0{\hat {z}} + \vec{u}$ and 
$P = P_0 + P_1$, where the perturbation in each variable $f$ obeys 
$f_1<<f_0$. To study the growth of instabilities, we decompose each 
of the perturbed quantities into Fourier modes of the form 
\be 
\label{eq:perturbation_form}
f_1(x,y,z,t) = f_1(x)e^{i\left(k_yy+k_zz-\omega t\right)}.
\ee
{\no}In other words, the perturbations are travelling waves in 
the $yz$ plane with wave vector $\vec{k}=k_y \hat{y} + k_z \hat{z}$ 
and frequency $\omega$, and an arbitrary $x$ dependence.

\smallskip
By inserting these perturbations into \equs{continuity}-\equref{energy} 
and linearizing, we can derive algebraic relations between the $x$ dependent 
amplitudes of the pressure perturbation and its derivatives with respect to 
$x$ to those of all other perturbations\footnote{We could have chosen any of 
the 5 perturbed variables and expressed the other 4 in terms of it. However, 
the pressure is a convenient choice because it must always be continuous, while 
the other variables can in principle have discontinuities.}. Using the conventions 
$\partial f / \partial x = f'$, $k_z = k~{\rm cos}(\varphi)$, $k_y = k~{\rm sin}(\varphi)$, 
and $v_k = {\vec {v}}_{\rm 0} \cdot {\hat {k}} = v_{\rm 0}{\rm cos}(\varphi)$, 
we obtain: 
\be 
\label{eq:rho_1}
\rho_1 = -\frac{1}{k^2 \left(v_k-\frac{\omega}{k}\right)^2} \left[P_1''-\frac{2v_k'}{v_k-\frac{\omega}{k}}P_1' - k^2 P_1\right],
\ee
\be 
\label{eq:uz}
u_z = -\frac{{\rm cos}(\varphi)}{\rho_0 \left(v_k-\frac{\omega}{k}\right)} \left[\frac{v_k'}{k^2~{\rm cos^2}(\varphi)\left(v_k-\frac{\omega}{k}\right)}P_1' + P_1\right],
\ee
\be 
\label{eq:uy}
u_y = -\frac{{\rm sin}(\varphi)}{\rho_0 \left(v_k-\frac{\omega}{k}\right)}P_1,
\ee
\be 
\label{eq:ux}
u_x = \frac{i}{\rho_0 k\left(v_k-\frac{\omega}{k}\right)}P_1'.
\ee
{\no}Note that in \equs{rho_1}-\equref{ux} all fluid variables are functions of $x$.
In addition, we are left with a second order ordinary differential equation for $P_1(x)$: 
\be 
\label{eq:P1}
P_1'' - \left[\frac{2v_k'}{v_k-\frac{\omega}{k}}+\frac{\rho_0'}{\rho_0}\right]P_1' - k^2\left[1-\left(\frac{v_k-\frac{\omega}{k}}{c}\right)^2\right]P_1=0.
\ee

\smallskip
\Equ{P1} is an eigenvalue equation. Given profiles along $x$ for the unperturbed 
density and velocity and boundary conditions for $P_1$, solutions exist only for 
certain combinations of $\omega$ and $k$, which define the dispersion relation, 
$\omega(k)$. Since \equ{P1} depends only on the component of the velocity parallel 
to the perturbation wave vector, $v_k$, with no explicit dependence on the propagation 
angle $\varphi$, we restrict our analysis to perturbations where $k_y=0$ so that 
$v_k=v_0$. 

\subsubsection{Temproal vs Spatial Stability Analysis}
\label{sec:Temp_Spat}
There are in general two types of stability analyses, \textit{temporal} 
and \textit{spatial}. In the former, the wavenumber $k$ is real while the 
frequency $\omega$ is complex. Physically, this represents seeding the entire 
system with a \textit{spatially-oscillating} perturbation and studying its 
\textit{temporal growth}. In the latter, $\omega$ is real while $k$ is complex. 
This represents seeding a \textit{temporally-oscillating} perturbation at the 
stream origin and studying its downstream \textit{spatial growth}. This distinction 
is particularly important when performing numerical simulations, as they change 
the required boundary and initial conditions. Studies of the stability of jets 
whose source has intrinsic variability (such as an AGN or a gamma ray burst) often 
employ spatial stability analyses. However, the cosmic web streams we are studying 
do not have a well defined variable source outside the halo, but rather experience 
perturbations from the halo throughout their extent. Therefore, we perform a temporal 
stability analysis, envisioning a stationary stream suffering some perturbation 
across its extent, and asking how much the perturbation will grow in a virial crossing 
time.

\subsection{The Planar Sheet}
\label{sec:sheet}

\smallskip
We first consider the classic case of two fluids separated at $x=0$. 
For consistency with later sections when we discuss a confined \textit{stream} 
with finite thickness in a \textit{background}, we label the two 
fluids with subscripts `b' and `s', for $x>0$ and $x<0$ respectively. 
We assume each fluid to have initially uniform density and velocity. 
This problem is often referred to as the ``vortex sheet" and was first 
addressed for two compressible fluids by \cite{Landau44}. In this case, 
the second term in \equ{P1} vanishes for all $x \ne 0$. The equation must 
be solved separately in the regions $x>0$ and $x<0$, subject to the boundary 
conditions that the pressure perturbation vanishes at infinity and is continuous 
across the boundary at $x=0$. The solution is
\be 
\label{eq:P1_sheet}
P_1=\left\{\begin{array}{c c}
Ae^{-\qb x}&x>0\\
Ae^{\qs x}&x<0
\end{array}\right. ,
\ee
{\no}where $A$ is a constant of integration that depends on the initial 
conditions, and we have defined the generalized wavenumbers $\qb$ and $\qs$ 
by
\be 
\label{eq:q_bs}
\qbs = k\left[1 - \left(\frac{\omega - k\vbs}{k\cbs}\right)^2\right]^{1/2}.
\ee
{\no}Since $\omega$ is in general complex, $\qbs$ is the 
square root of a complex number, forcing us to choose a branch 
cut in the complex plane. We have chosen to define ${\rm Re}(\qbs)> 0$, 
which ensures that the amplitude of perturbations decays exponentially 
away from the interface between the two fluids. These are therefore 
known as ``surface modes". A somewhat technical discussion of the 
meaning and justification of this branch cut can be found in appendix 
\se{branch_cut}.

\smallskip
To proceed, we require a fourth boundary condition. This is achieved 
by realizing that the velocity perpendicular to the interface between 
the fluids causes a spatial displacement in the interface position, from $x=0$ 
to $x=h$, and this displacement must be the same when approaching the 
interface from either side. This is often called ``the Landau condition". 
Expanding $h$ in the same Fourier modes as the other perturbed quantities, 
this results in the first-order equation 
\be 
\label{eq:boundary_h}
u_x|_{x=0} = \frac{\partial h}{\partial t} + \left(\vec{v}\cdot\vec{\nabla}\right)h = ik\left(v_0|_{x=0}-\frac{\omega}{k}\right)h.
\ee 

\smallskip
Inserting \equ{ux} and \equ{P1_sheet} into \equ{boundary_h} from both 
sides of the interface and then dividing out $h$ yields the dispersion 
relation
\be 
\label{eq:dispersion_sheet}
\frac{\left(\omega - k\vb\right)^2}{\left(\omega - k\vs\right)^2} = -\frac{\rhos}{\rhob}\frac{\qb}{\qs}.
\ee

\smallskip
In the incompressible limit, the speed of sound goes to infinity 
in both media, and therefore from \equ{q_bs} $q_{\rm b,s}\rightarrow k$. 
In this limit, \equ{dispersion_sheet} reduces to the familiar form 
of the classical Kelvin-Helmholtz dispersion relation \citep[e.g.][]{Chandrasekhar61}
\be 
\label{eq:KKHI}
\omega_{\pm} = \frac{\rhob \vb + \rhos \vs}{\rhob + \rhos}k \pm i \frac{\sqrt{\rhos \rhob}|\vs-\vb|}{\rhob + \rhos}k.
\ee

\smallskip
Since the growth rate cannot depend on the frame of reference, we analyse 
the general case in the frame where the background is static, $\vb=0$, and 
the stream is moving with velocity $\vs=V$. Furthermore, we define unitless 
variables 
\begin{equation} 
\label{eq:unitless}
\varpi \equiv \frac{\omega}{kV},\:\:\delta \equiv \frac{\rhos}{\rhob},\:\:\Mbs \equiv \frac{V}{\cbs}.
\end{equation}
{\no}Here, $\varpi$ is the phase velocity in units of the stream velocity, 
$\delta$ is the density contrast between the stream and the background 
and $\Mbs$ are the Mach number of the stream velocity with respect to 
the background and the stream itself, respectively. Since pressure 
equilibrium is assumed, $\Ms = \sqrt{\delta}\Mb$. In this notation, the 
dispersion relation for the incompressible sheet, \equ{KKHI}, becomes
\be 
\label{eq:KKHI2}
\varpi_{\pm} = \frac{\delta}{1+\delta} \pm i \frac{\delta^{1/2}}{1+\delta}.
\ee 
{\no}By further defining
\be 
\label{eq:Z_def}
Z \equiv -\frac{1}{\delta}\left(\frac{\varpi}{\varpi-1}\right)^2\left(\frac{1-\delta \Mb^2(\varpi-1)^2}{1-\Mb^2\varpi^2}\right)^{1/2}, 
\ee
{\no}the dispersion relation for the compressible sheet, \equ{dispersion_sheet}, 
becomes 
\be 
\label{eq:unitless_dispersion_sheet}
Z = 1, 
\ee
{\no}an algebraic equation for the unknown $\varpi$. We learn 
from the equation that $\varpi$ depends only on $\delta$ and 
$\Mb$, with no dependence on $k$. This implies that $\omega\propto Vk$, 
which could have been predicted from dimensional analysis, since 
the only length scale in the problem is the perturbation wavelength. 

\smallskip
By squaring both sides of \equ{unitless_dispersion_sheet}, inserting 
\equ{Z_def} and rearranging, we get a sixth degree polynomial equation 
in $\varpi$ that can be factored as the product of a quadratic with a 
quartic
\be 
\label{eq:unitless_dispersion_sheet_2}
\left[\delta\left(\varpi-1\right)^2-\varpi^2\right]\cdot\left[\delta \left(\varpi-1 \right)^2\left(\Mb^2\varpi^2-1\right)-\varpi^2\right]=0.
\ee

\smallskip
A detailed analysis of this equation is presented in \se{sheet_deriv}. 
We summarize the main points below. The two roots of the quadratic part 
and two of the four roots of the quartic part are always real, and do 
not solve \equ{unitless_dispersion_sheet}. Rather they are solutions to 
the equation $Z=-1$, arising from the fact that we squared 
\equ{unitless_dispersion_sheet}. At low Mach numbers, the two remaining 
roots of the quartic part are complex conjugates, representing a growing 
unstable mode and a decaying mode. Both of these are indeed solutions to 
\equ{unitless_dispersion_sheet}, with $Z=1$. \Equ{unitless_dispersion_sheet} 
thus admits only two solutions, a growing and a decaying mode, as in the 
incompressible limit (\equnp{KKHI2}). However, above a critical Mach number 
these two complex roots become real as well (while still solving $Z=1$), and 
the dispersion relation admits only \textit{stable} solutions. For $\Mb>>1$ 
the solutions converge to 
\be 
\label{eq:phi_infiity}
\varpi_{\rm \infty} = \Mb^{-1},\,\,1-\Ms^{-1},
\ee
{\no}which represent waves with phase velocities $\omega/k=\cb,\,V-\cs$.

\smallskip
The critical Mach number for stability is given by 
\be 
\label{eq:Mcrit}
M_{\rm crit} = \left(1 + \delta^{-1/3}\right)^{3/2}.
\ee
{\no}This generalizes the result of \citet{Landau44}, who showed 
that for identical fluids, with $\delta=1$, the flow is stable 
above $M_{\rm crit}=\sqrt{8}\simeq 2.83$. For $\delta=10$ and $100$, 
$M_{\rm crit}\sim 1.77$ and $1.34$ respectively. 

\begin{figure}
\centering
\includegraphics[trim={0.35cm 0.8cm 0.2cm 1.0cm}, clip, width =0.45 \textwidth]{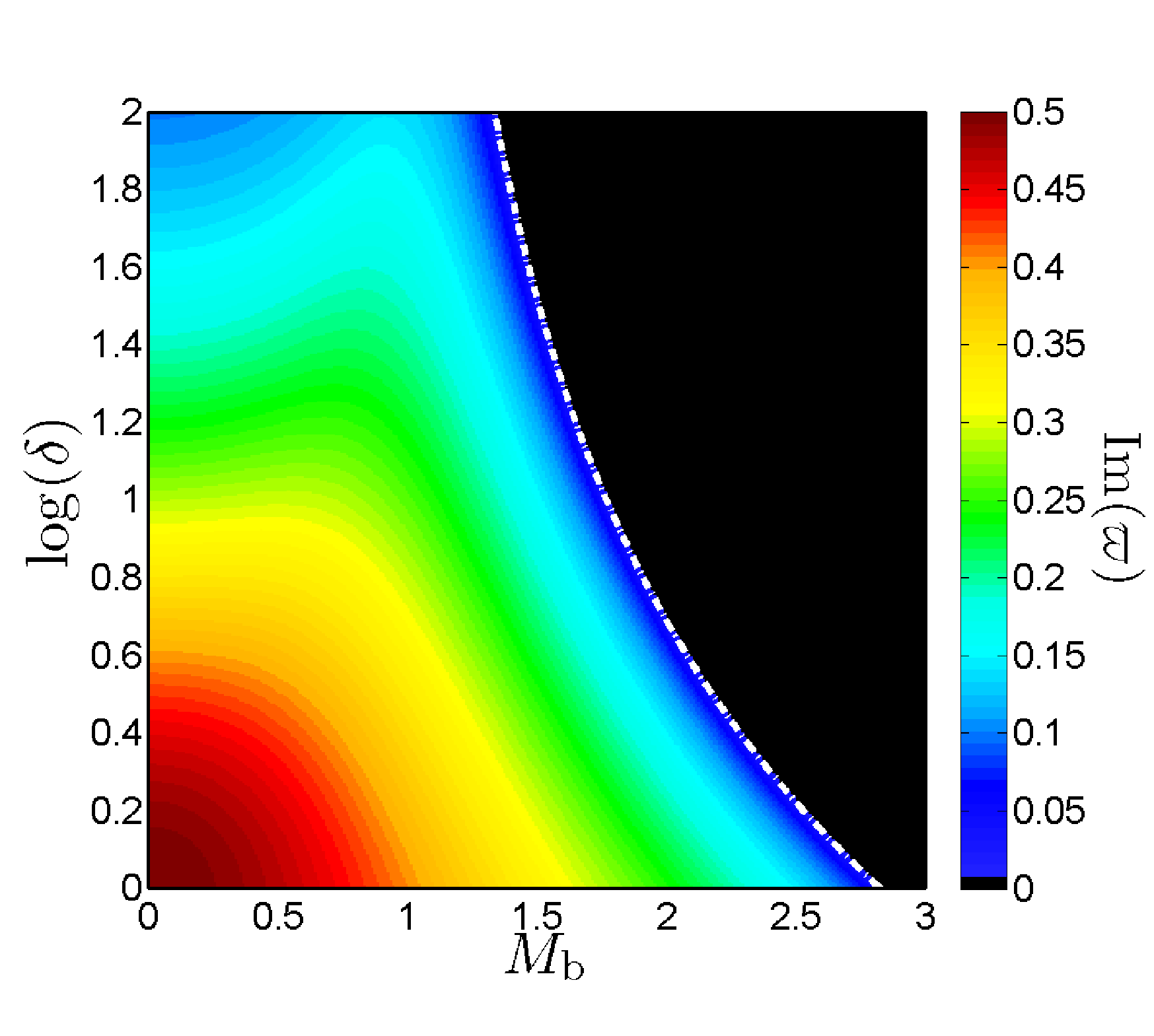}
\caption{Growth rate for unstable modes of the sheet as a function of the 
Mach number, $\Mb$, and the density contrast, $\delta$. Colour represents 
the imaginary part of $\varpi=\omega/(kV)$. The black region at high Mach 
numbers shows the stable zone where linear perturbations do not grow. The 
white dash-dotted line shows the analytic expression for $M_{\rm crit}$, the 
critical Mach number above which the sheet is stable, given by \equ{Mcrit}.}
\label{fig:planar_growth_rate} 
\end{figure}

\smallskip
The analytic expression for the growing mode solution as a function 
of $\Mb$ and $\delta$ can be found by finding the roots of the quartic 
polynomial in \equ{unitless_dispersion_sheet_2} and picking the complex 
root with the positive imaginary part. However, the full expression is 
very long and intractable. We show the growth rate of the instability, 
${\rm Im}(\varpi)$, as a function of $\Mb$ and $\delta$ in \fig{planar_growth_rate}. 
As $\Mb\rightarrow 0$, the growth rate converges to the solution for 
an incompressible sheet (\equnp{KKHI2}). For fixed $\delta$, raising 
$\Mb$ from 0 to a relatively small value causes the growth rate to become 
larger, meaning the system becomes more unstable. However, raising $\Mb$ 
further to larger values causes the growth rate to decline, until it reaches 
zero at $M_{\rm crit}$, shown by the white curve.

\begin{figure}
\centering
\includegraphics[trim={0.35cm 0.8cm 0.2cm 1.0cm}, clip, width =0.45 \textwidth]{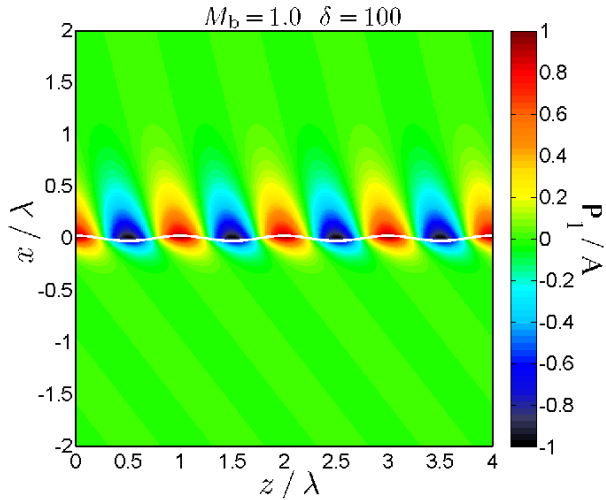}
\caption{Pressure perturbation in a sheet with $\Mb=1.0$ and $\delta=100$, 
from \equ{P1_sheet}, with the dense fluid on the bottom. The white line 
represents the perturbation in the interface height, which has an amplitude 
of $h=0.025\lambda$. This is a \textit{surface mode}, which decays exponentially 
with distance from the interface, because $\qbs$ are nearly real. The wave 
penetrates deeper into the fluid with lower density, and the angle of wave 
propagation breaks at the interface between the fluids.
}
\label{fig:P1_sheet_fig} 
\end{figure} 

\smallskip
Some intuition as to why the sheet becomes stable to linear perturbations 
at high Mach numbers can be gained by considering what happens when the 
initially flat interface is perturbed with a sinusoidal displacement. 
Upstream of each ``crest", the fluids are set to collide, creating a high 
pressure area, while downstream the fluids are set to separate, creating 
a low pressure area (see \fig{P1_sheet_fig}). The flow that develops in 
response to this pressure perturbation tends to increase the perturbation 
amplitude. The typical timescale for this process to occur is the sonic 
time across a perturbation wavelength, $\lambda/c$. However, if the flow 
is sufficiently fast with a high Mach number, this becomes very long compared 
to the relevant timescale for the steady state flow, $\lambda/V$. In this 
case, the fluid upstream does not have time to react to the displacement 
of the interface, colliding with the crests rather than flowing around them 
and suppressing the instability.

\smallskip
By inserting the growing mode solution into \equ{q_bs} and then into 
\equ{P1_sheet} we obtain the spatial form of the pressure perturbation. 
This is shown in \fig{P1_sheet_fig} for the case $\Mb=1.0$ and $\delta=100$. 
We have normalized the perturbation by its maximum amplitude $A$, so that it 
is unity at $x=0$. For reference we also show the expected form of the 
perturbed interface, with an amplitude $h=0.025\lambda$ where $\lambda$ 
is the perturbation wavelength. The pressure perturbation decays rapidly 
with distance from the interface because $\qbs$ are nearly real. This is 
a general feature of \textit{surface modes}. The differences in penetration 
depth and propagation angle between the two fluids are caused by differences 
in the real and imaginary parts of $q$ respectively between the two fluids. 
In the language of acoustic waves, this is caused by a change in the acoustic 
impedance of the two fluids.

\subsection{The Planar Slab}
\label{sec:slab}

\smallskip
We now consider a three zone problem, which we refer to as the \textit{slab}. 
The slab is confined to the region $|x|<\Rs$ and is infinite in the $y$ and 
$z$ directions. We refer to the fluid at $|x|>\Rs$ as the background, and 
assume it to be the same fluid on either side of the slab. As before, we 
assume each unperturbed medium to have uniform density and velocity. Following 
the same procedure as in \se{sheet}, we solve \equ{P1} in each region subject 
to the boundary conditions that $P_1$ vanishes at infinity and is continuous 
across both slab interfaces. The result is 
\be 
\label{eq:P1_slab}
P_1=\left\{\begin{array}{c c}
A~e^{-\qb (x-\Rs)}&x>\Rs\\
\\
\dfrac{A~{\rm sinh}\left(\qs \left[x+\Rs\right]\right) - D~{\rm sinh}\left(\qs \left[x-\Rs\right]\right)}{{\rm sinh}\left(2\qs \Rs\right)}&|x|<\Rs\\
\\
D~e^{\qb (x+\Rs)}&x<-\Rs
\end{array}\right. ,
\ee
{\no}where $A$ and $D$ are two constants of integration.

\begin{figure}
\centering
\subfloat{\includegraphics[width =0.45 \textwidth]{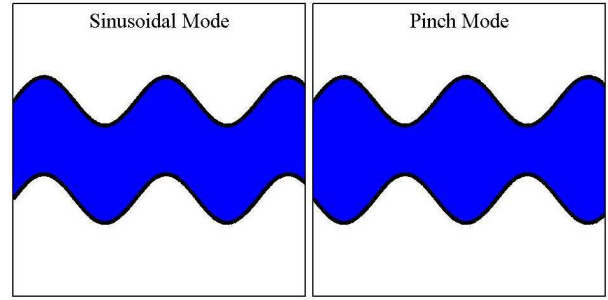}}
\caption{Schematic representation of the two instability modes in 
the planar slab. The left panel represents the anti-symmetric 
\textit{Sinusoidal} (S) mode while the right panel represents the 
symmetric \textit{Pinch} (P) mode.}
\label{fig:modes} 
\end{figure} 

\smallskip 
By applying the Landau condition (\equnp{boundary_h}) at the 
interfaces, $x=\pm \Rs$, we learn that they are not independent. 
A self consistent solution where both $A$ and $D$ are non-zero 
requires $A=\pm D$, which in turn gives a relationship between 
the displacement of the two interfaces from equilibrium $h_{\Rs} = \mp h_{-\Rs}$. 
The case $A=D,\:h_{\Rs} = -h_{-\Rs}$ corresponds to a symmetric 
perturbation of the pressure and is called the \textit{Pinch mode}, 
hereafter P mode. The other case corresponds to an anti-symmetric 
perturbation of the pressure and is called the \textit{Sinusoidal 
mode}, hereafter S mode. These are shown schematically in \fig{modes}. 

\smallskip
Inserting these two solutions into \equ{P1_slab} gives for the 
pressure perturbation within the slab 
\be 
\label{eq:P1_slab2}
P_{1,s}(x)=A~\frac{S\left(\qs x\right)}{S\left(\qs \Rs\right)},
\ee
{\no}where $S(x)={\rm sinh}(x)$ or ${\rm cosh}(x)$ for S modes or P modes respectively. 
The corresponding dispersion relations are 
\be 
\label{eq:dispersion_slab}
\frac{\left(\omega - k\vb\right)^2}{\left(\omega - k\vs\right)^2} = -\frac{\rhos}{\rhob}\frac{\qb}{\qs}T(\qs \Rs),
\ee
{\no}where $T(x)={\rm tanh}(x)$ or ${\rm coth}(x)$ for S modes or P modes respectively.

\smallskip
To simplify \equ{dispersion_slab}, we again move into the frame where 
the background is static and the slab velocity is $\vs=V$, and 
rewrite the equation in unitless form, using \equ{unitless} and 
\be 
\label{eq:Xdef}
K=k\Rs . 
\ee
{\no}The result is 
\be
\label{eq:dispersion_slab_unitless}
Z = T\left(\left[1-\delta \Mb^2(\varpi-1)^2\right]^{0.5}K\right),
\ee
{\no}where $Z$ is defined in \equ{Z_def}. This should be compared to the 
dispersion relation for the sheet, $Z=1$.

\smallskip
The dispersion relations for S and P-modes can be written as a single equation 
by inverting \equ{dispersion_slab_unitless} and writing $K$ as a function 
of $\varpi$
\begin{subequations}
\label{eq:inverted_dispersion_slab}
\begin{equation}
\label{eq:inverted_dispersion_slab_a}
K = 0.5\left[1-\delta \Mb^2(\varpi-1)^2 \right]^{-0.5} \left(\alpha +i\beta\right)
\end{equation}
\begin{equation}
\alpha = {\rm ln}\left(|1+Z|\right)-{\rm ln}\left(|1-Z|\right)
\end{equation}
\begin{equation}
\beta = {\rm arg}\left(1+Z\right)-{\rm arg}\left(1-Z\right)+n\pi
\end{equation}
\end{subequations}
{\no}where $n$ is any whole number, odd for P-modes and even 
for S-modes, and ${\rm arg}\left(1\pm Z\right)$ is between 
$-\pi$ and $\pi$ due to our chosen branch cut (see \se{branch_cut}). 

\smallskip
\Equs{dispersion_slab}, \equref{dispersion_slab_unitless} and \equref{inverted_dispersion_slab} 
can be used interchangeably as the dispersion relation for the compressible 
slab. From \equs{inverted_dispersion_slab}, we learn that the slab solutions 
exhibit a qualitatively different behaviour than the sheet, for two reasons. 
Firstly, in the sheet, $\varpi$ was independent of $k$ which resulted in the 
scaling $\omega\propto k$. On the other hand, in the slab, $\varpi$ depends 
explicitly on $K$. This is due to the additional length scale in the problem, 
the slab width, and will lead to a non-trivial dependence of the growth rate 
on wavenumber. Secondly, while \equ{dispersion_sheet} admitted only one solution 
for the growing mode $\omega(k)$, in slab geometry there can be an infinite number 
of modes for a fixed wavenumber $k$, each corresponding to a different value of 
$n$ in \equs{inverted_dispersion_slab}, arising from the periodicity of tanh for 
complex arguments. We will discuss this in detail in the following sections, where 
we begin by examining various limits of the dispersion relation.

\subsubsection{Incompressible Limit}
\label{sec:incomp_slab}

\smallskip
In the incompressible limit, when in \equ{q_bs} $\qbs\rightarrow k$, 
the slab dispersion relation, \equ{dispersion_slab}, reduces to 
\be 
\label{eq:dispersion_slab_incomp} 
\hspace{-0.1cm}
\omega_{\pm} = \frac{\rhob \vb + T(K)\rhos \vs}{\rhob + T(K)\rhos}k \pm i \frac{\sqrt{T(K)\rhos \rhob}|\vs-\vb|}{\rhob + T(K)\rhos}k.
\ee
{\no}It is straightforward to see that this converges to \equ{KKHI} 
for short wavelengths, $K>>1$. In practice, convergence is achieved 
for wavelengths $\lambda\lsim 3\Rs$. At long wavelengths, $K<<1$, both 
modes have $\omega \propto k^{1.5}$, meaning that the growth rate for 
the slab decays more rapidly than for the sheet as $k\rightarrow 0$. 
The dashed lines in \fig{numerical_solution} show the growth rates 
(${\rm Im}(\omega)$, left) and oscillation frequencies (${\rm Re}(\omega)$, 
right) as a function of $K$ for the incompressible slab with $\delta=100$, 
in comparison to the compressible slab discussed below. 

\subsubsection{Long Wavelength Limit}
\label{sec:long_slab}

\smallskip
In the long wavelength limit, as $K\rightarrow 0$, we show in appendix 
\se{long_comp_slab} that $\qbs \rightarrow 0$ as well. Therefore, 
${\rm tanh}(\qs \Rs)\simeq 1/{\rm coth}(\qs\Rs) \simeq \qs \Rs$. 
This can be used to simplify \equ{dispersion_slab_unitless} and 
expand $\varpi$ in a power series in $K$. The result is that both 
S and P-modes are unstable at long wavelengths for any $\Mb$ and 
$\delta$. For this reason, the long wavelength modes are referred 
to as \textit{fundamental modes}. To leading order in $K$, the 
dispersion relations for the S and P-modes are (see \se{long_comp_slab} 
for the derivation) 
\begin{subequations}
\label{eq:fundamental_slab}
\begin{equation}
\label{eq:fundamental_slab_a}
\begin{array}{c c}
\varpi_{\rm S,\,f} \simeq \delta K \pm i(\delta K)^{1/2} &{\rm (S)},
\end{array}
\end{equation}
\begin{equation} 
\label{eq:fundamental_slab_b}
\begin{array}{c c}
\varpi_{\rm P,\,f} \simeq 1 \pm i\delta^{-1/2}(1-\Mb^2)^{-1/4}K^{1/2} &{\rm (P)}.
\end{array}
\end{equation}
\end{subequations}

\smallskip
As $K\rightarrow 0$, the fundamental S and P-modes approach $\varpi=0$ 
and $1$ respectively, which result in $(1+Z)/(1-Z)=1$ and $-1$. These 
modes thus correspond to $n=0$ and $n=-1$ in \equ{inverted_dispersion_slab}. 

\smallskip
It is instructive to compare these solutions to the long-wavelength 
limit of the incompressible slab (\equnp{dispersion_slab_incomp}). 
To leading order in $K$, the fundamental S-mode is identical to the 
incompressible case. Corrections dependent on Mach number are all 
higher order in $K$. 
On the other hand, the growth rate of the fundamental P-mode is 
multiplied by a factor\footnote{In the special case of $\Mb=1$, 
the fundamental P-mode has a slightly different form, where 
${\rm Im}(\omega)\propto k^{7/5}$. See \se{long_comp_slab} for 
details.}  $(1-\Mb^2)^{-1/4}$ compared to the incompressible 
case. For $\Mb<<1$, the compressible growth rate is enhanced 
by a factor $\sim (1+0.25\Mb^2)$, while for $\Mb>>1$ it is 
suppressed by a factor $\sim \sqrt{2\Mb}$. Thus, for sufficiently 
high Mach numbers the instability is suppressed, in qualitative 
similarity to the compressible sheet. 

\subsubsection{Short Wavelength Limit}
\label{sec:short_slab}

\smallskip
At short wavelengths, $K>>1$, the slab solution converges to the sheet solution, 
but it does so in different ways depending on the Mach number. We summarize the 
main points below, providing more details in \se{short_comp_slab}. We begin by 
searching for solutions to \equs{inverted_dispersion_slab} where $K\rightarrow \infty$ 
and the right-hand-side of \equ{inverted_dispersion_slab_a} is real, since $K$ 
is real by definition in the temporal stability analysis we are performing.

\smallskip
At low Mach numbers, $\Mb<<1$, when the sheet is unstable, we have 
${\rm Im}(\qs)<<{\rm Re}(\qs)$, and the solution is given by $Z=1$, 
which is the dispersion relation for the sheet (\equnp{Z_def}). These 
are \textit{surface modes}, decaying exponentially with depth in the 
slab. It is unsurprising that such modes resemble the sheet, since in 
the limit $\lambda << R_s$ we expect the perturbations not to be affected 
by the slab geometry. 

\smallskip
At high Mach numbers, $\Mb>M_{\rm crit}$, when the sheet is 
stable, we have ${\rm Im}(\qs)>>{\rm Re}(\qs)$, and the asymptotic 
solution is given by $1-\delta \Mb^2(1-\varpi)^2 = 0$. This leads to 
$\varpi = 1 - \Ms^{-1} = \varpi_\infty$, which is the high-Mach 
number (stable) limit of the growing mode in the sheet 
(\equnp{phi_infiity}). Physically, modes with ${\rm Im}(\qs)>>{\rm Re}(\qs)$ 
are \textit{body modes}, which traverse the width of the slab 
without decaying, and bring the two interfaces into causal contact. 
As $\varpi\rightarrow \varpi_\infty$, $Z$ goes to zero (\equnp{Z_def}), 
so \equs{inverted_dispersion_slab} can be expanded to derive an expression 
for the asymptotic transverse wavenumber within the slab 
\be 
\label{eq:standing2}
\qs \Rs = \left[1-\delta \Mb^2(\varpi-1)^2\right]^{1/2} K \simeq i\frac{(n+2)\pi}{2}.
\ee
{\no}Note that there is an extra $2\pi$ here compared to \equs{inverted_dispersion_slab}, 
because as $K$ is increased from $0$ to $\infty$, $(1+Z)/(1-Z)$ completes a full 
revolution about the origin in the complex plane, while its argument is defined 
in the range $(-\pi,\pi)$ (\se{Marginal2}). These represent standing waves within 
the slab, with wavelengths $\lambda_{\perp} = 4\Rs/(n+2)$. So the $n$-th mode has 
$n+1$ nodes across the slab width of $2\Rs$, which can be seen qualitatively in 
\fig{P1_slab_fig}. The slab acts as a waveguide for these modes, each of which have 
phase velocity $\omega/k = V-\cs$. This is a qualitatively new phenomenon compared 
to the sheet. In \se{reflected}, we show that each of these modes is unstable at 
finite wavelengths, and characterize the instability. Therefore, while each individual 
mode (each individual $n$) converges to the vortex sheet solution at short wavelengths 
for all Mach numbers, the appearance of higher order unstable modes at shorter and 
shorter wavelengths renders the slab unstable at all Mach numbers, unlike the sheet 
which is stable at high $\Mb$. 

\subsubsection{Unstable Body Modes}
\label{sec:reflected}

\smallskip
At long wavelengths, solutions to the dispersion relation (\equsnp{inverted_dispersion_slab}) 
exist only for $n=-1,\,0$. In the incompressible limit, $\Mb<<1$, these are the only two 
solutions at any wavelength. However, at high Mach numbers, there are an infinite number of 
\textit{body mode} solutions at short wavelengths. The questions we need to address are when 
do these modes appear, whether they are unstable, and what their growth rate is. These are 
answered in detail in \se{Marginal1} - \se{Marginal4}, and we 
summarize the main results below.

\smallskip
From \equs{Z_def} and \equref{inverted_dispersion_slab}, we see that when $\varpi=0$ the 
wavenumber of the $n$-th mode is $K_{\rm n,0}=n\pi/(2\sqrt{\Ms^2-1})$. Since $K$ 
must be real, such solutions are only possible for $n\ge 1$ if $\Ms>1$. This was 
incorrectly identified by previous authors \citep[e.g.][]{Payne_Cohn85,
Hardee_Norman88a} as the condition for unstable body modes. However, for every 
$n\ge 1$ the solution $(\varpi,K)=(0,K_{\rm n,0})$ is \textit{stable}, meaning 
that solutions to the dispersion relation with $K\gsim K_{\rm n,0}$ have real 
$\varpi$. A necessary and sufficient condition for body modes to be unstable 
is not $\Ms>1$, but rather (\se{Marginal1}) 
\be 
\label{eq:reflected_conditon}
M_{\rm tot} = \frac{V}{\cs+\cb} = \frac{\sqrt{\delta}}{1+\sqrt{\delta}}\Mb>1.
\ee
{\no}If $M_{\rm tot}<1$, only the fundamental modes with $n=-1,\,0$ are 
unstable, and these modes are surface modes\footnote{Note that when $\delta<<1$, 
which was the regime studied by \citet{Payne_Cohn85,Hardee_Norman88a}, 
$M_{\rm tot}\sim \Ms$.}.

\smallskip
When $M_{\rm tot}>1$, the smallest unstable wavenumber (corresponding to 
the longest unstable wavelength, and hereafter referred to as marginal 
stability) for the $n$-th body mode is well approximated by (\se{Marginal2}) 
\be 
\label{eq:marginally_stable_text}
K_{\rm n} \simeq \frac{n \pi}{2\sqrt{\delta(\Mb-1)^2-1}}.
\ee 

\smallskip
Defining $\kappa\equiv K-K_{\rm n}$, the growth rate of the $n$-th body mode 
near marginal stability scales as (\se{Marginal3})
\be 
\label{eq:marginal_growth1}
{\rm Im}(\varpi_{\rm n})\propto \delta^{1/4} \Mb^2 n^{-3/2} \kappa^{1/2}.
\ee
{\no}This growth rate diverges strongly with Mach number, which is in contrast to 
the fundamental modes. Recall that near marginal stability at $K=0$, the growth rate 
of the fundamental S-mode was independent of $\Mb$, while the growth rate of the 
fundamental P-mode scaled as $\Mb^{-1/2}$ (\equsnp{fundamental_slab}). 

\begin{figure*}
\centering
\subfloat{\includegraphics[trim={0.3cm 0.2cm 1.0cm 0.3cm}, clip, width =0.475 \textwidth]{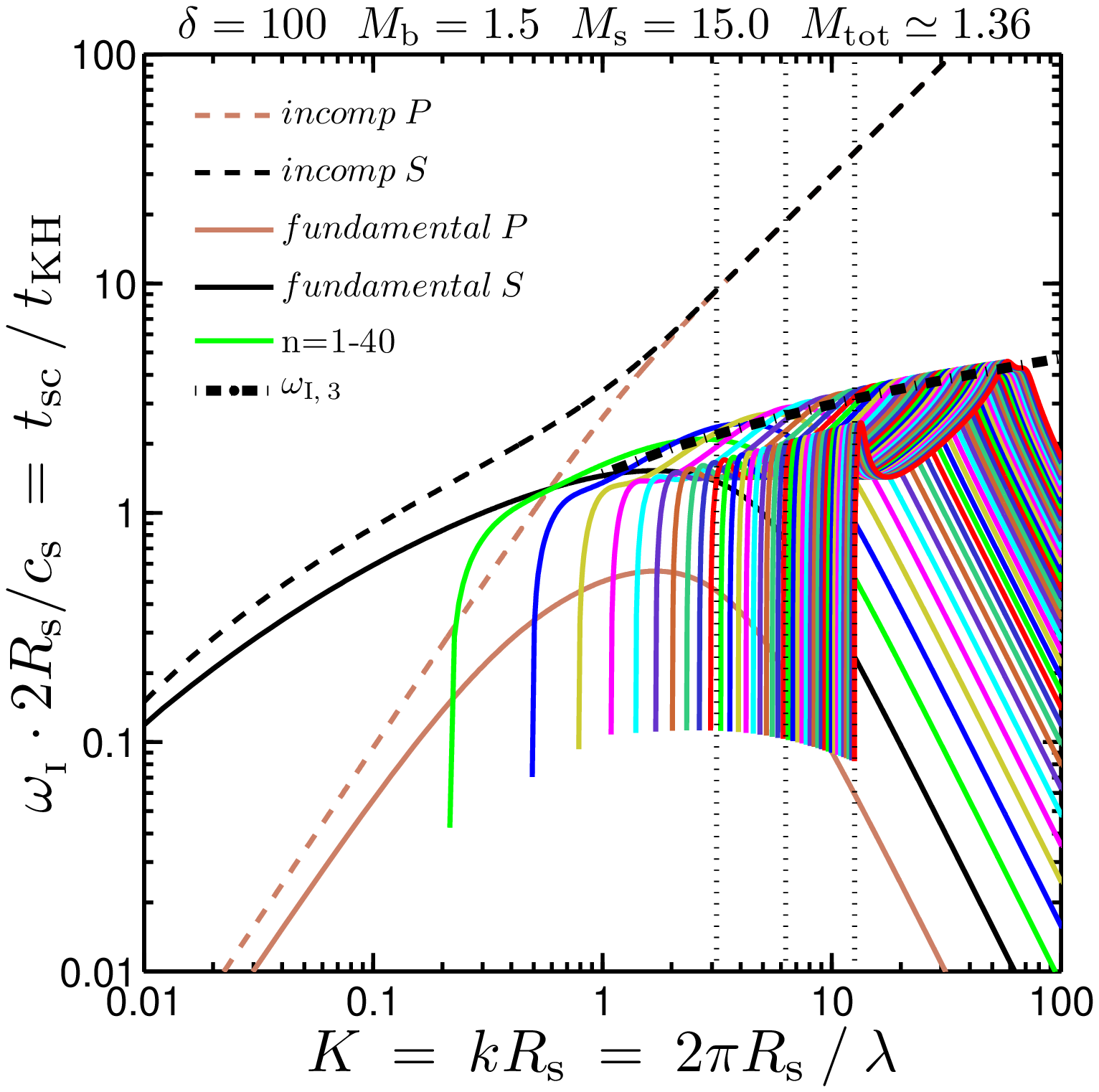}}
\subfloat{\includegraphics[trim={0.3cm 0.2cm 1.0cm 0.3cm}, clip, width =0.475 \textwidth]{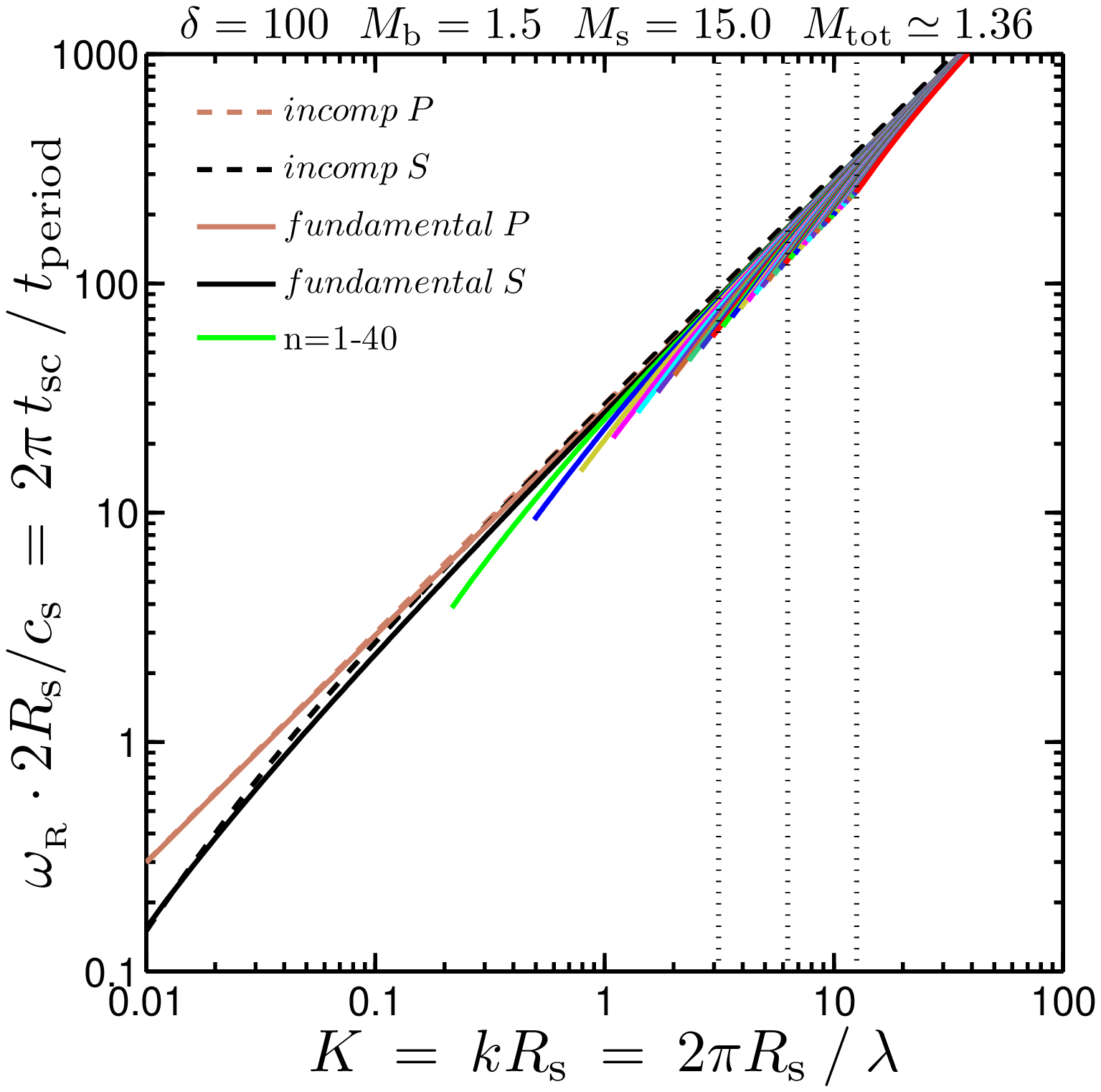}}
\caption{Numerical solution to the slab dispersion relation, \equ{dispersion_slab_unitless}, 
for $\Mb=1.5$ and $\delta=100$. For this choice of $\Mb$ and $\delta$, the sheet is stable 
(\fig{planar_growth_rate}), while the slab is unstable through body modes. The left-hand 
panel shows the growth rate, $\Oi$, normalized by the inverse of the sound crossing time 
in the slab, $\tsc^{-1}=\cs/(2\Rs)$, i.e. the ratio of the sound crossing time to the 
Kelvin-Helmholtz time. The right-hand panel shows the oscillation frequency of the wave, 
$\Or$, normalized by $\tsc^{-1}$. The solid black and beige lines show the fundamental 
(compressible) S and P-modes respectively, while the dashed lines show the corresponding 
solutions for the incompressible slab (\equnp{dispersion_slab_incomp}). The coloured lines 
show the $n=1-40$ modes ($n=1$ in green, $n=2$ in blue, and so on, with even/odd $n$ 
representing S/P-modes respectively). At long wavelengths, the two fundamental modes are 
similar to their incompressible counterparts. However, at $K>>1$ the growth rates for the 
fundamental modes decay while the incompressible modes diverge as $\Oi\propto k$. Modes 
with $n\ge 1$ are excited at finite wavenumbers that scale linearly with $n$, reach a 
maximum growth rate at resonance, and then decay at large $K$. The vertical dotted lines 
mark, from left to right, wavelengths of $\lambda=2\Rs$, $\Rs$ and $0.5\Rs$, where the 
dominant modes are $n=2$, $4$ and $9$ respectively. The ridge line connecting the maximal 
growth rates of each mode diverges logarithmically, and is well fit by $\omega_{\rm I,\,3}$ 
from \equ{resonant_growth_text}, shown by the thick, dash-dotted line. As each mode stabilizes, 
its phase velocity converges to $\Or/k=v-\cs$ (right panel).}
\label{fig:numerical_solution}
\end{figure*}

\begin{figure*}
\centering
\subfloat{\includegraphics[width =0.95 \textwidth]{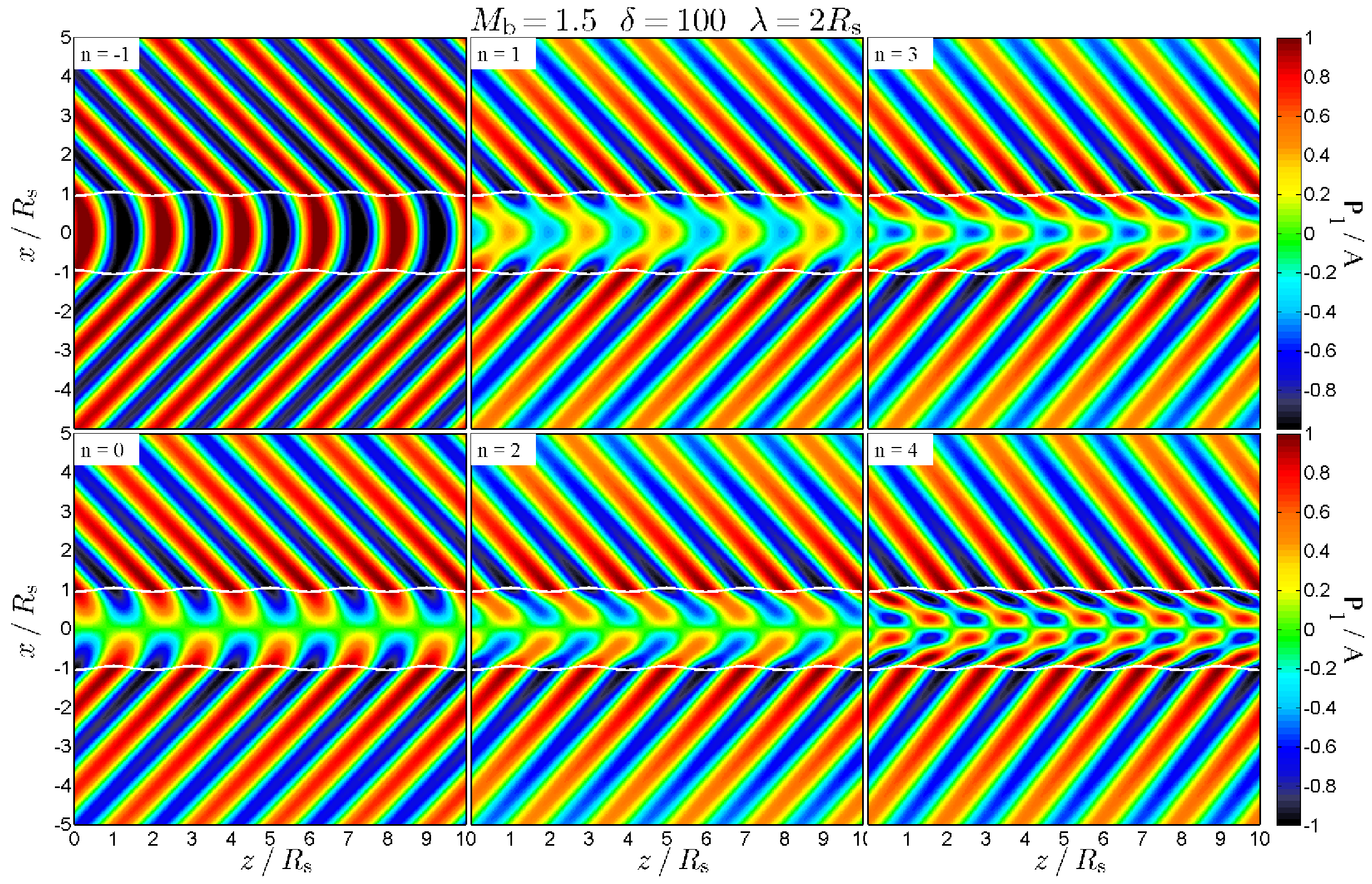}}
\caption{Pressure perturbation in a slab with $\Mb=1.5$ and $\delta=100$ 
normalized by its maximal value at the interfaces, $A$ (\equnp{P1_slab}), for the 
first 6 unstable modes. The longitudinal wavelength of the perturbation 
(along $z$) is equal to the slab diameter, $\lambda=2\Rs$, and the amplitude 
of fluid displacement at the interfaces is $h=0.025\lambda$, which is shown 
by the white curves. The top row shows the first three P-modes: $n=-1$ (the 
fundamental mode, left), $n=1$ (centre) and $n=3$ (right). The bottom row 
shows the first three S-modes: $n=0$ (the fundamental mode, left), $n=2$ 
(centre) and $n=4$ (right). All unstable modes for this case are \textit{body 
modes}, that penetrate to large depths in both the slab and the background. 
The number of transverse nodes within the slab (along $x$) is $(n+1)$, creating 
a more complex standing wave pattern as $n$ increases.}
\label{fig:P1_slab_fig}
\end{figure*}

\smallskip
Since the growth rate of each mode goes to zero as $K\rightarrow\infty$ (\se{short_slab}), 
it must reach a maximum at some intermediate $K$, hereafter the mode \textit{resonance}. 
The resonant wavenumber is well approximated by (\se{Marginal4})
\be 
\label{eq:resonant_X_text}
K_{\rm n,\,res} \simeq \frac{n\pi}{2M_{\rm tot}}.
\ee 
{\no}At a given wavelength, the effective growth rate of the slab is determined by the 
mode with the largest growth rate at that wavelength (see \fig{numerical_solution}). 
This growth rate, $\Oi = {\rm Im}(\omega)$, can be written as the limit of an 
infinite sequence of functions (\se{Marginal4})
\begin{subequations}
\label{eq:resonant_growth_text}
\be
\omega_{_{\rm I,\,1}} = \tsc^{-1}~{\rm ln}\left(4M_{\rm tot}\frac{\sqrt{\delta}}{1+\sqrt{\delta}}K\right)
\ee
\be 
\omega_{_{\rm I,\,j}} = \omega_{_{\rm I,\,1}} - \tsc^{-1}~{\rm ln}\left(\tsc\omega_{_{\rm I,\,j-1}}\right),
\ee 
\end{subequations}
{\no}where $\Oi={\rm lim_{j\rightarrow \infty}}~\omega_{_{\rm I,\,j}}$ and
\be 
\label{eq:tsc}
\tsc=2\Rs/\cs 
\ee
{\no}is the slab sound crossing time. 
In practice, the sequence converges by $j=3$ even for relatively low values of $K$ 
(\fig{ImPres}). In the asymptotic limit $K \rightarrow \infty$, $\Oi \propto {\rm ln}(K)$. 
Hence, the effective growth rate for KHI in a compressible slab diverges logarithmically 
with wavenumber. In contrast, the growth rate for an incompressible sheet or slab diverges 
linearly with wavenumber, $\Oi\propto k$, while the compressible sheet becomes stable at 
high Mach numbers, $\Oi=0$. The scaling of $\Oi\propto {\rm ln}(k)$ for the effective 
growth rate of the compressible slab is in some sense a compromise between these two 
extremes, though recall that each individual mode does stabilize as $k\rightarrow \infty$ 
(\se{short_slab}).

\smallskip
At resonance, $\qb \simeq \qs$ (\se{Marginal4}). This means that the penetration 
depth of the perturbation, $\Delta_{\rm b,s} = 1/{\rm Re}(\qbs)$, is comparable 
in both the background and the slab. Furthermore, the propagation angle of the 
perturbation wave with respect to the normal to slab, given by 
${\rm cot}(\theta_{\rm b,s}) = {\rm Im}(\qbs)/k$, is the same in both media 
(see \fig{P1_slab_fig}, panels marked $n=1$ and $n=2$). At resonance we have 
${\rm sin}(\theta_{\rm b,s})\simeq M_{\rm tot}^{-1}$, commonly referred to as 
\textit{the Mach angle}.

\smallskip
\textbf{The physical origin of body modes in a slab can be understood in the following way.} 
When a perturbation is excited, the two interfaces between the slab and the background initially 
behave as independent sheets, only coming into causal contact once a slab sound crossing time 
has elapsed. At high Mach numbers, surface modes are stable and the perturbation does not grow 
in amplitude, but rather results in acoustic waves propagating between the two interfaces, being 
reflected off of and transmitted through them. The pressure perturbation within the slab 
(\equnp{P1_slab2}) can be written as the sum of an incident and a reflected wave, with wavenumber 
$\qs$, while the pressure perturbation in the background can be thought of as a transmitted wave, 
with wavenumber $\qb$. At certain critical incident angles, there is constructive interference 
between waves emanating from different points along the slab, assumed to have infinite extent. 
It can be shown that the acoustic impedances of the two fluids are equal when $\qb=\qs$, which 
is roughly the case at the resonance of body modes (\se{Marginal4}). The equal impedances cause 
the reflectance and transmission coefficients of the system to diverge, which causes the perturbation 
amplitude to grow. For further details see \citet{Payne_Cohn85} and \citet{Hardee_Norman88a}. 
These authors estimated the resonant growth rates of body modes (which they call \textit{reflected 
modes}) by associating these singularities in the reflectance and transmission coefficients with 
unstable solutions to the dispersion relation, rather than deriving them directly from the dispersion 
relation as we do. 

\subsubsection{Numerical Solution}
\label{sec:slab_summary}

\smallskip
We here summarize, in \fig{numerical_solution} and \fig{P1_slab_fig}, all the features 
of unstable body modes in the slab derived above. \Fig{numerical_solution} shows a numerical 
solution to the slab dispersion relation, \equ{dispersion_slab_unitless}, for $\delta=100$ 
and $\Mb=1.5$. We show as a function of wavenumber, $K=k\Rs$, the growth rate of the 
perturbation, $\Oi = {\rm Im}(\omega) = \tkh^{-1}$ (left), and the oscillation frequency 
of the wave, $\Or = {\rm Re}(\omega) = 2\pi t_{\rm period}^{-1}$ (right). In both cases, 
we normalise the frequency by the inverse sound crossing time in the slab, $\tsc^{-1}$. 
We show the two fundamental modes ($n=-1,0$) and the modes $n=1-40$ (odd/even for P/S 
modes). We also show the solutions for the incompressible slab for comparison. Note that 
for this choice of $\Mb$ and $\delta$, the sheet is stable (\fig{planar_growth_rate}), 
while the slab is unstable due to body modes. 

\smallskip
At long wavelengths, $K<<1$, only the fundamental modes are unstable, and their 
behaviour is similar to the corresponding incompressible solutions. Higher order 
unstable modes are gradually excited at shorter and shorter wavelengths, according 
to \equ{marginally_stable_text}. Each mode reaches a maximal growth rate at a 
resonance wavelength (\equnp{resonant_X_text}), and these dominate over the fundamental 
modes at intermediate and short wavelengths. At short wavelengths, each mode stabilizes 
as $\varpi \rightarrow \varpi_{\rm \infty} = 1-\Ms^{-1}$, so that $\omega \rightarrow (V-\cs)k$. 
However, the ridge line formed by the peak resonant growth rates acts as an effective 
growth rate for the slab, which is always unstable. The effective growth rate of this 
ridge line is well fit by $\Oi\propto \tsc^{-1}{\rm ln}(K)$ (\equnp{resonant_growth_text}).

\smallskip
\Fig{P1_slab_fig} shows the spatial structure of the pressure perturbation 
in the $xz$ plane, $P_1$, for the first 6 unstable modes with $\delta=100$ 
and $\Mb=1.5$. The longitudinal perturbation wavelength is equal to the slab 
diameter, $\lambda=2\Rs$, and the displacement amplitude of the fluid interfaces 
is $h=0.025\lambda$ (shown in white). To compute the transverse wavenumbers, 
$\qbs$, we insert the numerical solutions to the dispersion relation (\fig{numerical_solution}) 
into \equ{q_bs}. The top row shows the first three P-modes: $n=-1$ (the fundamental 
mode), $n=1$ and $n=3$ from left to right. The bottom row shows the first three S-modes: 
$n=0$ (the fundamental mode), $n=2$ and $n=4$ from left to right. Since sheet is stable 
for these values of $\Mb$ and $\delta$, surface modes such as shown in \fig{P1_sheet_fig} 
are stable. All unstable modes, including the fundamentals, are \textit{body modes}, whose 
exponential decay length in the transverse direction is comparable to or larger than the 
slab width. The $n$-th mode has $n+1$ nodes across the slab width, creating a more complex 
standing wave pattern as $n$ increases. For $\lambda=2\Rs$, the $n=1$ and $n=2$ modes are 
near resonance, so the angle of wave propagation is nearly the same in the slab and the 
background. For the other modes, the pattern breaks at the slab interfaces.

\subsection{The Cylindrical stream}
\label{sec:cylinder}

\smallskip
We now consider a flow with cylindrical, rather than planar symmetry. Using 
the standard cylindrical coordinates, $(r,\varphi,z)$, we assume an equilibrium 
configuration where the density and flow velocity depend only on $r$, the flow 
is in the ${\hat{z}}$ direction, and the pressure is constant: 
$\rho_0(r)$, $v_0(r){\hat {z}}$, $P_0$. By rewriting the hydrodynamic equations 
(\equs{continuity} - \equref{energy}) in cylindrical coordinates, inserting 
perturbations of the form $f(r){\rm exp}[i(kz+m\varphi-\omega t)]$ where $m$ is 
an integer, and linearising, we obtain analogous expressions to \equs{rho_1} - \equref{ux} 
that relate the perturbations in density and velocity to the pressure perturbation. 
We also obtain a second order differential equation for the pressure perturbation, 
analogous to \equ{P1}: 
\be 
\label{eq:P1_cyl}
\begin{array}{c}
P_1'' - \left[\dfrac{2v'}{v-\omega/k}+\dfrac{\rho_0'}{\rho_0}-\dfrac{1}{r}\right]P_1' -\\
 \\
k^2\left[1-\left(\dfrac{v-\omega/k}{c}\right)^2+\left(\dfrac{m}{kr}\right)^2\right]P_1=0
\end{array},
\ee
{\no}where $f'=\partial f/\partial r$. This is identical to \equ{P1}, except for the 
geometrical terms $1/r$ and $m/(kr)$. 

\begin{figure}
\centering
\subfloat{\includegraphics[width =0.45 \textwidth]{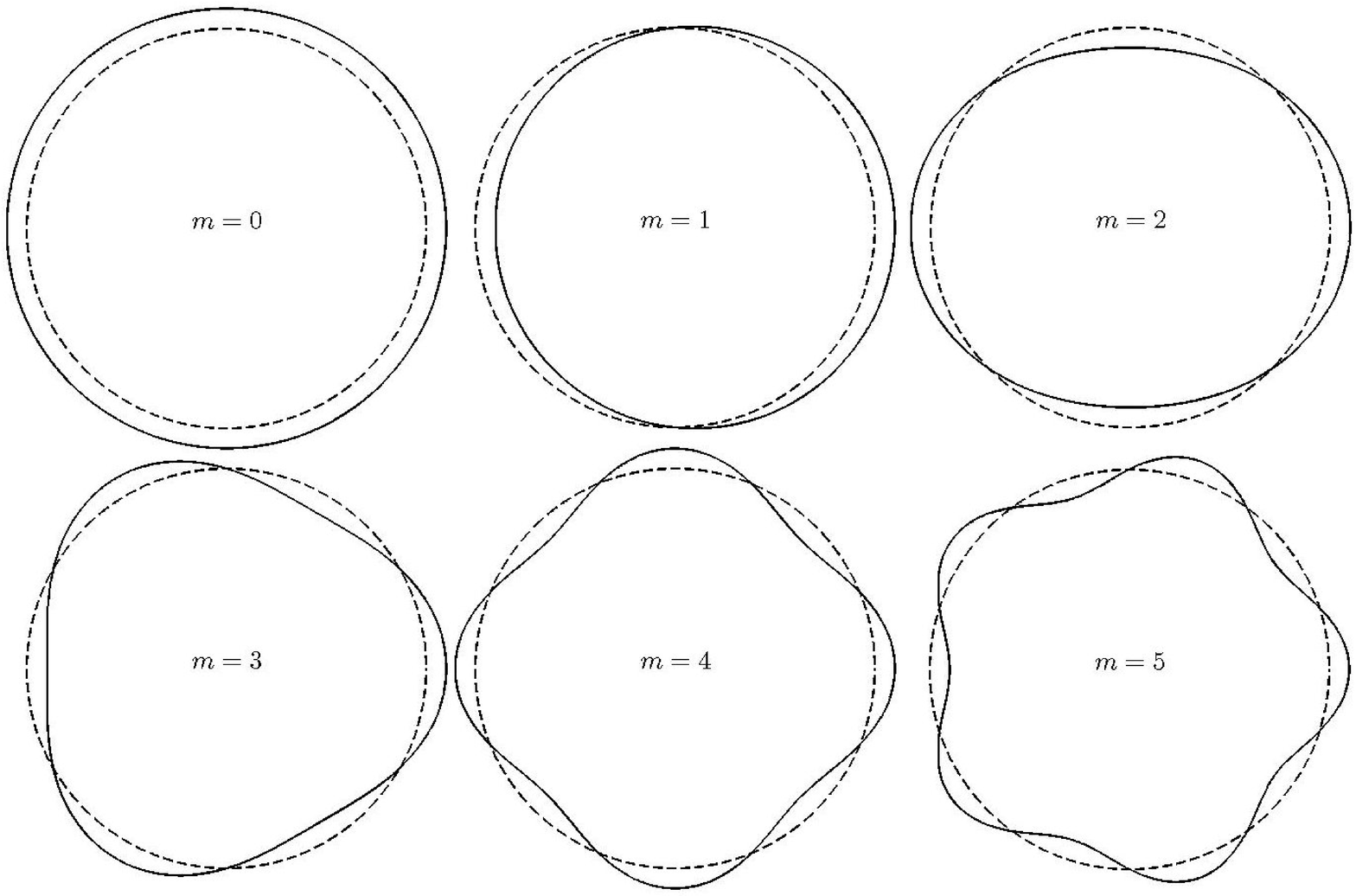}}
\caption{Schematic representation of the first six azimuthal modes for 
a cylindrical stream, $m=0-5$. Shown is a slice through the $xy$ plane. 
The dashed circle in each panel represents the unperturbed cylinder while 
the solid curve represents the perturbed cylindrical surface. The $m=0,1$ 
modes are analogous to the P,S modes in the slab case, respectively.
}
\label{fig:azimuthal} 
\end{figure} 

\smallskip
We consider an infinitely long cylindrical stream of radius $r=\Rs$, centred on 
the $z$ axis, with density and velocity $\rho=\rhos$ and ${\vec{v}}=V{\hat{z}}$. 
The background, at $r>\Rs$, has $\rho=\rhob$ and ${\vec{v}}=0$. \Equ{P1_cyl} 
reduces to two modified Bessel equations, for $r<\Rs$ and $r>\Rs$. Using the boundary 
conditions that the pressure perturbation is finite at $r=0$ and $r\rightarrow \infty$ 
and is continuous at $r=\Rs$, we obtain the solution 
\be 
\label{eq:P1_cyl_sol}
P_1(r)=\left\{\begin{array}{c c}
A~\dfrac{\mathcal{I}_{\rm m}\left(\qs r\right)}{\mathcal{I}_{\rm m}\left(\qs \Rs\right)}& r<\Rs,
\vspace{+2mm}\\
A~\dfrac{\mathcal{K}_{\rm m}\left(\qb r\right)}{\mathcal{K}_{\rm m}\left(\qb \Rs\right)} & r>\Rs.
\end{array}\right.
\ee
{\no}$\mathcal{I}_{\rm m}$ and $\mathcal{K}_{\rm m}$ are the $m$-th order modified Bessel functions 
of the first and second kind respectively and $A$ is a constant of integration. By applying the Landau 
condition at the stream boundary, in analogy to \equ{boundary_h}, we obtain the dispersion relation 
\be 
\label{eq:dispersion_cyl}
Z=-\frac{\mathcal{I}_{\rm m}\left(\sqrt{1-\delta\Mb^2(\varpi-1)^2}K\right)}{\mathcal{I}'_{\rm m}\left(\sqrt{1-\delta\Mb^2(\varpi-1)^2}K\right)}\frac{\mathcal{K}'_{\rm m}\left(\sqrt{1-\Mb^2\varpi^2}K\right)}{\mathcal{K}_{\rm m}\left(\sqrt{1-\Mb^2\varpi^2}K\right)}.
\ee

\smallskip
Comparing \equs{P1_cyl_sol} and \equref{dispersion_cyl} to the corresponding equations 
for the slab, \equs{P1_slab} and \equref{dispersion_slab_unitless}, we see one qualitative 
difference between the two configurations. While the slab admitted only two symmetry 
modes, the symmetric P-modes and the antisymmetric S-modes (\fig{modes}), the cylinder 
admits infinitely many symmetry modes, represented by the index $m$. Through 
the Landau condition, which introduces the perturbation to the cylinder surface, we 
learn that $m$ is the number of azimuthal nodes on this surface. This is shown schematically 
in \fig{azimuthal}, where we show a slice through the $z=0$ plane for the first six 
symmetry modes, $m=0-5$. The $m=0$ modes are axisymmetric \textit{pinch modes}, analogous 
to the P-modes in the slab. The $m=1$ modes are antisymmetric \textit{helical modes}, 
analogous to S-modes in the slab. Modes with $m>1$ are \textit{fluting modes} with no 
direct analogue in the slab, but as we shall see they do not qualitatively change the 
growth of instabilities at short wavelenghts.

\subsubsection{Long Wavelength Behaviour}
\label{sec:long_cyl}

\smallskip
At long wavelengths, $K\rightarrow 0$, we use the asymptotic form 
of the modified Bessel functions for $\xi<<1$ \citep{Bessel} 
\be 
\label{eq:Im_small}
\mathcal{I}_{\rm m}(\xi)\simeq\left\{\begin{array}{c c}
1-0.25\xi^2 & m=0\\
\left(2^m m!\right)^{-1}~\xi^m & m\ge 1,
\end{array}\right.
\ee
\be 
\label{eq:Km_small}
\mathcal{K}_{\rm m}(\xi)\simeq\left\{\begin{array}{c c}
1.27 - {\rm ln}(\xi) & m=0\\
2^{m-1}(m-1)!~\xi^{-m}& m\ge 1.
\end{array}\right.
\ee

\smallskip
Inserting these into \equ{dispersion_cyl} results in the leading order 
dispersion relations
\be 
\label{eq:dispersion_cyl_long}
\varpi \simeq \left\{\begin{array}{c c}
1 + i\dfrac{1}{\sqrt{2\delta}}K\sqrt{{\rm ln}\left(\dfrac{1}{\sqrt{\left|\Mb^2-1\right|}K}\right)} & m=0
\vspace{+2mm}\\
\dfrac{\delta}{1+\delta} + i\dfrac{\sqrt{\delta}}{1+\delta} & m\ge 1.
\end{array}\right.
\ee
{\no}These modes are unstable at all wavenumbers and for all values of $\delta$ and $\Mb$. 
They thus represent the \textit{fundamental modes} for the cylinder. As in the slab, each 
symmetry mode has one fundamental mode. Modes with $m\ge1$ all have the same growth rate 
which is independent of Mach number, similar to the fundamental S-mode in the slab 
(\equnp{fundamental_slab_a}). It is fascinating to note that this is exactly the 
dispersion relation for the incompressible sheet (\equnp{KKHI}). The fundamental 
$m=0$ mode is suppressed at large Mach numbers for a given $K$, qualitatively similar 
to the fundamental P-mode in the slab (\equnp{fundamental_slab_b}). 

\subsubsection{Short Wavelength Behaviour}
\label{sec:short_cyl}
\smallskip
At short wavelengths, when $K >> m+1$, we use the asymptotic form 
of the modified Bessel functions for $|\xi|>>m+1$ \citep{Bessel} 
\be 
\label{eq:Im_large}
\hspace{-4mm}\mathcal{I}_{\rm m}(\xi)\propto\left\{\begin{array}{c c}
\hspace{-2mm}\xi^{-1/2}~e^{\xi} &\hspace{-3mm} {\rm Re}(\xi)>0
\vspace{+2mm}\\
\hspace{-2mm}|\xi|^{-1/2}~{\rm cos}\left(|\xi|-\dfrac{(2m+1)\pi}{4}\right) &\hspace{-3mm} \xi=i|\xi|
\end{array}\right.
\ee
\be 
\label{eq:Km_large}
\hspace{-4mm}\mathcal{K}_{\rm m}(\xi)\propto \xi^{-1/2}~e^{-\xi}.
\ee
{\no}Note that these approximations become valid at shorter wavelengths 
for larger $m$. Inserting these into \equ{dispersion_cyl}, we obtain 
the asymptotic form of the dispersion relation\footnote{Using ${\rm coth}(ix)=-i~{\rm cot}(x)$ 
and ${\rm tanh}(x)={\rm coth}(x-i\pi/2)$.}
\be 
\label{eq:dispersion_cyl_short}
\hspace{-2mm}Z \simeq \left\{\begin{array}{c c}
\hspace{-2mm}1 & \hspace{-3mm} {\rm Re}(\qs)>>{\rm Im}(\qs)\hspace{-3mm}
\vspace{+2mm}\\
\hspace{-2mm}{\rm tanh}\left(\qs\Rs-i\dfrac{(2m-1)\pi}{4}\right)&\hspace{-3mm} {\rm Re}(\qs)<<{\rm Im}(\qs)\hspace{-3mm}
\end{array}\right.
\ee

\smallskip
As for slab and sheet geometries, ${\rm Re}(\qs)>>{\rm Im}(\qs)$ at low Mach 
numbers, when surface modes are unstable. In this case, the dispersion relations 
for the sheet, the slab and the cylinder all converge to $Z=1$. At higher 
Mach numbers when surface modes are stable, ${\rm Re}(\qs)<<{\rm Im}(\qs)$, 
and the dispersion relation for the cylinder becomes very similar 
to that of the slab. Except for the extra $-i\pi/4$ in the argument 
of the \textbf{${\rm tanh}$}, \equ{dispersion_cyl_short} with $m=0$ 
and $m=1$ corresponds exactly to \equ{dispersion_slab_unitless} for 
P and S modes respectively. We conclude that for any given $m$, at 
short enough wavelengths, $\lambda<<2\pi\Rs/(m+1)$, shorter than any 
features on the cylinder surface, we are not sensitive to the geometry 
(planar or cylindrical) and the dispersion relation for the cylinder 
converges to that of the slab. 

\smallskip
Given the similarity of \equ{dispersion_cyl_short} and \equ{dispersion_slab_unitless} 
for the short wavelength behaviour of body modes, we can apply our analysis of the slab 
(\se{reflected} and \se{Marginal1} - \se{Marginal4}) to the 
cylinder as well. It is straightforward to see that our results for the marginally 
stable wavenumbers (\equnp{marginally_stable_text}), the resonant wavenumbers 
(\equnp{resonant_X_text}), and the assymptotic transverse wavelength across the stream 
(\equnp{standing2}), can all be applied to the cylinder under the transformation 
$n_{\rm slab} \rightarrow 2n_{\rm cyl}+m-1/2$. As in the slab case, the mode number $n$ 
represents the number of nodes of the perturbation along the stream width (\fig{P1_slab_fig}). 
Most importantly, the effective growth rate of instabilities in the stream, given by the 
maximal growth rate at each wavenumber (\equsnp{resonant_growth_text}), is the same in both 
the slab and the cylinder.

\begin{figure*}
\centering
\subfloat{\includegraphics[width =0.95 \textwidth]{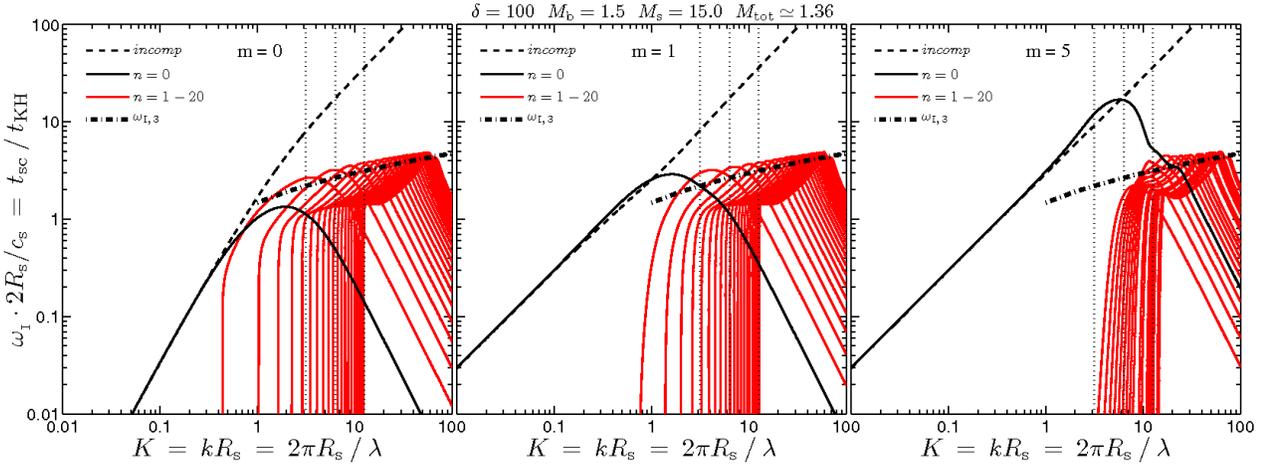}}
\caption{Growth rates for the cylinder. Shown are numerical solutions to the cylinder 
dispersion relation, \equ{dispersion_cyl}, for $\Mb=1.5$ and $\delta=100$, the same 
case shown in \fig{numerical_solution} for the slab. The $y$ axis shows the growth rate, 
$\Oi$, normalized by the inverse of the sound crossing time in the stream, $\tsc^{-1}=\cs/(2\Rs)$, 
as in the left-hand panel of \fig{numerical_solution}. The $x$ axis shows the normalized 
wavenumber $K=k\Rs$. We show solutions for $m=0$ (left), $m=1$ (centre), and $m=5$ (right). 
Dashed black lines show the incompressible solutions. Solid black lines show the fundamental 
modes, $n=0$ (one mode for each $m$). Solid red lines show the $n=1-20$ modes for each $m$. 
Surface modes, where $\Oi \propto K$, are unstable at wavenumbers $K\lsim m+1$ and the 
corresponding growth rates are very similar to the incompressible case. At shorter wavelengths, 
body modes dominate the instability, and the effective growth rate due to the ridge line connecting 
the mode resonances is very similar to the slab case, well fit by $\omega_{\rm I,\,3}$ from \equ{resonant_growth_text} (thick dash-dotted line). Modes with $m>1$ are expected to be stable 
for most physical scenarios (see text), so we expect body modes to dominate the instability for 
real streams. 
}
\label{fig:numerical_cyl} 
\end{figure*} 

\smallskip
\Fig{numerical_cyl} shows a numerical solution to the cylindrical dispersion relation, 
\equ{dispersion_cyl}, for $\delta=100$ and $\Mb=1.5$. We show as a function of wavenumber, 
$K=k\Rs$, the growth rate of the perturbation, $\Oi = {\rm Im}(\omega) = \tkh^{-1}$ 
normalised by the inverse sound crossing time in the stream, $\tsc^{-1}$. This can 
be directly compared to the corresponding solution for the slab case, shown in the 
left-hand panel of \fig{numerical_solution}. The three panels address different 
azimuthal modes, $m=0,1$ and $5$ as marked. For each $m$ we show the fundamental mode, 
$n=0$, and the modes $n=1-20$, together with the corresponding incompressible solution. 
At long wavelengths, $K<m+1$, only the fundamental modes are unstable, and their 
behaviour is similar to the corresponding incompressible solutions. Higher order 
unstable modes are gradually excited at shorter and shorter wavelengths, and their 
overall behaviour is similar to the slab case. Most importantly, the effective growth 
rate for the cylinder at short wavelengths, defined by the ridge line of peak growth rates 
of each $n\ge1$ mode, is well fit by the same formula as for the slab, namely 
$\Oi\propto \tsc^{-1}{\rm ln}(K)$ (\equnp{resonant_growth_text}).

\subsubsection{Surface Modes vs. Body Modes}
\label{sec:reflected_cyl}

\smallskip
In planar geometry, surface modes become stable when $\Mb>M_{\rm crit}$ (\equnp{Mcrit}), 
while body modes in the slab become unstable when $M_{\rm tot}>1$ (\equnp{reflected_conditon}). 
However, recall that the Mach number was defined using only the component of the velocity 
parallel to the perturbation wavevector, $v_k = {\vec {v}}\cdot{\hat {k}}$ (\equnp{P1}). 
On the surface of the cylinder, the wavevector is ${\vec {k}}=k{\hat {z}}+(m/R){\hat {\varphi}}$, 
resulting in $v_k = V[1+(m/K)^2]^{-1/2}$. Therefore, the effective value of $\Mb$ which is 
relevant for determining whether surface modes are stable, is reduced by a factor 
$[1+(m/K)^2]^{-1/2}$, which depends both on the azimuthal wavenumber $m$, and on the 
perturbation wavelength through $K$. As a result, at a given wavenumber surface modes 
will be unstable for azimuthal modes $m>K[(\Mb/M_{\rm crit})^2 -1]^{1/2}$. This means 
that surface modes are formally \textit{always} unstable for the cylinder, whatever the 
value of $\Mb$, for large enough $m$. While this may seem fundamentally different from 
the slab, it is actually very similar. Recall that we limited our analysis in \se{sheet} 
and \se{slab} to perturbations where ${\vec {k}}||{\vec {v}}$, so that $v_k=V$. In principle, 
perturbations in a slab can assume any angle $\varphi$ with respect to the flow velocity, 
and surface modes will be unstable so long as ${\rm cos}(\varphi)<M_{\rm crit}/\Mb$. 

\smallskip
However, surface tension can stabilize modes with $m\ge 2$, where the stream surface 
is highly perturbed with many small scale features and the surface to volume ratio is high 
(see \fig{azimuthal}). Previous studies have found that the inclusion of magnetic fields parallel 
to the flow, which act as a form of surface tension, stabilizes surface modes with $m\ge 2$ 
\citep[e.g][]{Ferrari81,Birkinshaw90}. Furthermore, perturbations with $m=0-2$ are likely the 
dominant modes in cold streams in galactic haloes, plausibly seeded by gravitational tidal 
interactions with satellite galaxies located either inside or outside the streams. We will 
therefore focus hereafter on low-$m$ modes, where body modes dominate the instability for 
wavelengths $\lambda \lsim Rs$. 

\section{Simulation Results} 
\label{sec:sim} 

\smallskip
In this section we use numerical simulations to study the growth of perturbations 
due to KHI in the linear regime. Guided by \se{linear}, we do this in two stages. 
First, we study the evolution of \textit{eigenmode} perturbations, whose initial 
spatial structure obeys the linearized equations of hydrodynamics as derived in 
\se{linear}. This corresponds to a pressure perturbation obeying \equ{P1_sheet} 
for a sheet, \equs{P1_slab} and \equref{P1_slab2} for a slab, or \equ{P1_cyl} for a cylinder, 
together with perturbations in the density and velocity obeying \equs{rho_1} - \equref{ux} 
in planar geometry, or the corresponding equations in cylindrical geometry. 
In the second stage, we study the evolution of arbitrary \textit{non-eigenmode} 
perturbations.

\subsection{Numerical Method} 
\label{sec:Numeric} 
\smallskip
We use the Eulerian code \texttt{RAMSES} \citep{Teyssier02}, with a piecewise-linear 
reconstruction using the MonCen slope limiter \citep{vanLeer77} and a HLLC approximate 
Riemann solver \citep{Toro94}. Since perturbations with wavelengths $\lambda \lsim \Rs$ 
in a slab and a cylinder should behave similarly in the linear regime, which is what 
interests us here, we limit our current analysis to 2D slab simulations. This allows 
us to achieve higher resolution than would be possible in 3D simulations of cylindrical 
streams. As highlighted below, the slab simulations also allow us to test our predictions 
for the sheet.

\smallskip
The simulation domain is a square of side $L=1$, representing the $xz$ plane, extending 
from $0$ to $1$ in the $z$ direction and from $-0.5$ to $0.5$ in the $x$ direction. 
The slab is centred at $x=0$ with a radius of $\Rs=1/160$, and extends the full domain 
in the $z$ direction. We use periodic boundary conditions at $z=0$ and $1$, and outflow 
boundary conditions at $x=\pm 0.5$ (such that gas crossing the boundary is lost from the 
simulation domain). The slab and the background are both ideal gasses with adiabatic 
index $\gamma=5/3$, and initial uniform pressure $P_0=1$. The background, at $|x|>\Rs$, 
is initialized with density $\rhob=1$ and velocity ${\vec {v}}_{\rm b,\,0}=0$. The slab, 
at $|x|<\Rs$, is initialized with $\rhos=\delta$ and ${\vec {v}}_{\rm s,\,0}=\Mb\cb{\hat {z}}$, 
where $\cb$ is the sound speed in the background, $(5/3)^{1/2}$ in simulation units. We 
simulate several different combinations of $\delta$ and $\Mb$ (\tab{sims}). 

\smallskip
A characteristic time common to all our simulations is the sound crossing time in the 
background, $T_{\rm box}=L/\cb\sim 0.775$ in our simulation units. The characteristic 
time for growth of perturbations in the linear regime is the \textit{Kelvin-Helmholtz 
time}, $\tkh$. This is the inverse of the imaginary part of the frequency, 
$\tkh = \Oi^{-1} = (kV\Pi)^{-1}$, and can be expressed as 
\be 
\label{eq:tkh}
\tkh = \left[2K\delta^{1/2}\Mb~\Pi\right]^{-1}\tsc,
\ee
{\no}with $\tsc=2\Rs/\cs$ the sound crossing time in the slab (\equnp{tsc}). We run each simulation 
for at least $5\tkh$ with 20 outputs per $\tkh$. Our use of a thin slab with $\Rs<<L$ ensures 
that $\tkh<<T_{\rm box}$ in all of our simulations (\tab{sims}). The boundary conditions at 
$x=\pm 0.5$ are thus unimportant as the boundary and the slab are not in causal contact at 
any point during the simulation.

\smallskip
In the setup described above, the density and velocity are discontinuous at the slab boundaries, 
$x=\pm\Rs$. While this is the case we solved analytically in \se{linear}, such a setup is 
problematic to simulate as it leads to numerical noise at the grid scale\footnote{One source of 
noise comes from trying to capture a sinusoidal shape of the slab boundary using a finite 
Cartesian grid, which leads to inaccuracies on the grid scale.} \citep[e.g][]{Robertson10}, 
causing artificial small-scale perturbations. Since shorter wavelength perturbations grow faster, 
these can quickly dominate over the seeded perturbation. Increasing the resolution decreases the 
wavelengths of the numerical noise and increases its growth rate, thus making the problem worse. 
To get around this, we smooth the density and velocity using a ramp function
\be 
\label{eq:ramp}
\begin{array}{c}
f(x) = f_{\rm b} + 0.25(f_{\rm s}-f_{\rm b}) \times \\
\\
\left[1+{\rm tanh}\left(\dfrac{\Rs-x}{\sigma}\right)\right]\left[1+{\rm tanh}\left(\dfrac{\Rs+x}{\sigma}\right)\right]. 
\end{array}
\ee
{\no}This creates a finite shearing layer between the slab and the background, with a width 
of $\sim 3\sigma$ for $95\%$ convergence, which suppresses the growth of perturbations with 
wavelengths comparable to or smaller than the shearing layer \citep{Robertson10}. In order 
to effectively suppress the grid noise we find that we require $\sigma>2\Delta$, where $\Delta$ 
is the cell size. 

\smallskip
While \texttt{RAMSES} has adaptive mesh refinement (AMR) capabilities, we instead use a 
statically refined grid. The region $|x|<5\Rs$ has the highest resolution, with cell size 
$\Delta$, and the cell size increases by a factor of 2 every $5\Rs$ in the $x$ direction 
until a maximal cell size. In all our simulations, $\Delta=2^{-15}\simeq \Rs/205$, and the 
maximal cell size was $2^{-9}$. 

\smallskip 
In order for our analytical solution of a sharp discontinuity to be valid, the eigenmode 
structure must be well resolved. For surface modes, this means resolving the exponential 
decay length of the perturbation. Since the perturbation decays more rapidly in the denser 
fluid (\fig{P1_sheet_fig}) the eigenmode is resolved if $\sigma<<1/{\rm Re}(\qs)$, which for 
$\delta\sim 10-100$ corresponds to $\sigma<<0.1-0.3~\lambda$. Thus, properly resolving the 
eigenmodes while at the same time suppressing artificial perturbations requires 
$\Delta < 0.5\sigma << 0.1\lambda$. In practice, we find that our results are well 
converged for $\lambda > 30\sigma > 60\Delta$. Body modes are easier to resolve. The smallest 
length scale we must resolve is the transverse wavelength within the slab, 
$2\pi/{\rm Im}(\qs) \sim 4\Rs/(n+2)$ (\equnp{standing2}). For $\lambda \lsim \Rs$, 
$\delta\sim 10-100$ and $\Mb\lsim 2$, typical of cold streams in haloes (\se{tkh}), 
the fastest growing mode has $n\lsim 10$ (\equnp{resonant_X_text}), yielding the 
requirement $\sigma << 0.3~\lambda$. \tab{sims} compares $\Delta$, $\sigma$, and 
$\lambda$ for each simulation.

\subsection{Eigenmode Simulations} 
\label{sec:eigen} 
\smallskip
\textit{Eigenmodes} are simultaneous perturbations of all the fluid variables that 
self-consistently solve the linearised equations of hydrodynamics. To find the true 
eigenmodes of the problem, we would have to insert the smoothed density and velocity 
profiles given by \equ{ramp} into \equ{P1}, solve the ODE to find the form of the pressure 
perturbation, and then insert this into \equs{rho_1} - \equref{ux} to find the 
corresponding perturbations in density and velocity. However, there is no analytic 
solution to \equ{P1} with the profiles given by \equ{ramp}, and this would not offer 
a direct test of the growth rates derived in \se{linear}. Instead, we approximate the 
eigenmodes of the smoothed profile as smoothed versions of the eigenmodes corresponding 
to the ``step-function" profile. More precisely, we take the pressure perturbation given 
by \equs{P1_slab} - \equref{P1_slab2}, and insert this into \equs{rho_1} - \equref{ux} 
assuming constant density and velocity within each region (slab and background) to 
find the corresponding density and velocity perturbations. As these are also discontinuous 
at $x=\pm \Rs$, we smooth them using \equ{ramp} as well. This approximation for the eigenmodes, 
and subsequently the general discussion of a step-function slab, will be judged by how well 
the simulations match the predicted evolution of eigenmodes, namely that they grow exponentially 
in amplitude from $t=0$ while maintaining their spatial structure.

\smallskip
Based on the analysis in \se{linear}, there are three regimes of instability 
for a slab, depending on the values of $M_{\rm tot}$ and $\delta$. To see this, 
it is useful to define 
\be 
\label{eq:Mtc}
M_{\rm tot,\,crit} \equiv M_{\rm crit}\frac{\delta^{1/2}}{1+\delta^{1/2}} = \frac{\left(1+\delta^{1/3}\right)^{3/2}}{1+\delta^{1/2}},
\ee
{\no}where $M_{\rm crit}$ is the critical value of $\Mb$ above which surface modes become stable 
(\equnp{Mcrit}). $M_{\rm tot,\,crit}$ is the corresponding critical value of $M_{\rm tot}$. 
It is straightforward to show that $M_{\rm tot,\,crit}>1$ for any finite $\delta$. The three 
regimes of instability are thus:

\begin{enumerate}
\smallskip
\item If $M_{\rm tot}<1$, then \textit{surface modes are unstable}, 
while \textit{body modes are stable}. The only unstable modes are the two fundamental 
modes, $n=-1,\,0$.

\smallskip
\item If $1<M_{\rm tot}<M_{\rm tot,\,crit}$, then \textit{both surface modes and body 
modes are unstable}. The fundamental modes correspond to unstable surface modes while 
modes with $n\ge 1$ correspond to unstable body modes.

\smallskip
\item If $M_{\rm tot}>M_{\rm tot,\,crit}$, then \textit{surface modes are stable} 
while \textit{body modes are unstable}. All modes including the fundamentals correspond 
to unstable body modes.
\end{enumerate}

\begin{table}
\centering
\textbf{Parameters of the simulations}
\begin{tabular}{ccccccccc}
\hline
$\delta$ & $\Mb$ & $M_{\rm tot}$ & $S/B$ & $n$ & $\lambda/\Rs$ & $\lambda/\sigma$ & $\lambda/\Delta$ & $\tkh/T_{\rm box}$  \\
\hline
\hline
1   & 1.5 & 0.75 & $S$ & 0 & 2 & 102 & 410 & 0.004 \\
10  & 1.5 & 1.13 & $S$ & 0 & 2 & 102 & 410 & 0.009 \\
10  & 1.5 & 1.13 & $B$ & 2 & 2 & 25  & 410 & 0.038 \\
100 & 1.5 & 1.36 & $B$ & 2 & 2 & 102 & 410 & 0.053  \\
1   & 5.0 & 2.50 & $B$ & 4 & 2 & 102 & 410 & 0.006  \\
\hline
\hline
1   & 1.5 & 0.75 & $S$ & -1 & 1 & 102 & 205 & 0.002 \\
10  & 1.5 & 1.13 & $S$ & -1 & 1 & 102 & 205 & 0.004 \\
1   & 5.0 & 2.50 & $B$ & 5  & 1 & 102 & 205 & 0.005  \\
\end{tabular}
\caption{The top five entries correspond to the eigenmode runs (\se{eigen}), 
the bottom three to the non-eigenmode runs (\se{non_eigen}). For each simulation 
we list the values of $\delta$, $\Mb$ and $M_{\rm tot}$; whether it corresponds 
to a surface ($S$) or body ($B$) mode; the mode number, $n$; the ratio of the 
perturbation wavelength, $\lambda$, to the slab radius, $\Rs$, the smoothing 
scale, $\sigma$ (\equnp{ramp}), and the smallest cell size, $\Delta$; and the 
KH time, $\tkh$ (\equnp{tkh}), in units of the box sound-crossing time, $T_{\rm box}$. 
For eigenmodes, $n$ and $\tkh$ correspond to the seeded mode. For the non-eigenmode 
runs, they correspond to the fastest growing mode for the given $\delta$, $\Mb$, 
$\lambda$ and mode symmetry (they are all symmetric P-modes).
}
\label{tab:sims}
\end{table}

\smallskip
To explore the three regimes, we ran a total of five eigenmode simulations. Relevant parameters 
of these simulations are listed in \tab{sims}. The case $(\delta,\Mb)=(1,1.5)$ represents the 
first regime where only surface modes are unstable. The case $(\delta,\Mb)=(10,1.5)$ represents 
the second regime where both surface and body modes are unstable, and we simulate one of each: 
the $n=0$ surface mode and the $n=2$ body mode. The cases $(\delta,\Mb)=(100,1.5)$ and 
$(\delta,\Mb)=(1,5.0)$ represent the third regime where only body modes are unstable, and 
we simulate the $n=2$ and $n=4$ modes respectively\footnote{The first of these two cases 
has $\Mb$ very close to $M_{\rm crit}$ (see \fig{planar_growth_rate}) while the second is deep 
within the third regime.}. Note that all simulated modes correspond to S-modes with even mode 
number, $n$. We normalize all perturbation amplitudes by setting $A=0.05$ in \equs{P1_slab} - \equref{P1_slab2} 
and set the perturbation wavelength equal to the slab width, $\lambda=2\Rs$, yielding $K=\pi$. 

\smallskip
The predicted complex frequency corresponding to each simulated mode, $\varpi$, is found 
by numerically solving the dispersion relation, \equ{dispersion_slab_unitless}. For surface 
modes, these are nearly identical to the corresponding values in the sheet given by \equ{unitless_dispersion_sheet_2}. The Kelvin-Helmholz time, $\tkh$, is then calculated 
using \equ{tkh}, and listed in \tab{sims}. Each simulation was run until time $t=5\tkh$, 
which corresponds to a different time for each mode. \tab{sims} also shows the ratio of 
the wavelength to the smoothing scale, $\sigma$ from \equ{ramp}. In most cases we use 
relatively narrow smoothing, $\sigma=\lambda/102$, in order not to deviate too far from 
the step-function slab. However, for the $\delta=10$ body mode we require a larger smoothing 
scale, $\sigma=\lambda/25$, to suppress artificial surface modes which have much faster growth 
rates than the body mode. This is less of an issue in the third regime where surface modes 
are intrinsically stable.

\begin{figure}
\centering
\subfloat{\includegraphics[trim={0.4cm 0.2cm 0.4cm 1.04cm}, clip, width =0.45 \textwidth]{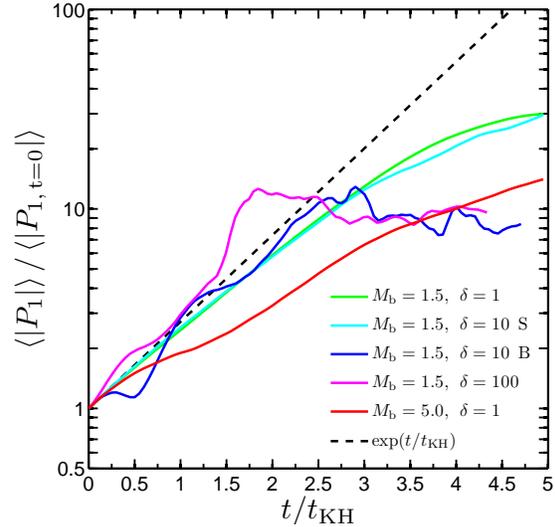}}
\caption{Growth rate of eigenmode perturbations in the numerical simulations, 
represented by the five solid curves. The x axis shows time normalized by the 
respective KH time for each mode according to \tab{sims}. Each simulation was 
run for $5\tkh$, corresponding to a longer time for higher $\delta$ and for body 
modes compared to surface modes. The y axis shows the pressure perturbation 
amplitude normalized by its value at $t=0$. The dashed line marks the expected 
exponential growth. All simulations match the analytically predicted growth rate 
to within $\sim 20\%$ for a period of between $\sim 1.5-3\tkh$, after which the 
growth rate saturates due to non-linear effects.
}
\label{fig:eigen} 
\end{figure}

\smallskip
In each snapshot, we estimate the perturbation amplitude by calculating the average of 
$|P_1|=|P-P_0|=|P-1|$ in the high resolution region, $|x|<5\Rs$. We experimented with 
varying the region within which we average the perturbation between $|x|<\Rs$ and $|x|<5\Rs$, 
calculating the root-mean-squared value of $P_1$ rather than the averaged absolute value, 
and using the transverse velocity or the displacement of the slab interface\footnote{This 
was calculated by using a passive scalar, called ``color", to differentiate the slab material 
from the background.} rather than the pressure to estimate the perturbation amplitude. These 
variations change our growth rates by less than $\sim 10\%$. 

\smallskip
\Fig{eigen} shows the perturbation amplitude as a function of time, normalized by 
the respective value of $\tkh$ from \tab{sims}, for the five eigenmode simulations. 
The analytical prediction is that eigenmode perturbations grow exponentially from 
time $t=0$, their amplitude scaling as ${\rm exp}(t/\tkh)$, which is shown by the 
dashed line. All simulated modes match the predicted growth rate to within $\sim 20\%$ 
for a period of between $1.5-3\tkh$. The largest deviation occurs for the case 
$(\delta,\Mb)=(1,5.0)$, where the measured growth rate is $\sim 20\%$ below the 
predicted value. This is presumably caused by numerical diffusion, more severe 
for higher Mach number flows \citep{Robertson10}, as this mode has the highest 
value of $M_{\rm tot}$ by nearly a factor of 2. For the $(\delta,\Mb)=(100,1.5)$ 
mode, the growth rate increases sharply at $t\sim1.5\tkh$. This is due to mixing 
of the slab and the background which causes the effective density contrast and Mach 
number to decrease slightly, rendering the configuration unstable to artificial 
(numerical) surface modes, as the initial configuration was already very close to the 
critical Mach number. In the other three simulations, the growth rate is well behaved 
until $t\sim 3\tkh$, at which point non-linear effects cause the amplitude to 
saturate. A detailed study of the quasi-linear and non-linear phases of the instability, 
including this saturation, will be the subject of a forthcoming paper (Padnos et al., 
in preparation). 

\smallskip
We have rerun all simulations with twice higher and twice lower resolution while keeping 
$\sigma/\lambda$ fixed, and found no noticeable effect on the results. We also reran the 
simulations with the same resolution while varying $\sigma/\Delta$ between 1 and 16. Larger 
values of $\sigma$ result in slower growth rates compared to the predicted values, especially 
for surface modes where eigenmodes are harder to resolve (\se{Numeric}). Smaller values of 
$\sigma$ bring the simulated growth rates into better agreement with the predictions at early 
times, but lead to artificial noise dominating the perturbation amplitude and hence the growth 
rate before the non-linear saturation of the initial seeded mode. 

\subsection{Non-Eigenmode Simulations} 
\label{sec:non_eigen} 
\smallskip
We now study the evolution of general perturbations, that are not eigenmodes of the problem. 
We initialize pressure perturbations which are harmonic along the slab axis and decaying 
perpendicular to it:
\be 
\label{eq:NE}
\begin{array}{c}
P_1 = A~{\rm cos}(kz)\times\\
\\
\left[{\rm exp}\left(-\dfrac{(x-\Rs)^2}{2\Sigma^2}\right) +{\rm exp}\left(-\dfrac{(x+\Rs)^2}{2\Sigma^2}\right)\right]
\end{array},
\ee
{\no}with $A=0.05$ and $\Sigma=5\sigma$. The precise value of $\Sigma$, the width of the 
perturbation, is not important as long as it is larger than $\sigma$, the width of the 
smoothing layer. We also tried initializing perturbations in the transverse velocity 
component, and found no qualitative difference in our results. We prefer to focus here 
on the pressure perturbations, because all simulations, eigenmode and non-eigenmode, 
have initially uniform pressure in the entire domain, $P_0=1$, whereas the slab velocity 
and sound speed vary between different simulations. 

\begin{figure}
\centering
\subfloat{\includegraphics[trim={0.4cm 0.2cm 0.4cm 1.04cm}, clip, width =0.45 \textwidth]{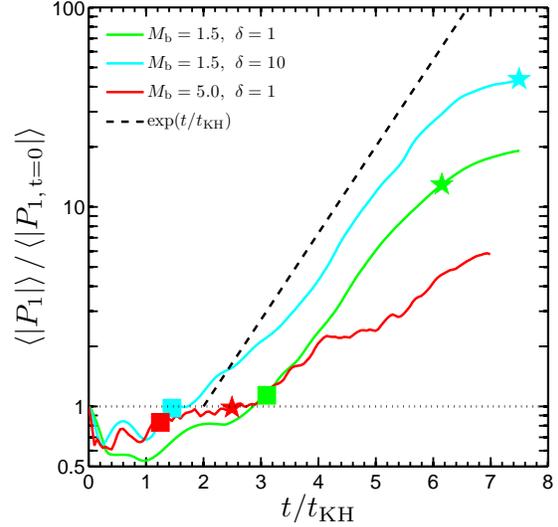}}
\caption{Growth rate of non-eigenmode perturbations in the simulations with wavelengths 
$\lambda=\Rs$, represented by the three solid curves. The x axis shows time normalized 
by the KH time of the predicted fastest growing mode given $\Mb$, $\delta$, and $K$, 
according to \tab{sims}. For the two cases shown with $\Mb=1.5$ (green and cyan curves) 
the fastest growing mode is a surface mode, while for the case with $\Mb=5$ (red line) 
it is a body mode. The y axis shows the perturbation amplitude normalized by its value 
at $t=0$, marked by the dotted line. The slope of the dashed line marks the expected 
exponential growth (the zero-point has been shifted for clarity). Squares mark the 
wavelength sound crossing time within the background, $t_{\rm \lambda}=\lambda/\cb$, 
after which eigenmodes develop and unstable surface modes begin to grow exponentially. 
Stars mark the slab sound crossing time, $\tsc=2\Rs/\cs$, after which the two sides of 
the slab come into causal contact and unstable body modes begin to grow exponentially. 
}
\label{fig:non_eigen} 
\end{figure}

\begin{figure*}
\centering
\subfloat{\includegraphics[width =0.95 \textwidth]{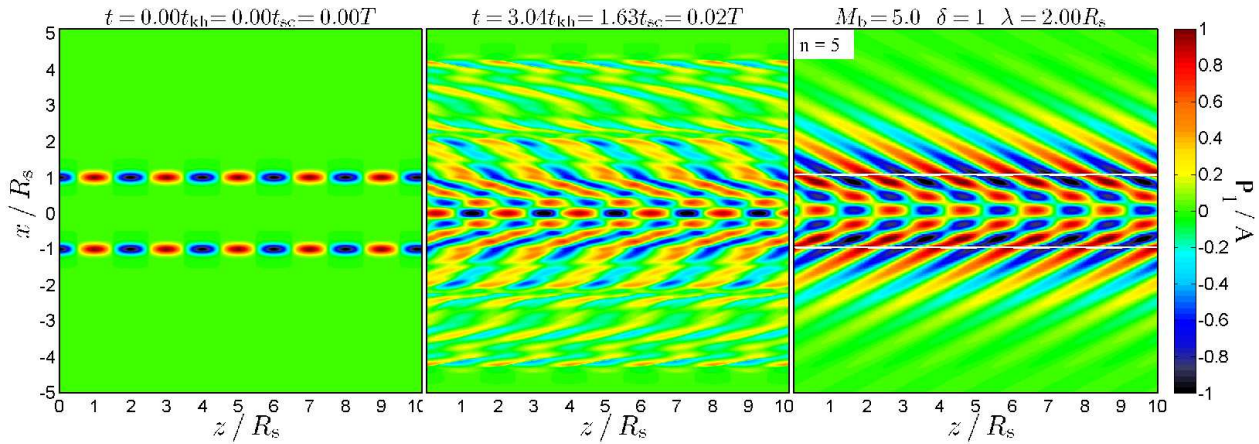}}
\caption{Transition of a general initial perturbation into the fastest growing eigenmode. The left and 
centre panels are taken from a simulation of a slab with $\Mb=5.0$, $\delta=1$, and $\lambda=2\Rs$, 
showing the pressure perturbation, $P_1$, at time $t=0$ (left) and $t \sim 1.63\tsc$ (centre), after 
the slab has become coherent. In both panels we list the time in units of the Kelvin-Helmholz time of 
the fastest-growing mode, $\tkh$, the slab sound-crossing time, $\tsc$, and the box sound-crossing time, 
$T_{\rm box}$. The color scale has been normalized by $A$, the maximal value of the perturbation on the slab 
interface at $t=0$. The right-hand panel shows the analytic form of the fastest-growing mode, the 
$n=5$ P-mode, with the amplitude normalized to unity at the slab interface. The resemblance of the 
simulated perturbation to the analytic mode is striking.
}
\label{fig:sim_comp} 
\end{figure*} 

\smallskip
According to the analysis in \se{linear}, growing modes are eigenmodes. However, eigenmodes 
\textit{do not} span the full range of perturbations, and an arbitrary initial perturbation 
cannot be decomposed into a linear combination of eigenmodes. Each eigenmode contains perturbations 
in all fluid variables whose amplitudes are linearly related to one another, while different 
eigenmodes have different transverse wavelengths. Thus, no linear combination of eigenmodes 
can result in a perturbation in only one fluid variable (such as the pressure) while leaving 
the others unperturbed. 
Before an arbitrary initial perturbation in one of the fluid variables can begin to grow, 
corresponding perturbations in the other variables will develop. The resulting set of 
perturbations can be decomposed into a linear combination of eigenmodes, both growing and 
decaying modes, and a residual perturbation. The residual propagates as a sound wave away 
from the slab boundary, the decaying eigenmodes decay away, and eventually the growing 
eigenmodes dominate. The minimal timescale over which the fluid can ``arrange" itself into 
eigenmodes is the wavelength sound-crossing time within the hot medium, $t_{\rm \lambda}=\lambda/\cb$. 
At $t>t_{\rm \lambda}$, after eigenmodes have developed, their amplitudes will begin to grow 
according to their corresponding growth rates until the fastest growing mode will eventually 
dominate the instability. 

\smallskip
An additional relevant timescale is the slab sound-crossing time, $\tsc$. 
At $t<\tsc$, the slab is not coherent, and information regarding a perturbation 
on one edge will not have reached the opposite edge. In our analysis of the 
dispersion relation for the slab (and the cylinder), we implicitly assumed both 
boundaries to be in causal contact, so these solutions can only be applied at 
$t>\tsc$. At earlier times, each boundary must behave as an independent sheet. 
In the regime where surface modes are unstable, the slab and sheet are practically 
identical anyway, and perturbations begin to grow at $t\gsim t_{\rm \lambda}$. 
Body modes, on the other hand, can only begin to grow once the slab is coherent, 
at $t\gsim\tsc>t_{\rm \lambda}$. These are triggered by sound waves reverberating 
between the slab boundaries, which is why they have been referred to as \textit{reflected 
modes} in the literature.

\smallskip
We simulated several configurations, with $(\delta,\Mb)=(1,1.5),\,(10,1.5),\,(1,5.0)$, 
similar to our eigenmode runs. Each configuration was run once with a perturbation 
wavelength $\lambda=2\pi/k=2\Rs$ (as in the eigenmode runs), and once with $\lambda=\Rs$. 
We focus mainly on the runs with $\lambda=\Rs$, summarized in \tab{sims}. There we list 
the mode number and $\tkh$ of the fastest growing mode into which the perturbation can 
decay given $\Mb,\delta$ and $\lambda/\Rs$ (surface modes for the first two and a body 
mode for the third). Since the pressure perturbation we are initiating (\equnp{NE}) is 
symmetric, it can only decay into P-modes. 

\smallskip
\Fig{non_eigen} shows the perturbation amplitude as a function of time, similar to \fig{eigen}. 
The time has been normalized by the $\tkh$ corresponding to the fastest growing mode (\tab{sims}). 
As in \fig{eigen} we estimate the perturbation amplitude by the average of $|P_1|$, though we 
calculate the average in a smaller region, $|x|<2\Rs$, since the initial perturbations were 
localized on the slab boundaries. However, our results are not strongly dependent on the size of 
this region, or on whether we use pressure or transverse velocity to estimate the amplitude. For 
each configuration we have marked the corresponding $t_{\rm \lambda}$ with squares and $\tsc$ with 
stars. As expected, the two surface modes begin to grow in amplitude at $t\sim t_{\rm \lambda}$ 
while the body mode does not grow until $t\sim \tsc$. During the growth phase, the growth rates 
match the predicted growth rate of the fastest growing mode to within $\sim 20\%$, similar to the 
eigenmode runs. 

\smallskip
\Fig{sim_comp} shows the pressure perturbation for the $(\delta,\Mb,\lambda)=(1,5.0,2\Rs)$ 
simulation at $t=0$ and at $t\sim 1.6\tsc$, shortly after the perturbation begins to grow. 
For comparison, we show the analytic form of the corresponding fastest growing P-mode, $n=5$, 
normalized to the same amplitude. The fastest growing mode can be found from \equ{resonant_X_text} 
with $K=k\Rs=\pi$ and $M_{\rm tot}=2.5$. The resemblance of the simulation result and the analytic 
prediction is striking, illustrating that the initial perturbation has evolved into eigenmodes and 
the fastest growing mode, $n=5$, dominates.

\smallskip
We also simulated the case $(\delta,\Mb)=(100,1.5)$. Despite surface modes being formally 
stable for this configuration (\fig{planar_growth_rate}), it lies very close to the 
$M_{\rm crit}$ boundary and small numerical errors seed artificial surface modes. Due to 
the high density contrast, $\tsc>>t_{\rm \lambda}$ and these artificial modes dominate the 
instability before the body mode has a chance to grow.

\section{Linear Stability of Cold Flows in Hot Haloes} 
\label{sec:tkh} 

\smallskip
In this section we evaluate the potential importance of KHI in the evolution of cold 
streams that feed massive galaxies at high redshift. This depends on the ratio of the 
total time a perturbation can grow before the stream joins the central galaxy, $\tg$, 
and the Kelvin-Helmholtz time, $\tkh$. In the linear regime, the amplitude of perturbations 
grows as ${\rm exp}(t/\tkh)$, so the number of e-foldings in the perturbation growth is 
\be 
\label{eq:efold}
N_{\rm e\:folding} \equiv \tg/\tkh
\ee
{\no}If $N_{\rm e\:folding}=1,~3,$ or $10$, a small perturbation will grow to roughly $2.7,~20,$ 
or $2.2\times10^4$ times its initial amplitude. However, as we saw in \se{sim}, the exponential 
growth does not continue indefinitely. Once the perturbation becomes quasi-linear, the amplitude 
saturates before continuing to grow linearly with time in a self-similar way (Padnos et al., in 
preparation). For our purposes here, we loosely refer to cases where $N_{\rm e\:folding}\gsim 3$ 
as being quasi-linear and to cases where $N_{\rm e\:folding}\gsim 10$ as being non-linear. However, 
without knowledge of the initial perturbation amplitudes, these thresholds are somewhat arbitrary. 
All we can say is that if the perturbation begins small, significant growth in the linear regime is 
a necessary condition for significant growth overall.

\smallskip
We begin by evaluating $\tg$. For haloes of $\Mv \sim 10^{12}\msun$ at $z\sim 2$, the virial 
shock radius is roughly the halo virial radius, $\Rv$ \citep[e.g.][]{db06,Dekel09}, though 
we note that in more massive clusters the shocked region can extend to several times $\Rv$ 
(Zinger et al., in preparation). Cosmological simulations indicate that the stream velocity 
is roughly constant during infall \citep{Dekel09,Goerdt10,Goerdt15a}, comparable to the halo 
virial velocity, 
\be 
\label{eq:vvir}
V \simeq \Vv = \sqrt{\frac{G\Mv}{\Rv}}.
\ee
The travel time of the stream through the shock-heated medium is thus roughly the 
virial crossing time
\be 
\label{eq:tv}
t_{\rm infall} \simeq \tv=\Rv/\Vv.
\ee
In the EdS regime, valid at $z\gsim 1$, this is a constant fraction of the cosmological 
time, $\tv\simeq 0.14t_{\rm Hubble}$ \citep[e.g.][]{Dekel13}. 

\begin{figure*}
\centering
\subfloat{\includegraphics[trim={0.1cm 0.8cm 3.0cm 0}, clip, width =0.33 \textwidth]{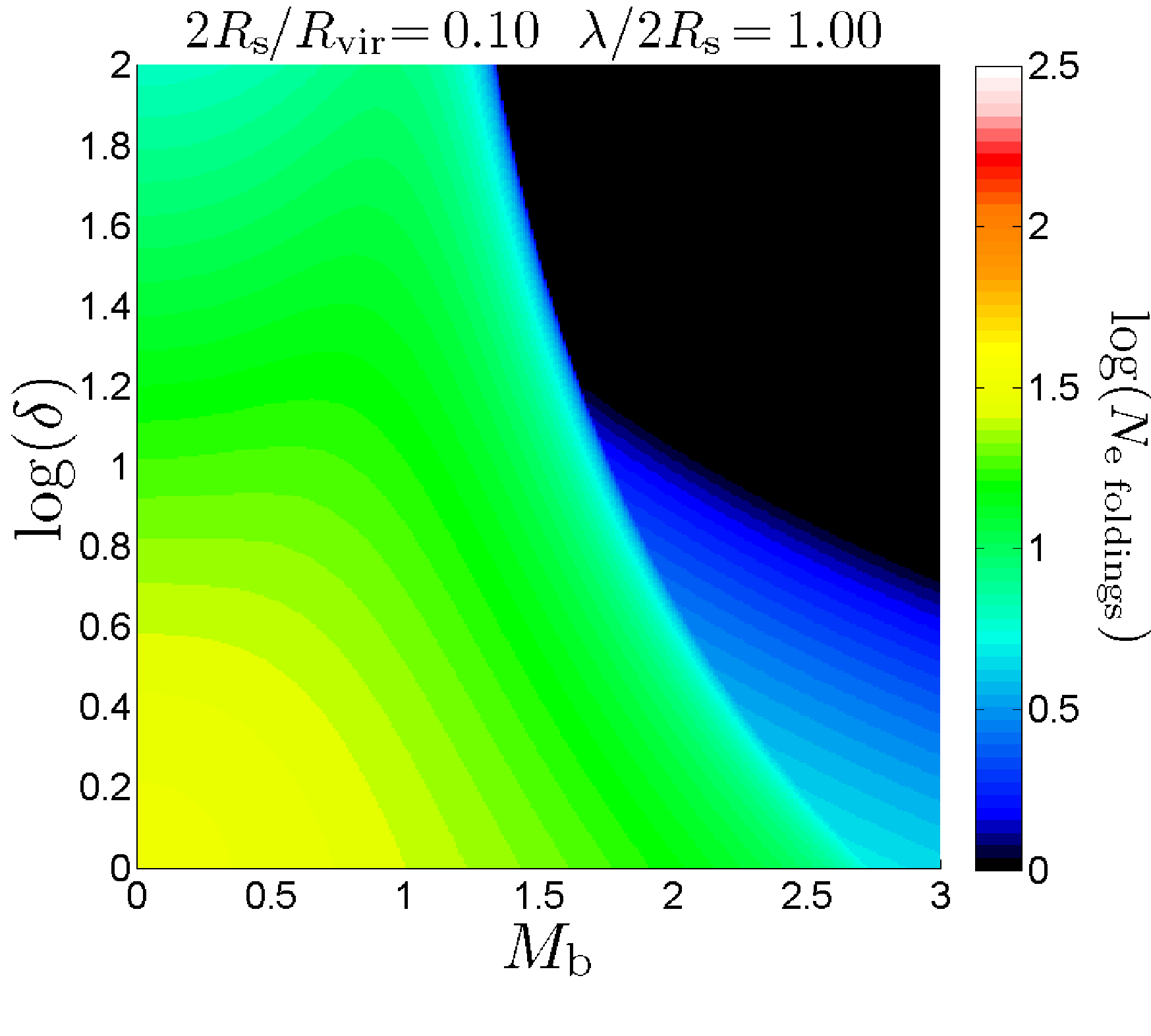}}
\subfloat{\includegraphics[trim={2.0cm 0.8cm 3.0cm 0}, clip, width =0.2875 \textwidth]{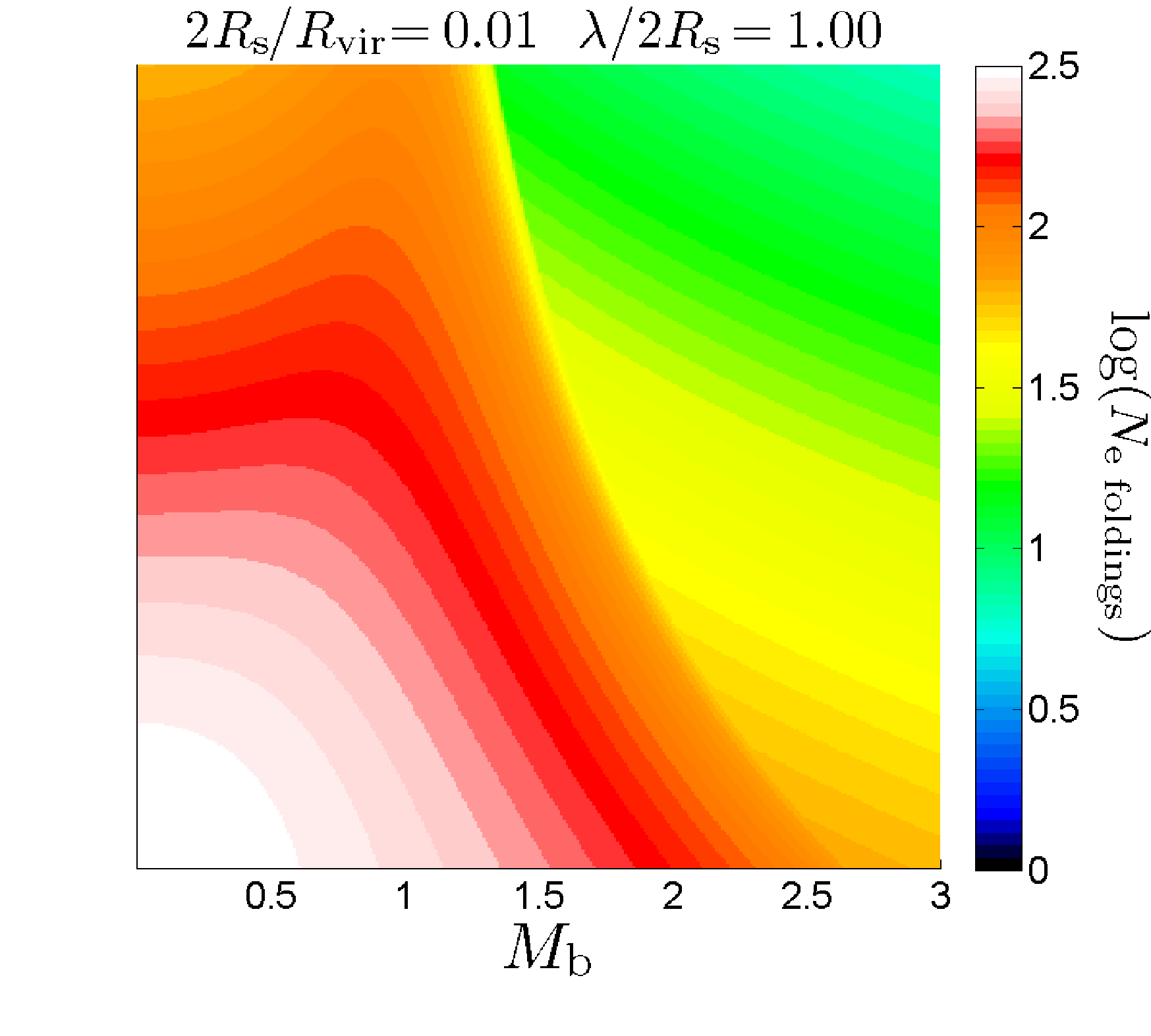}}
\subfloat{\includegraphics[trim={2.0cm 0.8cm 0 0}, clip, width =0.355 \textwidth]{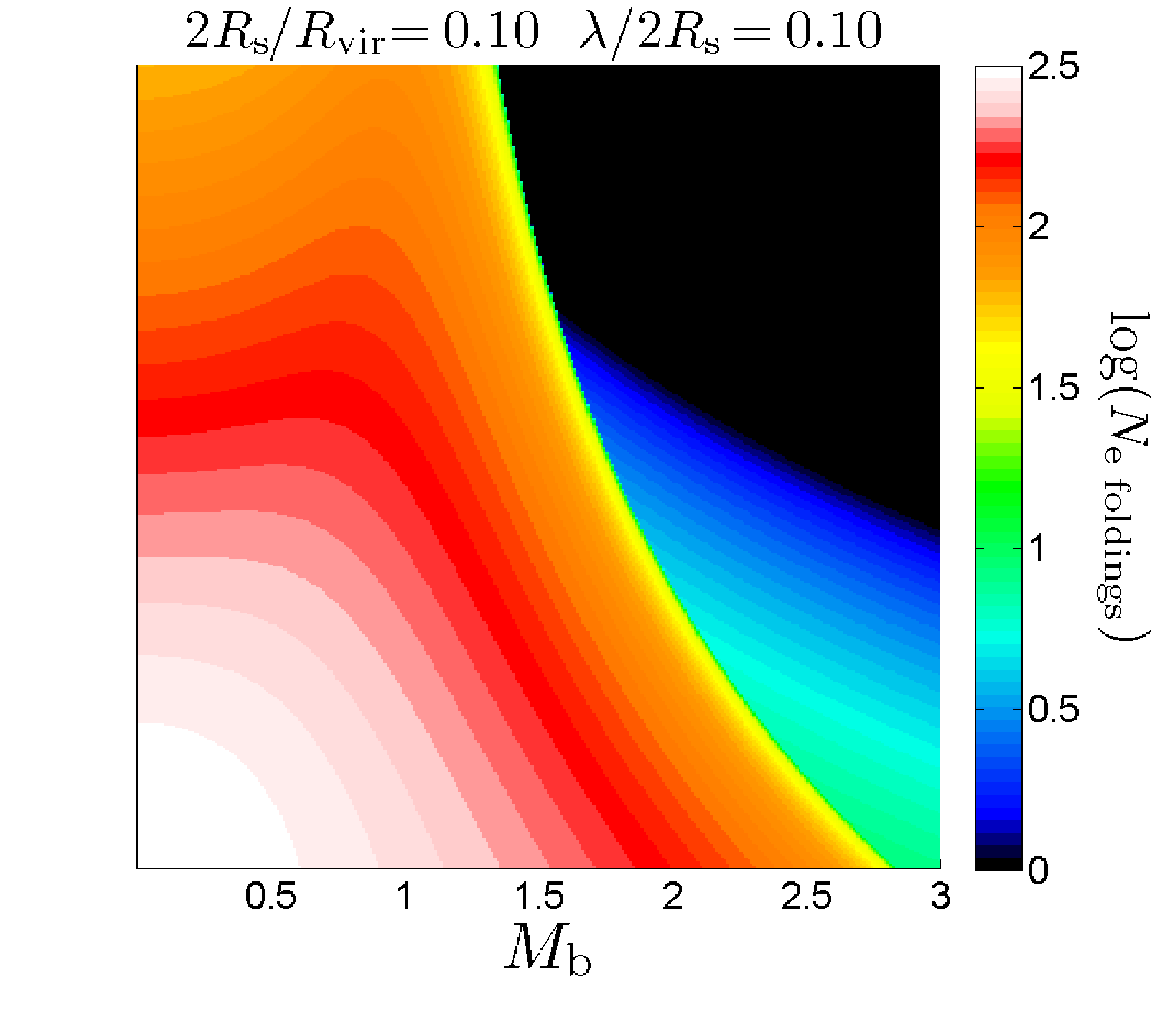}}
\caption{Linear growth of KHI in a virial crossing time, as the stream 
penetrates from the dark-matter halo virial radius to the central galaxy. 
Shown in color is the number of e-foldings of growth in the perturbation 
amplitude, in the linear regime, as a function of the stream parameters, 
$\delta$ and $\Mb$. The different panels are for different ratios of stream 
radius to virial radius and perturbation wavelength to stream radius, as 
listed at the top of each panel. When the sheet is unstable, at $\Mb<M_{\rm crit}$ 
(\equnp{Mcrit}), we used the growth rate for surface modes in the sheet 
(\fig{planar_growth_rate}) and allowed the perturbations to grow for a full 
virial crossing time, $\tv=\Rv/\Vv$. When the sheet is stable, at $\Mb>M_{\rm crit}$, 
we used the effective resonant growth rate in the slab/cylinder 
(\equsnp{resonant_growth_text} and \fig{numerical_solution}). In this regime, 
perturbations begin to grow only after a stream sound-crossing time has elapsed, 
so they only grow for a time $(\tv-\tsc)$. The figure shows that realistic values 
for the stream parameters, $\Mb\sim 0.75-2.25$ and $\delta\sim 10-100$, put them 
near the phase transition between rapid growth of surface modes and slow growth 
of body modes. While certain regions of parameter space are stable to linear KHI, 
with $\lsim 3$ e-foldings of growth, a wide range of allowed parameters do result 
in quasi-linear or non-linear perturbations, with $N_{\rm e\:foldings}\gsim 3$ and 
$10$ respectively.
}
\label{fig:phases} 
\end{figure*} 

\smallskip
As we saw in \se{sim}, unstable surface modes begin to grow after the initial perturbation 
has decayed into eigenmodes, roughly at time $t_{\rm \lambda}=\lambda/\cb$. We assume 
$\lambda\lsim \Rs<<\Rv$ and $\cb \sim \Vv$ (see below), so $t_{\rm \lambda}<<\tv$ and 
we can assume that surface modes begin to grow instantaneously, $\tg\sim\tv$. On the other 
hand, unstable body modes can only grow after the stream becomes coherent, at time $\tsc$. 
Since $\cs$ can be much slower than $\cb\sim\Vv$ for a dense stream, $\tsc$ can be a significant 
fraction of $\tv$, and $\tg\sim \tv-\tsc$. The relation between $\tg$ and $\tv$ thus 
depends on whether surface modes or body modes dominate the instability. 

\smallskip
Combining our estimate of $\tg$ for surface/body modes with \equ{tkh} for $\tkh$, we obtain for 
the number of e-folding times 
\be 
\label{eq:Nefolding}
\begin{array}{c}
N_{\rm e\:folding}=2K\delta^{1/2}\Mb\times\\
\\
{\rm max}\left[\varpi_{_{I,\,sheet}}\dfrac{\tv}{\tsc},\: \varpi_{_{I,\,body}}\left(\dfrac{\tv}{\tsc}-1\right)\right].
\end{array}
\ee
{\no}This depends on four parameters: $\Mb$, $\delta$, $\tv/\tsc$ and $K=k\Rs$. We estimate 
each of these in turn below.

\smallskip
To estimate $\Mb$ for typical cold streams, we make the simplifying assumption 
that the halo CGM is isothermal. The sound speed in the halo is thus approximately
\be 
\label{eq:sound_speed}
\cb = \sqrt{\frac{\gamma K_{\rm B}\Tv}{\mu m_{\rm p}}},
\ee
{\no}where $K_{\rm B}$ is the Boltzmann constant, $\Tv$ is the virial temperature, 
$\mu m_{\rm p}$ is the mean particle mass, and $\gamma=5/3$ is the adiabatic index 
of the gas. The temperature can be found from virial equilibrium 
\be 
\label{eq:Tvir}
\frac{3}{2}K_{\rm B}\Tv \simeq \frac{1}{2}\frac{G\Mv \mu m_{\rm p}}{\Rv} = \frac{1}{2}\mu m_{\rm p} \Vv^2.
\ee
{\no}Inserting \equ{Tvir} into \equ{sound_speed} gives for the Mach number of the 
stream relative to the background $\Mb\simeq \Vv/\cb \simeq 1.34$. In practice, we 
assume values in the range $\Mb\sim 0.75-2.25$. 

\smallskip
To estimate $\delta$, we again make the simplifying assumption that both the halo CGM 
and the stream are isothermal. The temperature in the halo is $T=\Tv$ (\equnp{Tvir}), 
which for a $\sim 10^{12}\msun$ halo is a few times $10^6{\rm K}$. However, due to 
efficient cooling in the high density streams, they do not support a stable shock at 
the virial radius \citep{db06}, and their temperature is set by the cooling curve for 
low metallicity gas, at a few times $10^4{\rm K}$. Assuming pressure equilibrium 
between the streams and the hot CGM at any given halo-centric radius, this leads to a 
density contrast of $\delta\sim 10-100$, consistent with cosmological simulations 
\citep[e.g.][]{Ocvirk08,Dekel09,Goerdt10}. 

\smallskip
The ratio $\tv/\tsc$, can be related to the ratio of the stream width to the virial radius, 
\be 
\label{eq:tsc_tv}
\tsc/\tv = \Mb~2\Rs/\Rv. 
\ee
A narrower stream has a shorter sound crossing time, leading to more rapid growth 
of body modes. Based on cosmological simulations, the cold streams we are discussing 
have characteristic widths of roughly $2\Rs/\Rv\sim 0.01-0.1$\footnote{In realistic haloes 
both the stream and the background become denser at smaller radii, making the stream 
of a conical, rather than cylindrical shape. This is discussed further in \se{disc}.}. 

\smallskip
The final ingredient is the ratio of the perturbation wavelength to the stream width, 
$\lambda/(2\Rs)$. Since shorter wavelengths grow faster for both surface and body modes, 
the instability will be dominated by the shortest wavelengths that can grow. In the 
idealized problem we studied in \se{linear}, perturbations with arbitrarily small wavelengths 
can be excited and their growth rate diverges. However, in a realistic stream, there is 
a finite transition layer at the interface between the stream and the background, similar 
to the smoothing layer introduced in our simulations (\se{sim}). Perturbations with wavelengths 
comparable to or smaller than this layer will be damped. The width of this layer is determined 
by processes such as thermal conduction and viscosity which are not incorporated in cosmological 
simulations. This is crudely addressed in \se{disc}, while a more rigorous study will be the topic 
of future work. For our purposes, we crudely consider a range of $\lambda/(2\Rs) \sim 0.1-1$, 
corresponding to wavenumbers $K\sim 3-30$. As we have seen, the growth rate scales linearly with 
wavenumber for surface modes, $\Oi \propto K$, but only logarithmically for body modes, 
$\Oi\propto {\rm ln}(K)$.

\smallskip
\Fig{phases} shows $N_{\rm e\:folding}$ as a function of $\Mb$ and $\delta$ 
for different values of $\Rs/\Rv$ and $\lambda/\Rs$. We show a wide stream 
with $2\Rs=0.1\Rv$ and a stream 10 times narrower. We show a long wavelength 
perturbation, $\lambda=2\Rs$, and one ten times smaller. When surface modes 
dominate the instability, we calculate the growth rates, $\Pi$, from the sheet 
dispersion relation (\equnp{unitless_dispersion_sheet_2}). These are practically 
identical to the growth rates of surface modes in the slab calculated numerically 
from \equ{dispersion_slab_unitless}. When body modes dominate the instability, 
we use the approximation for the fastest growing mode in a slab or a cylinder 
given by \equs{resonant_growth_text}. 

\smallskip
The allowed range for the stream parameters places them near 
the transition between surface modes and body modes. When surface modes 
dominate, the stream is highly unstable to KHI in a virial time, and even 
very small perturbations will become highly non-linear. However, only a very 
minor change in velocity pushes the stream into the region where body modes 
dominate and the growth rate is much slower. Wide streams are marginally unstable, 
with $N_{\rm e\:folding}<3$. However, narrow streams can be highly unstable even 
in this regime, with $N_{\rm e\:folding} \lsim 30$. Furthermore, recall that if 
$m>1$ in a cylinder, the effective Mach number for determining whether surface 
modes are still unstable is $\Mb[1+(m/K)^2]^{-1/2}$ (\se{reflected_cyl}). Thus 
for high $m$ modes, unstable surface modes will occupy a larger region of the 
parameter space, making \fig{phases} only a lower limit on $N_{\rm e\:folding}$. 
However, as stated in \se{reflected_cyl}, modes with $m>1$ are likely to be suppressed 
in realistic streams, making our estimates of $N_{\rm e\:folding}$ reasonable. Overall, 
we estimate that for the relevant range of parameters, $N_{\rm e\:folding} \sim 0.1-100$, 
indicating that KHI can in principle be important for the evolution of cold streams.

\section{Additional Physics}
\label{sec:disc} 

\smallskip
While the analysis presented here has been thorough and  accurate, with interesting 
new understanding of the body modes in the compressible regime, it has also been 
very simplistic. We have only studied the linear regime of adiabatic, purely 
hydrodynamic instabilities. This is an important first step, which allowed us to obtain 
a definitive result. However, as it stands, further, more detailed analysis is necessary.
In a forthcoming paper (Padnos et al., in preparation), we will present a detailed study 
of the non-linear evolution of idealized KHI, without cooling or gravity. Since we 
have seen that streams can find themselves in the non-linear regime for a wide range of 
plausible parameters, this is necessary to complete our study of the effect 
of KHI in the evolution of cold streams. Additionally, we are planning a series of papers 
in which we will account for additional physical processes one by one, both analytically 
as in this work, and numerically. Our ultimate goal is to build a comprehensive understanding 
of the evolution of cold streams, from the bottom up. Below, we outline the additional 
processes that we intend to address in future work, and their possible effects.

\begin{enumerate}

\smallskip
\item 
\textbf{Cooling:}
The importance of cooling in the linear regime of KHI depends on 
the relation between $\tkh$ and the cooling times in the hot and 
cold media, $t_{\rm h}$ and $t_{\rm c}$, respectively. We 
will always have $t_{\rm c}<t_{\rm h}$. In the case of cold 
streams in hot haloes, the cooling time in the hot halo is longer 
than the virial time, $\tv<t_{\rm h}$, which is what 
allows the presence of a stable virial shock to begin with. Since 
we are only interested in cases where $\tkh \lsim \tv$, we ignore 
the case $t_{\rm h}<\tkh$\footnote{This could come up in certain 
physical circumstances, e.g., near an X-ray source that is capable 
of keeping the hot gas hot.}. If $\tkh<t_{\rm c}<t_{\rm h}$, then 
cooling should not make any difference in the linear evolution of 
KHI. However, if $t_{\rm c}<\tkh<t_{\rm h}$, then the cold gas returns 
to an equilibrium temperature fast compared to the hydrodynamic growth 
timescale. This is exactly as if the gas were isothermal, and thus all 
of our analysis stays the same, except that $\gamma = 1$ in the cold 
medium while $\gamma=5/3$ in the hot medium. While our analysis implicitly 
assumed that both media had the same value of $\gamma$, this should only 
affect the ratio of sound speeds and it seems unlikely that this change 
will dramatically alter our results for the linear evolution. 
In the non-linear regime, cooling can enhance the instability by introducing 
thermal instabilities in addition to hydrodynamic ones. Once KHI sets 
in, regions of the stream will become denser and shocks may develop. 
The cooling rate in these overdense regions may increase causing 
them to become even denser and creating a runaway process. However, 
cooling can also weaken the non-linear evolution of the instability. 
If the instability is dominated by surface modes, then the non-linear 
evolution occurs via a shearing layer that forms between the hot and 
cold media, where the two fluids mix. If the cooling time in the cold 
stream is shorter than the sound crossing time, this region will be 
confined close to the stream boundary and will not consume the stream 
interior. This effect has been seen in simulations of cold gas clouds 
surrounded by a hotter confining medium within the ISM \citep{Vietri97}. 

\smallskip
\item 
\textbf{Thermal Conduction:}
This can suppress the growth of short wavelength perturbations, by forming 
a finite transition region between the stream and the background, similar to the smoothing 
region in our simulations. Thermal Conduction can be treated as a diffusive heat flow with 
a diffusion coefficient $D(T,\rho) = \kappa(T)/(\rho c_{\rm v})$, where $c_{\rm v}$ is the 
specific heat capacity and $\kappa(T)$ is the thermal conductivity. Using \citet{Spitzer56} 
to evaluate $\kappa(T)$ under the simplifying assumption that both fluids are pure hydrogen, 
we obtain $D \sim 6 \times 10^{28}~{\rm cm^2~s^{-1}} ~ T_6^{5/2} n_4^{-1}$, where $T_6$ 
is the gas temperature in units of $10^6 {\rm K}$ and $n_4$ is the gas density in units of 
$10^{-4} \cmc$.\footnote{Similarly, viscosity acts as diffusion in momentum space, but the 
corresponding diffusion coefficient is $\sim 100$ times smaller than for heat conduction 
\citep{Spitzer56}.} The characteristic time for diffusion to ``smear out" features with a 
characteristic lengthscale $L$ is $t_{\rm dif} \sim L^2/D$. To estimate the effect this will 
have on the growth of \textit{surface modes}, we compare $t_{\rm dif}$ over the perturbation 
wavelength, $L=\lambda$, to the e-folding time for perturbation growth, $\tkh$, as surface 
modes begin to grow immediately (\se{tkh}). For short wavelengths, 
$\lambda<\lambda_{\rm c} = D/(2\pi \Mb \cb \Pi)$, we have $t_{\rm dif}<\tkh$ 
and we thus expect the instability to be suppressed. For $\delta \sim 100$ and $M_b \sim 1$, we 
have $\Pi \sim 0.15$ (\fig{planar_growth_rate}). For a virial temperature of $\Tv\simeq 10^6{\rm K}$, 
we have $\cb \sim 1.2\times10^7~{\rm cm~s^{-1}}$, and the critical wavelength becomes 
$\lambda_{\rm c} \sim 1.7~\kpc ~ T_6^{5/2} n_4^{-1}$. In the cold streams we estimate 
$T_6 \sim 0.01$ and $n_4 \sim 100$, yielding $\lambda_{\rm c} \sim 1.7\times 10^{-4}~\pc$. 
For the shock-heated halo gas, we have $T_6 \sim n_4 \sim 1$, yielding 
$\lambda_{\rm c} \sim 1.7~\kpc$, comparable to the stream radius. However, in the hot gas 
the effective diffusion coefficient is likely to be reduced from the \citet{Spitzer56} estimate, 
since the maximum rate of conductive energy transport cannot exceed $\sim v_{\rm e}K_{\rm B}T$, 
where $v_{\rm e}$ is the electron thermal velocity, i.e. conduction can't move thermal energy 
any faster than a population of free-streaming electrons. This can decrease the critical wavelength 
by at least an order of magnitude. Regarding \textit{body modes}, since these are triggered 
by sound waves reverberating within the stream, the relevant timescale is the stream sound-crossing 
time. Comparing $\tsc$ to $t_{\rm dif}$ with $L=2\Rs$, $T_6 \sim 0.01$ and $n_4 \sim 100$, we find 
heat conduction can suppress the excitement of unstable body modes only if the stream radius is 
$\Rs\lsim 0.1\pc$. While more accurate conclusions await the the explicit inclusion of heat 
conduction in the analysis of KHI, it seems that in cold streams only wavelengths much smaller 
than the stream width will be suppressed. 

\smallskip
\item 
\textbf{External Gravity:}
We refer here to the underlying gravitational potential of the dark matter halo, which 
will create a density gradient, and hence a pressure gradient in the background gas halo. 
As the external pressure increases towards smaller radii, the stream will become narrower 
towards the halo centre, assuming a conical rather than cylindrical shape. However, we 
expect our linear analysis as presented in \se{linear} to remain valid for perturbations 
with wavelengths smaller than the length-scale over which the stream radius changes by 
order unity. If we make the simplifying assumptions that the halo is an isothermal sphere, 
with $\rho_{\rm b}(r)\propto r^{-2}$ and $T_{\rm b}(r)={\rm const}$, and that the background 
and stream are in pressure equilibrium locally at each radius $r$, than the stream is also 
isothermal and hence $\cs$ is constant. Therefore, the conical shape causes the stream sound 
crossing time to decrease towards the halo centre leading to more rapid growth of instabilities, 
as $\tkh\propto\tsc\propto\Rs$. Local pressure equilibrium and the isothermality of the stream 
imply that the density profile within the stream scales as $\rho_{\rm s}(r)\propto r^{-2}$, which 
in turn implies that the stream width scales as $\Rs\propto r$, i.e. a perfect cone with constant 
opening angle. If the stream velocity is roughly constant during infall, as indicated by cosmological 
simulations, this implies that $\tkh(r)\propto \Rs(r) \propto r \propto t_{\rm infall}(r)$. So at 
every radius, the number of e-foldings a perturbation can grow from its current state before the 
stream reaches the galaxy is constant. Integrated over the lifetime of the stream, this can lead to 
significantly more growth than predicted in \se{tkh}. In future work, we will address the dispersion 
relation of KHI in a conical stream, and perform simulations of streams in an external spherical 
potential.

\smallskip
\item 
\textbf{Self Gravity:}
Self gravity within the stream will have two effects. Firstly, it will 
result in gradients in density, pressure and perhaps velocity within 
the stream, as a function of distance from the stream axis. While these 
gradients will be much weaker than those in the transition zone between 
the stream and the background, it may still alter the growth rates of 
perturbations in an appreciable way. Secondly, self-gravity may enhance 
stream instability by causing overdense regions within the stream to 
become gravitationally unstable. It is worth noting that a crude estimate 
of the Jeans scale in the stream is $\lambda_{\rm J} \sim 2\kpc ~ c_{\rm 10} 
n_{\rm 0.1}^{-1/2}$, where $c_{\rm 10}$ is the sound speed in units of 
$10 \kms$, and $n_{\rm 0.1}$ is the gas density in units of $0.1\cmc$. 
For $T\sim 10^4$ and $n\sim 0.1$, relevant in the streams near 
the galaxy, the Jeans length is comparable to the stream width, and could 
become smaller due to overdensities caused by hydrodynamical and thermal 
instabilities. We may also witness a fragmentation into sub-filaments, similar 
to the multi-scale filamentary structure observed in Galactic molecular clouds. 

\smallskip
\item 
\textbf{Magnetic Fields:}
While the intra-cluster medium is known to be mildly magnetized \citep{Churazov08} 
little is known about the magnetization state of gas in galactic haloes. \citet{Dubois10} 
argued that initial magnetic fields in the ICM originate from winds outflowing from 
dwarf galaxies, whose IGM contain magnetic fields well below $1\mu G$ in amplitude. 
This indicated that the ratio of thermal to magnetic pressure (the plasma $\beta$ 
parameter) is $\sim 50-10^3$. Furthermore, we expect this to be only an upper-limit 
on the magnetization of the filament gas, which is in general more ``pristine" and 
has not been affected as much by galactic processes. However, some measurements indicate 
magnetic fields as high as $0.3\mu G$ in cosmic filaments feeding galaxy clusters 
\citep{Bagchi02}. The KH stability of mildly magnetized gas can be altered from its 
pure hydrodynamic analogue in three ways. The first is the modifications to the equation 
of state of the gas, though for a sub-dominant component, this effect is expected to be 
small. The second effect is surface tension that could possibly arise from the shearing 
of magnetized gas at the boundaries of the instability. The shearing layer is comprised 
of gas that has been entrained from the filament gas and from the ambient halo, and could 
be magnetized even when the filament gas is completely unmagnetized, if the halo is not. 
Through magnetic draping, the magnetic field lines at the shear layer will become aligned 
with the flow direction, and the amplitude of the magnetic fields there will increase. This 
process will stabilize the boundary and could strongly effect the results in the linear phases 
of the instability. We leave for future work a more robust comparison of the growth rate of 
magnetic fields in the shear layers compared to the growth rate of the instability and their 
dependence on the initial hydrodynamic perturbation and initial magnetic fields. The third 
aspect of magnetic fields is its effect on conduction perpendicular to the shearing layer and 
magnetic draping that is expected there, which could drastically reduce the coefficients of 
thermal conductivity and viscosity discussed above.

\smallskip
\item 
\textbf{Galaxy Formation:}
Even after accounting for all the effects mentioned above, the analysis 
will still be idealized. In order to address the stability of realistic cold 
streams feeding galaxies from the cosmic-web, we must account for the 
presence of additional merging galaxies within the streams, interaction 
between several streams within the same halo, interaction between streams 
and feedback induced outflows from the central galaxy, and possible star-formation 
and feedback within the streams themselves. The best way to account for 
all these effects is with a fully cosmological simulation. In parallel to 
the methodical work outlined above, we are experimenting with running cosmological 
zoom-in simulations using \texttt{RAMSES}, with refinement based on gradients 
in density and velocity, rather than the density based quasi-Lagrangian strategy 
typically used (Roca-Fabrega et al., in preparation). This will allow us to resolve 
the streams better than any existing cosmological simulation. As described in \se{Numeric}, 
to properly resolve the the instabilities in cold streams, the cell size must be at 
least $\sim 60$ times smaller than the perturbation wavelength. For $\lambda \sim \Rs \sim 1\kpc$, 
this corresponds to a resolution of $\lsim 15\pc$ within the cold streams, which will 
be challenging to achieve. Such simulations will also allow us to gauge the effect of 
initially large instabilities on the streams, such as caused by the collision of a stream 
with a satellite galaxy in the halo. An idealized study of such a collision suggests that 
while the stream is initially destroyed, it reforms within $\lsim 0.3$ virial crossing times 
\citep{Wang14}, after which our linear analysis is again relevant.
\end{enumerate}

\section{Summary and Conclusions}
\label{sec:conc} 

\smallskip
We have presented a detailed analysis of linear Kelvin-Helmholtz instabilities 
for fully compressible fluids in three different geometries: a sheet, a slab 
and a cylindrical stream, confirming our analytical predictions using numerical 
simulations with \texttt{RAMSES}. We then applied our results to the problem of 
cold streams that feed massive SFGs at high redshift, showing that KHI can be 
important in their evolution. For a large region of allowed parameters, the 
number of e-foldings of growth experienced by a linear perturbation is between 
10-100. However, the estimated range of parameters overlaps the phase transition 
between rapidly growing surface modes and more slowly growing body modes. As a 
result, perturbations may still remain linear by the time the stream reaches the 
central galaxy, with less than one e-folding of growth in a virial crossing time. 
The linear analysis of KHI in the adiabatic limit thus indicates that it can be 
relevant for the evolution of cold streams, but it cannot definitely asses its 
importance. Our main results can be summarised as follows: 

\begin{enumerate}
\smallskip
\item 
For a sheet, KHI is suppressed at high Mach numbers, 
$\Mb>M_{\rm crit}=(1+\delta^{-1/3})^{3/2}$, with 
$\Mb=V/\cb$ the fluid velocity normalized by the 
sound speed in the hot medium, and $\delta=\rho_{\rm s}/\rho_{\rm b}$ 
the density contrast between the fluids. At lower 
Mach numbers the growth rate for instabilities scales 
linearly with the wavenumber, $\Oi \propto k$, and the 
perturbations themselves decay exponentially with distance 
from the interface between the fluids. 

\smallskip
\item 
At low Mach numbers, when the sheet is unstable, \textit{surface modes} 
dominating the instability in a slab, which behaves similarly to a sheet 
for wavelengths comparable to or shorter than the slab width. 

\smallskip
\item
At high Mach numbers, $M_{\rm tot}=V/(\cb+\cs)>1$, the slab remains 
unstable despite surface modes having stabilized due to the appearance 
of \textit{body modes} at shorter and shorter wavelengths. These modes 
are triggered by waves being reflected off the slab boundaries and are 
qualitatively different than the surface modes present at low Mach numbers. 
They penetrate the width of the slab, and resemble standing waves propagating 
through a waveguide. There is an infinite set of these body modes, 
characterized by a mode number, $n=1,2,3,...$, representing the number 
of nodes in the transverse direction. Each mode becomes unstable at a 
finite wavelength, its growth rate increases towards a maximum at a resonant 
wavelength, and then decays to zero as $\lambda\rightarrow 0$. The effective 
growth rate of instabilities in the slab is given by the sequence of resonant 
growth rates for each mode. This is inversely proportional to the slab sound 
crossing time and scales logarithmically with wavenumber, 
$\Oi\propto \tsc^{-1}{\rm ln}(k\Rs)$.

\smallskip
\item 
The only qualitative difference between a slab and a cylinder is that a cylinder 
has an infinite sequence of symmetry modes, characterised by an azimuthal mode 
number $m$, while a slab only has two symmetry modes: symmetric Pinch modes and 
antisymmetric Sinusoidal modes. However, at wavelengths comparable to or shorter 
than the stream width, the growth rates of instabilities in cylindrical geometry 
are very similar to those in a slab. The slab is thus a good approximation to 
the cylinder at short wavelengths.

\smallskip
\item 
Simulations of the linear regime of KHI in a slab geometry reproduce the analytic 
results. When the slab is perturbed with an eigenmode of the problem, so that the 
initial conditions are self consistent with the linearized hydrodynamic equations, 
the growth rate matches the analytic prediction for both surface modes and body modes. 
When the slab is perturbed by an arbitrary, non-eigenmode perturbation, this must decay 
into eigenmodes before it can begin to grow. For surface modes, this process lasts for 
a wavelength sound-crossing time in the hot medium, $t_{\rm \lambda}=\lambda/\cb$. For 
body modes, we must wait until the stream becomes coherent, after a stream sound-crossing 
time, $\tsc=2\Rs/\cs$. Once the initial perturbation has decayed into eigenmodes, the fastest 
growing mode for the given $\delta$, $\Mb$ and $\lambda/\Rs$ dominates the instability.

\smallskip
\item 
The allowed range of parameters for cold streams in massive galaxies at high-$z$ 
is near the transition between surface modes and body modes. Thus, even minor variations 
in the stream parameters within their realistic ranges can have large affects on the 
growth rate of instabilities. For a realistic range of stream parameters, the number of 
e-foldings of growth within a virial crossing time can range from roughly 0.1 to 100. 
This implies that KHI could in principle have an important role in the evolution of cold 
streams, and a study of the non-linear phases of the instability and the effects of additional 
physical processes is well motivated. 

\smallskip
\item
When non-linear effects become important, the perturbation amplitude 
saturates at first, before continuing to grow linearly with time via 
the mergers of vortices. In an upcoming paper (Padnos et al., in 
preparation) we will address this process in detail, both analytically 
and using simulations.

\smallskip
\item 
Rough, order of magnitude estimates suggest that heat conduction and 
magnetic fields should not drastically alter the linear analysis presented 
here. On the other hand, the gravitational potential of the halo may 
enhance the instability by causing the stream to become narrower closer to 
the halo centre, decreasing the sound crossing time and thus increasing the 
growth rate of KHI. Cooling is unlikely to drastically affect the linear evolution, 
but can have significant effects in the non-linear regime. In future work we 
will address all these effects one by one in more detail, while in parallel 
we will study the effects of galaxy formation on cold streams using cosmological 
simulations with tailored refinement in the streams.

\end{enumerate}

\section*{Acknowledgments} 
We thank the referee, J. Xavier Prochaska, for helpful 
comments that improved the quality of this manuscript. 
We thank Romain Teyssier for making \texttt{RAMSES} 
publicly available. We thank Frederic Bournaud, John 
Forbes, Sharon Lapiner, Baruch Meerson, Santi Roca-Fabrega, 
Eva Ntormousi and Almog Yalinewitch for helpful discussions. 
The simulations were performed on the Astric cluster at 
HU. This work was supported by ISF grants 24/12, 1059/14 
and 1829/12, by BSF grant 2014-273, by the I-CORE Program 
of the PBC, by NSF grants AST-1010033 and AST-1405962, 
and by ARC grant DP160100695.

\bibliographystyle{mn2e}

\appendix 
\section{Branch Cuts of the Solutions}
\label{sec:branch_cut} 

\smallskip 
Since the generalized wavenumbers $\qbs$ (\equnp{q_bs}) are given 
by the square root of a complex number, we must define a branch cut 
in the complex plane. As mentioned in \se{sheet} we have chosen to 
have ${\rm Re}(\qbs)\ge 0$. This is equivalent to defining arguments 
of complex numbers between $-\pi$ and $\pi$. If the argument of $\qbs^2$ 
is $2\alpha \in (-\pi, \pi)$, then the argument of $\qbs$ is 
$\alpha \in (-\pi/2, \pi/2)$ and we have ${\rm Re}(\qbs) > 0$. This 
ensures that the amplitude of the perturbation decays, rather than 
grows, away from the interface between the fluids. However, in order 
to understand how waves generated by the instability propagate away 
from the interface, we must also determine the sign of ${\rm Im}(\qbs)$. 
We limit our discussion here to forward travelling growing modes, 
i.e. modes with $k,\:{\rm Re}(\omega),\:{\rm Im}(\omega)>0$. 

\smallskip 
We define $R\,{\rm exp}(i\theta) \equiv (\omega - kv)/c$. 
If ${\rm Re}(\omega)<kv$, then $\pi/2<\theta<\pi$ and 
$q^2 = \left(k^2-R^2{\rm cos}(2\theta)\right) - i\left(R^2{\rm sin}(2\theta)\right)$ 
has a positive imaginary part. Therefore, if the argument 
of $q$ is $\alpha$, we have $0<2\alpha<\pi\Rightarrow 0<\alpha<\pi/2$, 
which gives ${\rm Im}(q)>0$. On the other hand, if ${\rm Re}(\omega)>kv$, 
similar arguments yield $-\pi/2<\alpha<0$, so that ${\rm Im}(q)<0$. 

\smallskip 
In all cases presented in the text, we assumed that $\vb=0$ 
and $\vs=V$, and found $0<{\rm Re}(\omega)<kV$. This means 
that ${\rm Im}(\qb)<0$ and ${\rm Im}(\qs)>0$. 

\section{Solving the Dispersion Relation of the Compressible Sheet}
\label{sec:sheet_deriv} 

\smallskip
In this appendix we examine the 6 solutions to \equ{unitless_dispersion_sheet_2} 
and explain why only 2 of them are solutions to the dispersion relation of the 
compressible sheet(\equnp{unitless_dispersion_sheet}). 

\smallskip
The two solutions to the quadratic part of the equation are 
$\varpi^2 = \delta(1-\varpi)^2$. Both of these result in 
$\qb=\qs$ and in $Z=-1$. Therefore, neither of these are 
solutions to the dispersion relation, but only come about 
because \equ{unitless_dispersion_sheet} was squared.

\begin{figure*}
\centering
\subfloat{\includegraphics[width =0.95 \textwidth]{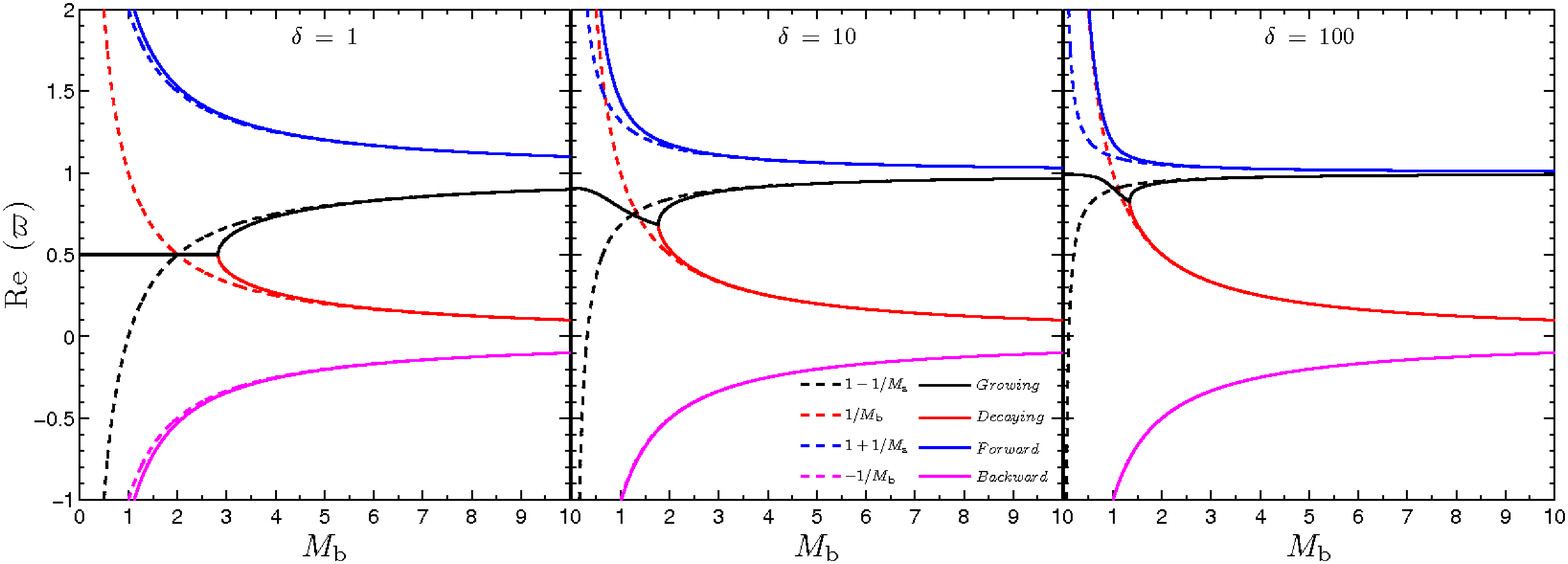}}
\caption{Real solutions to the quartic part of \equ{unitless_dispersion_sheet_2}. 
We show ${\rm Re}(\varpi)$ as a function of $\Mb$, for three different values of 
$\delta$, as marked in the panels. Solid lines show numerical solutions. At low Mach 
numbers, two of the solutions are complex conjugates and therefore have the same real 
part, shown by the black solid line at low Mach numbers. These are the unstable growing 
and decaying mode, and the imaginary part is shown in \fig{planar_growth_rate}. This 
line bi-furcates into the solid black and red lines at the critical Mach number, $M_{\rm crit}$ 
from \equ{Mcrit}, above which the system becomes stable and these modes become independent 
travelling waves. The other four lines are stable travelling waves, which are not actually 
solutions to the dispersion relation $Z=1$. The four dashed lines show the asymptotic high-Mach 
number solutions from \equ{real_phi}. The fit is excellent in the entire stable regime.}
\label{fig:planar_real} 
\end{figure*} 

\smallskip
Attempting to find tractable analytic expressions for the four 
solutions to the quartic part of \equ{unitless_dispersion_sheet_2} 
were unsuccessful. However, analytic insight can still be gained 
by examining its discriminant. Given the general quartic equation 
\be 
\label{eq:quartic}
f(x)=ax^4+bx^3+cx^2+dx+e,
\ee
{\no}the discriminant is given by 
\be 
\label{eq:disc_0}
\begin{array}{c}
\Delta_0= 256a^3e^3 - 192a^2bde^2 - 128a^2c^2e^2 \\
+ 144a^2cd^2e -27a^2d^4 + 144ab^2ce^2 - 6ab^2d^2e  \\
- 80abc^2de + 18abcd^3 + 16ac^4e - 4ac^3d^2 \\
- 27b^4e^2 + 18b^3cde - 4b^3d^3 - 4b^2c^3e + b^2c^2d^2
\end{array}.
\ee
{\no}When $\Delta_0<0$, the quartic has two real roots and two 
complex conjugate roots \citep{Quartic}. For the quartic part 
of \equ{unitless_dispersion_sheet_2} we have 
\be 
\label{eq:disc2_0}
\begin{array}{c}
\Delta_0=16\Mb^2\delta^{-4} \cdot [\delta^3\Mb^6 - 3\delta^2(1+\delta)\Mb^4 \\
\\
+ 3\delta(\delta^2-7\delta+1)\Mb^2 - (\delta+1)^3]
\end{array}.
\ee
{\no}It is straightforward to show that $\Delta_0<0$ if and only if 
$\Mb<M_{\rm crit}$, where $M_{\rm crit}$ is given by \equ{Mcrit}. 
Since the solution must converge to that of the incompressible sheet 
(\equnp{KKHI2}) for $\Mb<<1$, and since this admits only two complex 
conjugate solutions, we deduce that the two real solutions to the quartic 
part of \equ{unitless_dispersion_sheet_2} at $\Mb<M_{\rm crit}$ are not 
solutions to the dispersion relation, $Z=1$. Rather they must also be 
solutions to $Z=-1$, similar to the two solutions to the quadratic part. 
We discuss this further below. 

\smallskip
For $\Mb>M_{\rm crit}$ we obtain $\Delta_0>0$ and the nature of roots 
depends on the signs of two additional parameters,
\be 
\label{eq:disc_1}
\Delta_1=8ac-3b^2,
\ee
\be 
\label{eq:disc_2}
\Delta_2 = 64a^3e-16a^2c^2+16ab^2c-16a^2bd-3b^4.
\ee
{\no}If both $\Delta_1<0$ and $\Delta_2<0$ when $\Delta_0>0$, then 
all four roots are real and distinct. Otherwise the four roots are 
two pairs of complex conjugates \citep{Quartic}. For the quartic part 
of \equ{unitless_dispersion_sheet_2} we have 
\be 
\label{eq:disc2_1}
\Delta_1 = -\frac{4\Mb^2\left(2+\delta(2+\Mb^2)\right)}{\delta^2} < 0,
\ee
\be 
\label{eq:disc2_2}
\Delta_2 = -\frac{16\Mb^2(1+\delta)(1+\delta+2\delta\Mb^2)}{\delta^4} < 0.
\ee
{\no}Since $\Delta_1$ and $\Delta_2$ are always negative, all four roots 
are real when $\Mb>M_{\rm crit}$, and the sheet is stable. 

\smallskip
When $\Mb>>1$, asymptotic solutions can be found for the quartic part 
of \equ{unitless_dispersion_sheet_2}, 
\be 
\label{eq:real_phi}
\varpi \simeq \pm\frac{1}{\Mb}, \: 1\pm\frac{1}{\Ms},
\ee
{\no}where $\Ms=\delta^{1/2}\Mb$. Note that all four of these 
solutions result in $\qb$ and $\qs$ that are purely imaginary. 

\smallskip 
When $\varpi=-\Mb^{-1}$ we have $\omega<0$ and when $\varpi=1+\Ms^{-1}$ 
we have $\omega>kV$. Therefore, following the discussion in \se{branch_cut}, 
$\qb$ and $\qs$ will have the same sign, so $\qs/\qb>0$ and $Z<0$. Therefore, 
these cannot be solutions to $Z=1$, but must yield $Z=-1$. We conclude that 
these are the high-Mach number limits of the two roots that were real even 
for $\Mb<M_{\rm crit}$, which we already determined were not true solutions 
to the dispersion relation. The other two solutions, $\varpi=\Mb^{-1},\:1-\Ms^{-1}$ 
both result in $0<\omega<kV$ and therefore $\qs/\qb<0$ and $Z>0$, meaning that 
$Z=1$. Thus, these are the high-Mach number limits of the true solutions to 
the dispersion relation.

\smallskip
In \fig{planar_real} we show numerical solutions to the quartic part of 
\equ{unitless_dispersion_sheet_2} as a function of Mach number, $\Mb$, 
for different values of the density contrast, $\delta$. For the two complex 
conjugates at low Mach numbers, representing the growing and decaying unstable 
modes, we show only the real part which is the same for both solutions (the 
complex part is shown in \fig{planar_growth_rate}). At $\Mb>M_{\rm crit}$, 
these modes stabilize and become two independent travelling waves. This can 
be seen as the bifurcation point between the red and black lines in the figure. 
The other two solutions are always real, and can be thought of as a forward 
travelling wave with $\varpi>0$ and a backward travelling wave with $\varpi<0$. 
However, as discussed above, these are not solutions to the dispersion relation. 
At high Mach numbers, the four solutions to the quartic are well approximated by 
the asymptotic expressions in \equ{real_phi}, shown with dashed lines. These 
represent stable sound waves with phase velocities $\omega/k = v \pm \cs,\: \pm \cb$.

\section{Long Wavelength Limit of the Compressible Slab}
\label{sec:long_comp_slab} 

\smallskip
In this section we derive the growth rate of the compressible slab 
in the long wavelength limit, where $K<<1$. The first step is to 
verify that in this limit, $\qs\Rs<<1$ as well, as this is what is 
needed to simplify the dispersion relation in \equ{dispersion_slab_unitless}. 
We accomplish this by showing that as $k\rightarrow 0$, $\omega \rightarrow 0$ 
as well. Otherwise, if $k=0$ and $\omega\neq 0$ \equ{q_bs} reduces to 
$q_{\rm b,s}=\pm i \omega/\cbs$. On the one hand, we are only interested 
in growing modes where ${\rm Im}(\omega)>0$, and on the other hand we require 
${\rm Re}(q_{\rm b,s})>0$. Therefore we take $q_{\rm b,s}= -i \omega/\cbs$. 
Inserting this into \equ{dispersion_slab} yields 
\be 
\label{eq:large_k_finite_w}
T\left(-i \frac{\omega \Rs}{\cs}\right) = -\delta^{-1/2}.
\ee
{\no}Solving for $\omega$ results in 
\be 
\label{eq:large_k_finite_w2}
\omega = \frac{\cs}{2\Rs}\left[-n\pi +i{\rm ln}\left(\frac{\sqrt{\delta}-1}{\sqrt{\delta}+1}\right)\right],
\ee
{\no}where $n$ is any whole number, even for S-modes and odd for P-modes. 
For any $\delta>0$, \equ{large_k_finite_w2} results in ${\rm Im}(\omega)<0$, 
which is in contradiction to our original assumption. We conclude that there 
are no solutions with $\omega(k=0)\neq 0$. Therefore, $\qbs \rightarrow 0$ 
as $k\rightarrow 0$, and ${\rm tanh}(\qs \Rs)\simeq 1/{\rm coth}(\qs\Rs) \simeq \qs \Rs$. 
Inserting this into \equ{dispersion_slab_unitless} yields
\begin{subequations}
\label{eq:dispersion_slab_small_k}
\begin{equation} 
\label{eq:dispersion_slab_small_k_a}
\begin{array}{c c}
K = -\dfrac{\varpi ^2}{\delta \left(\varpi-1\right)^2\left(1- \Mb^2\varpi^2\right)^{1/2}} &{\rm (S)}
\end{array}
\end{equation}
\begin{equation} 
\label{eq:dispersion_slab_small_k_b}
\begin{array}{c c}
K = -\dfrac{\delta \left(\varpi-1\right)^2 \left(1- \Mb^2\varpi^2\right)^{1/2}}{\varpi^2 \left(1- \delta\Mb^2(\varpi-1)^2\right) } &{\rm (P)}
\end{array}
\end{equation}
\end{subequations}

\smallskip
\textbf{S-modes:} The solution to \equ{dispersion_slab_small_k_a} 
for $K=0$ is $\varpi_0=0$. For $K<<1$, the lowest order correction 
is $\varpi_1 \simeq \pm i\delta^{1/2}K^{1/2}$. To find the next 
order correction, we insert $\varpi = \varpi_1 + \varpi_2$ with 
$\varpi_2 << \varpi_1 << 1$. This results in $\varpi_2 \simeq \delta K$, 
which is the leading order term in ${\rm Re}(\varpi)$. 
Thus, at long wavelengths S-modes are unstable with the approximate 
dispersion relation 
\be 
\label{eq:omega_fundamental_s_mode}
\omega_{\rm S,\,f} \simeq \dfrac{V}{\Rs}\left[\delta K^2 \pm i\delta^{1/2}K^{3/2}\right].
\ee

\smallskip
\textbf{P-modes:} When $K=0$, there are two solutions to 
\equ{dispersion_slab_small_k_b}: $\varpi_0 = 1$ and $\varpi_0 = \Mb^{-1}$. 
In the vicinity of $\varpi_0=1$ the system is unstable. 
Inserting $\varpi=1+\varpi_1$ with $\varpi_1<<1$ into 
\equ{dispersion_slab_small_k_b} yields 
$\varpi_1\simeq \pm i K^{1/2}\delta^{-1/2}\left(1-\Mb^2\right)^{-1/4}$. 
So at long wavelengths, P-modes are unstable with the approximate dispersion 
relation 
\be 
\label{eq:omega_fundamental_p_mode}
\omega_{\rm P,\,f} \simeq \dfrac{V}{\Rs}\left[K \pm i\delta^{-1/2}(1-\Mb^2)^{-1/4}K^{3/2}\right].
\ee
{\no}For the special case of $\Mb = 1$, one gets instead 
$\varpi_1^{5/2}\simeq (iK)/(\sqrt{2}\delta)$ and the 
dispersion relation becomes, to lowest order in $K$, 
\be 
\label{eq:omega_fundamental_p_mode_M_1}
\omega_{\rm P,\,f} \simeq \dfrac{V}{\Rs}\left[K \pm (-2\delta^2)^{-1/5}K^{7/5}\right].
\ee

\smallskip
When $\Mb\neq1$, the second solution 
to \equ{dispersion_slab_small_k_b}, 
$\varpi_0 = \Mb^{-1}$, is stable. 
Inserting $\varpi=\Mb^{-1}(1+\varpi_1)$ 
with $\varpi_1<<1$ into 
\equ{dispersion_slab_small_k_b} yields 
\be 
\varpi_1 = -\frac{1}{2}\left[1-\frac{1}{\delta (\Mb-1)^2}\right]^2 K^2
\ee

\section{Short Wavelength Limit of the Compressible Slab}
\label{sec:short_comp_slab} 

\smallskip
In this section we discuss in detail the short wavelength limit 
of the slab, where $K>>1$. There are two solutions to the dispersion 
relation, \equs{inverted_dispersion_slab}, for $K\rightarrow\infty$. 
These are $Z=1$ and $1-\delta \Mb^2(\varpi-1)^2 = 0$. However, care 
must be taken to ensure that $K$ remain real, which must be the case 
in the temporal analysis we are discussing. By examining 
\equs{inverted_dispersion_slab}, we see that there are two different 
regimes, depending on the parameter 
\be 
\label{eq:planar_convergence}
\gamma \equiv \frac{{\rm Im}\left(\sqrt{1-\delta \Mb^2(\varpi-1)^2}\right)}{{\rm Re}\left(\sqrt{1-\delta \Mb^2(\varpi-1)^2}\right)} = \frac{{\rm Im}(\qs)}{{\rm Re}(\qs)}.
\ee

\begin{figure}
\centering
\includegraphics[trim={0.35cm 0.8cm 0.2cm 1.0cm}, clip, width =0.45 \textwidth]{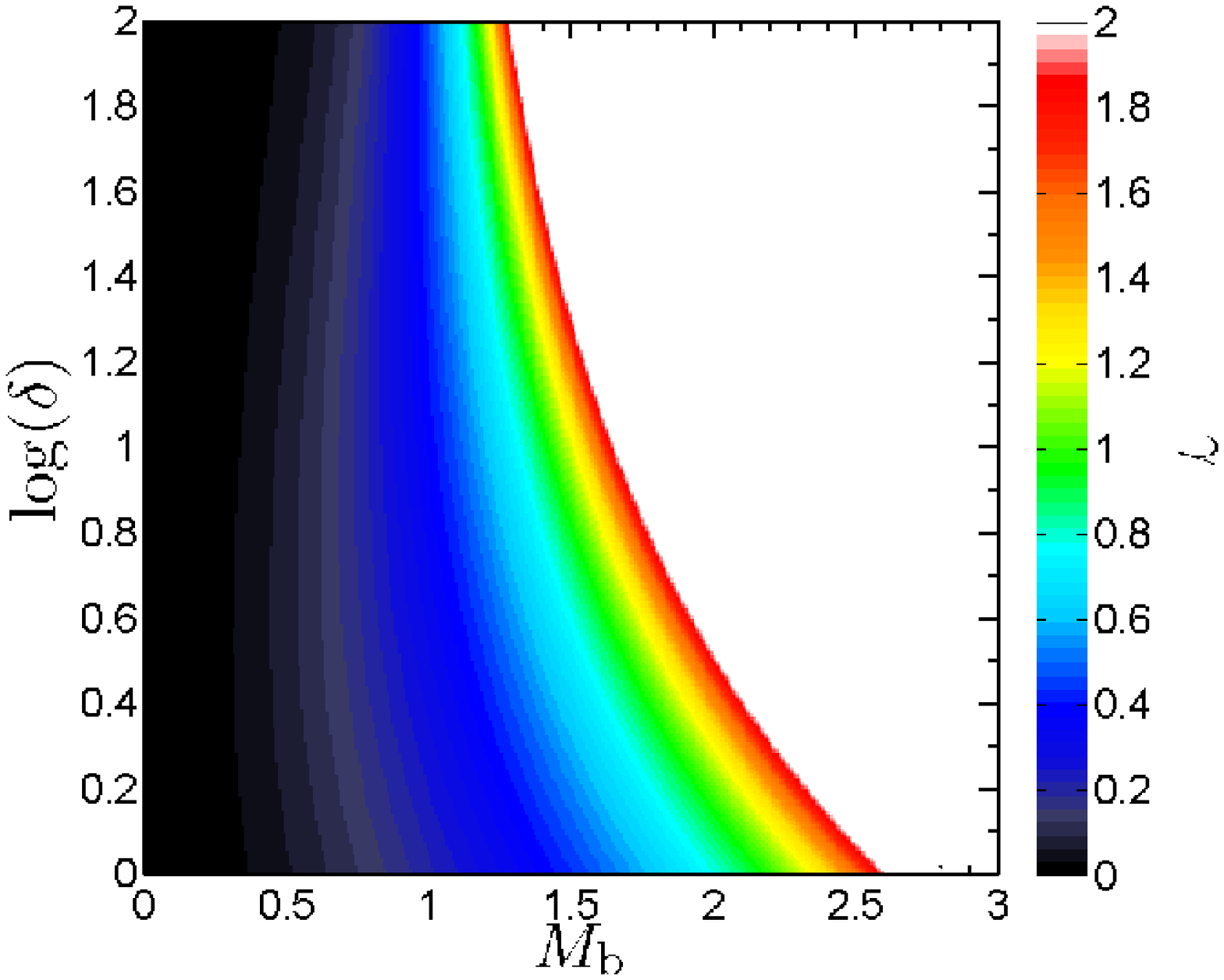}
\caption{The value of $\gamma={\rm Im}(\qs)/{\rm Re}(\qs)$ 
for the sheet solution. When $\gamma<<1$, the dispersion 
relation of the slab converges to that of the sheet, $Z=1$, 
at short wavelengths. This happens only for low Mach numbers, 
$\Mb<<1$. When $\gamma>>1$, which happens when the sheet is 
stable, the slab converges to the solution $\varpi=1-\Ms^{-1}$, 
which is the high-Mach number stable limit of the growing mode 
solution in the sheet.}
\label{fig:gamma} 
\end{figure}

\smallskip
If $\gamma<<1$, then $K$ will remain real as $Z\rightarrow 1$, in other 
words the slab dispersion relation converges to that of the sheet at short 
wavelengths, as expected. In this case, $\qs$ is very nearly real which means 
that these solutions represent \textit{surface modes}, similar to the sheet. 
These decay exponentially with depth in the slab and prevent the two surfaces 
from coming into causal contact. We examine the regime of validity of this 
solution in \fig{gamma}, where we show the value of $\gamma$ obtained for 
solutions to the sheet, i.e. $Z=1$, as a function of $\Mb$ and $\delta$. 
We see that $\gamma<<1$ only for low Mach numbers, $\Mb<<1$. At $\Mb\sim 1$, 
we have $\gamma\lsim 1$ and the slab will deviate somewhat from the sheet 
solution of $Z=1$ even at short wavelengths. As $\Mb$ approaches $M_{\rm crit}$ 
and the sheet becomes stable, $\qs$ becomes purely imaginary and thus 
$\gamma>>1$, so $Z=1$ is not a valid solution to the slab at short wavelengths. 

\smallskip
When $\gamma>>1$, the short wavelength limit of the slab is given by 
$1-\delta \Mb^2(\varpi-1)^2 \rightarrow 0$. In this case, 
$\varpi \rightarrow \varpi_\infty \equiv 1\pm 1/\Ms$, which as 
we saw is the high-Mach number asymptotic value of the growing mode 
solution in the sheet (\equnp{real_phi}). So either way, the slab 
converges to the sheet at short wavelengths. This leads to $Z<<1$, 
which ensures that $K$ is real in \equs{inverted_dispersion_slab}. 
Such solutions have $\qs$ is very nearly imaginary which means that 
they represent \textit{body modes}, which penetrate deep into the 
slab. In such a case, stable waves emanate from the interfaces and 
do not decay. These waves will be reflected off the slab boundaries, 
causing the two sides to come into causal contact.

\section{Stability and Marginal Stability in the Slab}
\label{sec:Marginal1} 

\begin{figure*}
\centering
\subfloat{\includegraphics[width =0.95 \textwidth]{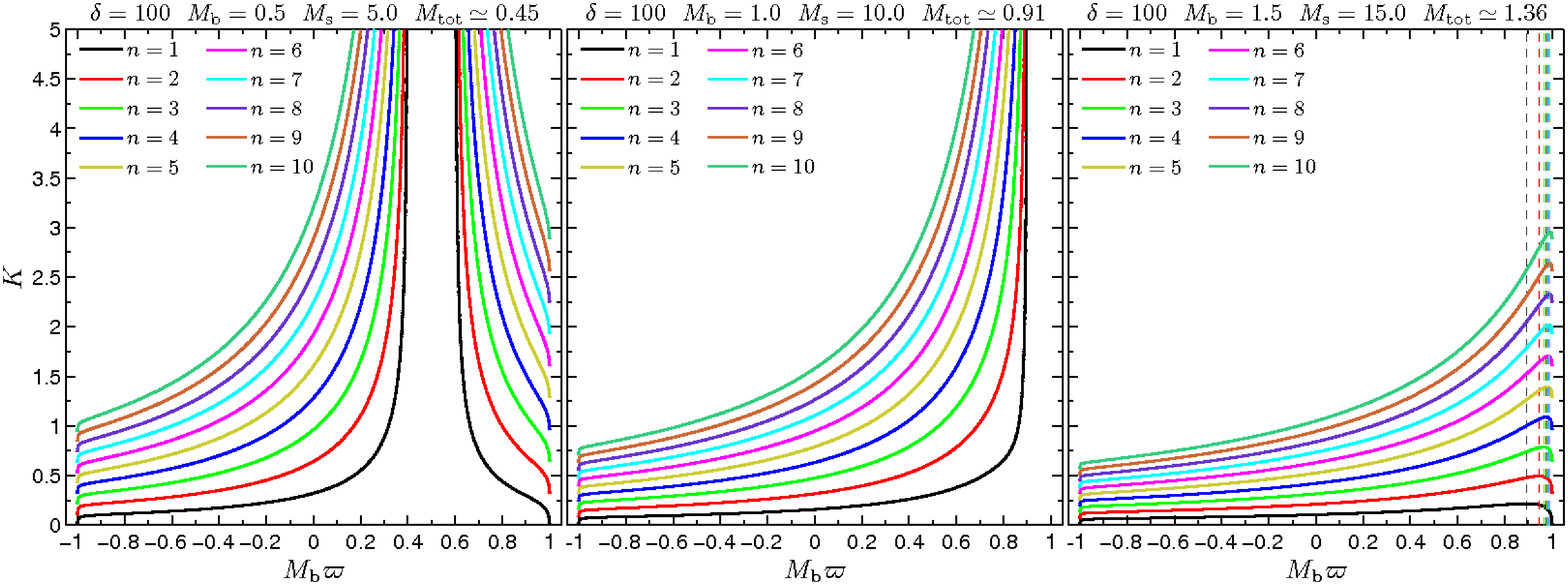}}
\caption{The first 10 modes of \equ{marginal_3}, for three different 
regimes of $\Mb$. The left hand panel represents low Mach numbers, 
$\Mb<1-1/\sqrt{\delta}$, and the middle panel represents intermediate 
Mach numbers, $1-1/\sqrt{\delta}<\Mb<1+1/\sqrt{\delta}$. In both these 
cases there are no marginally stable points and each mode is stable at 
all wavelengths. The right hand panel represents high Mach numbers, 
$1+1/\sqrt{\delta}<\Mb$, where the flow is super-sonic with respect to 
the sum of the two sound speeds $M_{\rm tot}=v/(\cb+\cs)>1$. In this 
case, each $K(\varpi)$ curve has a maximum in the range $0<\varpi<\Mb^{-1}$, 
marked with a dashed line. These maxima, $K_{\rm n}$, represent marginally 
stable points, since increasing $K$ beyond $K_{\rm n}$ introduces an imaginary 
component to $\varpi$ and triggers an instability.}
\label{fig:marginal}
\end{figure*}

\smallskip 
In this section we analytically find all stable 
solutions to the slab dispersion relation, i.e. 
solution to \equs{inverted_dispersion_slab} where 
both $K$ and $\varpi$ are real. This is a prelimenary 
step towards finding the \textit{marginally stable 
points} of the system. These are points in $(\varpi,K)$ 
space where both parameters are real, but in the vicinity 
of which one or both of them become complex. In our temporal 
analysis $K$ is real by definition, so a marginally stable 
point is a stable solution where an infinitesimal change in 
$K$ results in complex $\varpi$, which must correspond to an 
extremal point of the function $K(\varpi)$ where $\varpi$ is 
real. To see this, consider a solution with $\varpi=0,\:K(0)$, 
and imagine increasing $K$ (decreasing the wavelength $\lambda$). 
So long as $K(\varpi)$ is monotonic, there is still a solution with 
real $\varpi$, and therefore the solution remains stable. But when 
$K(\varpi)$ reaches an extremal point, decreasing the wavelength further 
requires extending $\varpi$ to the complex plane, indicating an instability 
(\fig{marginal}, discussed below). 

\smallskip
There are two main branches of stable solutions, depending on whether 
$\qs$ is real or imaginary. We will address each of these separately, 
as they define different families of solutions, for low and high Mach 
numbers.

\subsection{$\qs \in \mathbf{R} \Rightarrow 1-\delta\Mb^2(1-\varpi)^2>0 $}

\smallskip
Such solutions represent surface waves that decay exponentially 
within the slab, with a penetration depth of $\qs^{-1}$. Such modes 
behave similarly to the sheet at short wavelengths since the two 
sides of the slab are not in contact with each other (\se{short_comp_slab}). 
For $K$ to be real we require 
\begin{subequations}
\label{eq:marginal_1a}
\begin{equation} 
\label{eq:marginal_1aa}
K = {\rm Re}(K) = \frac{{\rm ln}\left(|1+Z|\right)-{\rm ln}\left(|1-Z|\right)}{2\sqrt{1-\delta \Mb^2(\varpi-1)^2}},
\end{equation}
\begin{equation} 
\label{eq:marginal_1ab}
{\rm Im}(K) = \frac{{\rm arg}(1+Z)-{\rm arg}(1-Z)+n\pi}{2\sqrt{1-\delta\Mb^2(\varpi-1)^2}} = 0.
\end{equation}
\end{subequations}

\smallskip
If $\qb\in\mathbf{R}$, then $1-\Mb^2\varpi^2\ge0$ and $Z=-|Z|$ is on the 
negative real axis. If $|Z|\le 1$, then  ${\rm arg}(1+Z)=0$ and \equ{marginal_1ab} 
can be satisfied with $n=0$. If $|Z|>1$, then ${\rm arg}(1+Z)=\pi$ and 
\equ{marginal_1ab} can be satisfied with $n=-1$. Either way, \equ{marginal_1aa} 
results in $K\le0$. Physically, we require $K=|k|\Rs\ge0$, so the only 
relevant solutions in this case are those with $K=0$. If $n=0$ then 
$-1\le Z \le 0$, so $K=0$ is only possible for $Z=0$. Since we have 
assumed here $\qs>0$, this in turn requires $\varpi=0$, which corresponds 
to the fundamental S-mode (\se{long_comp_slab}). If $n=-1$ then $Z<-1$, 
so $K=0$ is only possible for $Z=-\infty$. This in turn requires $\varpi=1$ 
(or $\Mb^{-1}$), which corresponds to the fundamental P-mode (\se{long_comp_slab}). 
Recall that at $K=0$, the system is unstable for $\varpi=0$ and $\varpi=1$, 
but stable for $\varpi=\Mb^{-1}$ (\se{long_comp_slab}). 

\smallskip
Note that our assumptions above, whereby $\qs>0$ and $\qb\ge0$, limit the range 
of $\varpi$ to $1-\Ms^{-1}<\varpi<1+\Ms^{-1}$ and $-\Mb^{-1}\le\varpi\le\Mb^{-1}$. 
The solutions $\varpi=0,\:1,\:\Mb^{-1}$ are only included in this range 
at low Mach numbers: $\Ms<1$, $\Mb\le 1$ and $1-1/\sqrt{\delta}<\Mb<1+1/\sqrt{\delta}$ 
respectively. 

\smallskip
On the other hand, if $\qb\in\mathbf{I}$ so that 
$1-\Mb^2\varpi^2 < 0$, then $\qb = -i\left|\qb\right|$ 
and $Z=-i|Z|$ is on the negative imaginary axis. 
Therefore $-\pi/2\le{\rm arg}(1 \pm Z)\le \pi/2$ and 
\equ{marginal_1ab} can be satisfied with $n=0$ only 
if $Z=0$, which would require $\varpi=0$, in contradiction 
to the assumption that $1-\Mb^2\varpi^2 < 0$. A solution 
with $n=-1$ is possible if $|Z|=\infty$ so that 
$(1 + Z)/(1 - Z) = -1$. In this case, provided $\Mb>1$, 
the solution is $K=0$ and $\varpi=1$, corresponding to 
the unstable fundamental (P) mode. The solution 
$\varpi=\Mb^{-1}$ is in contradiction to the assumption 
that $1-\Mb^2\varpi^2 < 0$. 

\subsection{$\qs \in \mathbf{I} \Rightarrow 1-\delta\Mb^2(1-\varpi)^2<0 $}

\smallskip
Such solutions represent waves that do not decay spatially 
within the slab. They travel from one interface to the other 
and are reflected off of and transmitted through the slab 
boundaries. These are \textit{body modes}, or \textit{reflected modes}. 
In this case, since the two sides of the slab can interact with 
each other through the reflection of waves, the slab will differ 
greatly from the sheet. In this regime $\qs=i|\qs|$ (\se{branch_cut}) 
and $\varpi$ must obey either $1+\Ms^{-1}<\varpi$ or $\varpi<1-\Ms^{-1}$. 
For $K$ to be real we require 
\begin{subequations}
\label{eq:marginal_2a}
\begin{equation} 
\label{eq:marginal_2aa}
K = {\rm Re}(K) = \frac{{\rm arg}(1+Z)-{\rm arg}(1-Z)+n\pi}{2\sqrt{\delta\Mb^2(\varpi-1)^2-1}},
\end{equation}
\begin{equation} 
\label{eq:marginal_2ab}
{\rm Im}(K) = -\frac{{\rm ln}\left(|1+Z|\right)-{\rm ln}\left(|1-Z|\right)}{2\sqrt{\delta \Mb^2(\varpi-1)^2-1}} = 0.
\end{equation}
\end{subequations}

\smallskip
If $\qb\in\mathbf{I}$ so that $1-\Mb^2\varpi^2 < 0$, then $\qb = -i\left|\qb\right|$ 
(\se{branch_cut}) and $Z=+|Z|$ is on the positive real axis. \Equ{marginal_2ab} can 
thus only be satisfied if $Z=0$, which requires $\varpi=0$, which is in contradiction 
to the assumption that $1-\Mb^2\varpi^2 < 0$. We therefore conclude that no such solution 
exists. 

\smallskip
On the other hand, if $\qb\in\mathbf{R}$ so that $1-\Mb^2\varpi^2 \ge 0$, 
then $Z=-i|Z|$ is on the negative imaginary axis. In this case $1-Z=1+{\bar{Z}}$ 
and \Equ{marginal_2ab} is always satisfied. Furthermore, \equ{marginal_2aa} can 
be rewritten as 
\begin{equation} 
\label{eq:marginal_3}
K = \frac{-{\rm arctan}(|Z|) + (n/2)\pi}{\sqrt{\delta\Mb^2(\varpi-1)^2-1}}.
\end{equation}
{\no}For $n < 0$ \equ{marginal_3} always results in $K < 0$, which is not 
relevant for our discussion. For $n=0$, \equ{marginal_3} results in $K \le 0$ 
so the only relevant solution is $K=0$, which requires $|Z|=0$, which in turn 
requires $\varpi=0$. Provided $\Ms>1$, this corresponds to the fundamental 
S-mode. 

\smallskip
For $n\ge 1$, \equ{marginal_3} results in $K \ge 0$ for all $\varpi$. 
However, our assumptions of $\qs^2<0$ and $\qb^2\ge 0$ limit the 
range of allowed $\varpi$ values to $-\Mb^{-1}\le \varpi \le \Mb^{-1}$ 
and either $1+\Ms^{-1}<\varpi$ or $\varpi<1-\Ms^{-1}$. For the $n$-th 
mode, $\varpi=0$ when $K=K_{\rm n,0} = n\pi/\left(2\sqrt{\Ms^2-1}\right)$, 
provided $\Ms>1$.

\smallskip
In \Fig{marginal} we show $K$ as a function of $(\Mb\varpi)$ from \equ{marginal_3}, 
for $n=1-10$. The three panels are representative of three different regimes of Mach 
number. The left panel represents low Mach numbers, $\Mb<1-1/\sqrt{\delta}$, corresponding 
to flow velocities of $V<\cb-\cs$. In this case, both branches of $\varpi$ are accessible. 
$K(\varpi)$ increases monotonically from $\varpi=-\Mb^{-1}$ until $\varpi=1-\Ms^{-1}$ where 
$K\rightarrow\infty$. It then decreases monotonically from $\varpi=1+\Ms^{-1}$, where again 
$K\rightarrow\infty$, until $\varpi=\Mb^{-1}$. The center panel represents intermediate Mach 
numbers, $1-1/\sqrt{\delta}<\Mb<1+1/\sqrt{\delta}$, corresponding to flow velocities 
$\cb-\cs<V<\cb+\cs$. In this case, only $\varpi<1-\Ms^{-1}$ is accessible, but the qualitative 
behaviour of $K(\varpi)$ in this regime is the same as in the previous case, shown in the left 
panel. In both these cases, there are \textit{no} marginally stable points with $K>0$. Solutions 
with $n\ge 1$ are stable at all wavelengths. Once the $n$-th mode has been excited, at 
$K=K_{\rm n,0}$, we can increase $K$ (decrease the wavelength) continuously and there will 
always be a solution to the mode with real $\varpi$. Since we never require complex $\varpi$, 
no instability is ever triggered.

\smallskip
The right hand panel of \fig{marginal} represents large Mach numbers, $1+1/\sqrt{\delta}<\Mb$, 
corresponding to flow velocities which are super-sonic with respect to the sum of the two sound 
speeds: $V>\cb+\cs\Rightarrow M_{\rm tot}>1$. In this case, the $K(\varpi)$ curve corresponding 
to each mode has exactly one maximum point in the range $0<\varpi<\Mb^{-1}$, as shown in the figure. 
We refer to these maxima as $K_{\rm n}$ and $\varpi_{\rm n}$. \textit{These are marginally stable 
points}.

\section{Marginally Stable Points for Large $n$}
\label{sec:Marginal2} 

\begin{figure*}
\centering
\subfloat{\includegraphics[width =0.95 \textwidth]{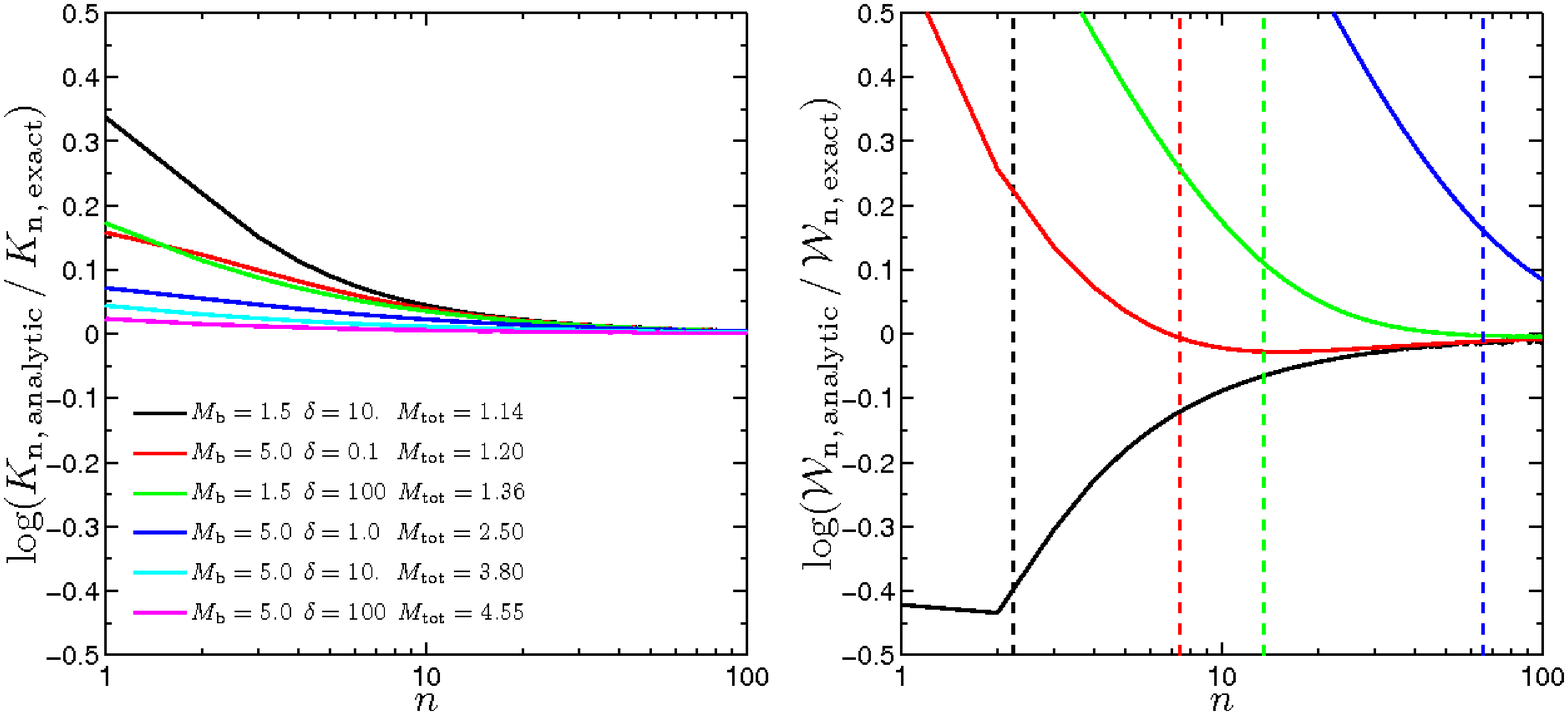}}
\caption{Comparison of our analytic approximation for the marginally stable points from 
\equs{marginal_analytic}, with the exact solution obtained by numerically finding the 
maxima of \equ{marginal_3}. The left panel compares $K_{\rm n}$ and the right panel 
compares $\mathcal{W}_{\rm n}$. Solid lines show the log of the ratio of the analytic 
estimate to the numerical solution for different combinations of $\Mb$ and $\delta$, 
corresponding to different values of $M_{\rm tot}$. In all cases examined, there is 
less than $10\%$ error in the approximation for $K_{\rm n}$ by $n\sim 10$, and for 
$M_{\rm tot}\gsim 2$ this is the case already at $n\sim 2$. On the other hand, the 
approximation for $\mathcal{W}_{\rm n}$ converges only for $n>n_{\rm 1}$ (\equnp{marginal_n}, 
dashed lines), which increases rapidly with $M_{\rm tot}$. The curves with $M_{\rm tot}=3.80$ 
and $4.55$ lie outside the bounds of the panel, converging only at very high $n$.
}
\label{fig:marginal_analytic}
\end{figure*}

\smallskip
The marginally stable points are given by $dK/d\varpi = 0$ with $K(\varpi)$ 
given by \equ{marginal_3}. This equation cannot be solved analytically in 
general, but in this section we derive an analytical approximation for large $n$. 

\smallskip 
Based on the discussion in the previous section, when $M_{\rm tot}>1$ 
\equ{marginal_3} is continuous in the range $-\Mb^{-1}\le\varpi\le\Mb^{-1}$. 
Since $0\le{\rm arctan}(|Z|)\le\pi/2$, for $n>>1$ \equ{marginal_3} can 
be approximated as $K \simeq (n\pi/2)\left(\delta\Mb^2(\varpi-1)^2-1\right)^{-1/2}$, 
which is a monotonically increasing function of $\varpi$. We conclude that 
for $n>>1$, $\varpi_{\rm n}$ converges to $\Mb^{-1}$ (see the right hand panel 
of \fig{marginal}). When $\varpi=\Mb^{-1}$ we get $|Z|=\infty$, so for 
$\varpi_{\rm n}=\Mb^{-1}(1-\mathcal{W})$, with $0<\mathcal{W}<<1$, we have 
$|Z|>>1$ and ${\rm arctan}(|Z|) = \pi/2 - |Z|^{-1} + O\left(|Z|^{-3}\right)$. 
To leading order in $\mathcal{W}$, 
\be 
\label{eq:Zn}
|Z|^{-1} \simeq \frac{\sqrt{2}\delta(\Mb-1)^2}{\sqrt{\delta(\Mb-1)^2-1}}\mathcal{W}^{1/2}.
\ee

\smallskip
In order to find $K_{\rm n}$, the maximum of \equ{marginal_3}, 
we expand $K$ to the second leading order in $\mathcal{W}$:  
\begin{subequations}
\label{eq:approx_Xn_equ}
\begin{equation} 
K \simeq K_0 + K_1\mathcal{W}^{1/2} + K_2\mathcal{W},
\end{equation}
\begin{equation} 
K_0 = \frac{n\pi}{2\sqrt{\delta(\Mb-1)^2-1}},
\end{equation}
\begin{equation} 
K_1 = \frac{\sqrt{2}\delta(\Mb-1)^2}{\delta(\Mb-1)^2-1},
\end{equation}
\begin{equation} 
K_2 = \frac{n\pi\delta(\Mb-1)}{2\left(\delta(\Mb-1)^2-1\right)^{3/2}}.
\end{equation}
\end{subequations}
{\no}It is now straightforward to find the maximum of \equs{approx_Xn_equ}, which 
approximates the marginally stable point of the $n$-th mode 
\begin{subequations}
\label{eq:marginal_analytic}
\begin{equation} 
\label{eq:marginal_analytic_phi}
\mathcal{W}_{\rm n} \simeq \frac{2\left(\Mb-1\right)^2\left[\delta\left(\Mb-1\right)^2-1\right]}{n^2\pi^2},
\end{equation}
\begin{equation} 
\label{eq:marginal_analytic_x}
K_{\rm n} \simeq \frac{n\pi}{2\sqrt{\delta(\Mb-1)^2-1}}.
\end{equation}
\end{subequations}
{\no}$\varpi_{\rm n}$ thus converges to $\Mb^{-1}$ as $n^{-2}$ 
while $K_{\rm n}$ grows linearly with $n$.

\smallskip
To ensure $\varpi_{\rm n}\le\Mb^{-1}$ we require $\mathcal{W}_{\rm n}\ge 0$. 
From \equ{marginal_analytic_phi}, this leads to
\be 
\label{marginal_condition}
\Mb^{-1}+\Ms^{-1}\le1 \Rightarrow v\ge\cb+\cs
\ee
{\no}This supports what was inferred from \fig{marginal} in \se{Marginal1}, 
namely that \textit{marginally stable points exist only for flows that are 
supersonic with respect to the sum of the two sound speeds}. All we assumed 
in deriving \equs{marginal_analytic} was that $K(\varpi)$ from \equ{marginal_3} 
was continuous for $|\varpi|\le\Mb^{-1}$, which requires only $v\ge\cb-\cs$. 
Such a flow is not necessarily supersonic at all, while the existence of unstable 
modes with $n\ge 1$ is a purely supersonic effect.

\smallskip
Our approximation for $\mathcal{W}_{\rm n}$ is only valid if $|Z|^{-1}<<\pi/2$, 
with $|Z|^{-1}$ given in \equ{Zn}. It is straightforward to verify that this 
condition also guarantees $\mathcal{W}_{\rm n}<<1$ in \equ{marginal_analytic_phi}. 
The value of $n$ for which $|Z|^{-1} \simeq 1$ is 
\be 
\label{eq:marginal_n}
n_1 \simeq \delta\left(\Mb-1\right)^3
\ee
{\no}We expect \equs{marginal_analytic} to converge for $n>>n_1$. 

\smallskip
In \fig{marginal_analytic} we compare \equs{marginal_analytic} to exact numerical 
solutions for the maxima of \equ{marginal_3}, for different values of $\Mb$ and 
$\delta$. The left-hand panel shows $K_{\rm n}$ and the right-hand panel shows 
$\mathcal{W}_{\rm n}$. The approximation for $K_{\rm n}$ improves with increasing $n$ 
and increasing $M_{\rm tot}$. For $M_{\rm tot}\gsim 2$, the error is less than $10\%$ already 
for $n\gsim 2$, while for lower $M_{\rm tot}$ this error is achieved for $n\gsim 10$. 
On the other hand, the approximation for the marginally stable frequency, \equ{marginal_analytic_phi}, 
only converges for $n>n_1$ as expected (\equ{marginal_n}). Unfortunately, for even 
moderately high Mach numbers and density contrasts, $n_1$ can be quite high. For example, 
for $\Mb=5$ and $\delta=1$, $n_1\simeq 65$.

\smallskip
Modes with $n\ge 1$ begin stable at $K=K_{\rm n,0},\,\varpi=0$, pass through marginal 
stability at $K=K_{\rm n},\,\varpi \simeq \Mb^{-1}$ and end at 
$K=\infty,\,\varpi_{\rm \infty}=1-\Ms^{-1}$. As a result, $|Z|$ begins near $0$, reaches 
large values $|Z|>>1$ and ends near $0$ again. Therefore, $(1+Z)/(1-Z)$ goes from $\sim 1$ 
to $\sim -1$ and back to $\sim 1$, completing a full revolution about the origin in the 
complex plane. This is the reason for the extra $2\pi$ in \equ{standing2} compared to \equs{inverted_dispersion_slab}. 
\section{Growth Rates near Marginal Stability}
\label{sec:Marginal3} 

\smallskip
Using our approximation for the behaviour of $K(\varpi)$ near marginal 
stability (\equsnp{approx_Xn_equ}), we can derive the growth rate of the 
$n$-th mode. We assume $K=K_{\rm n}+\kappa$ and $\varpi = \Mb^{-1}(1-\mathcal{W}_{\rm n}+\xi)$, 
with $\kappa<<K_{\rm n}$ and $|\xi|<<\mathcal{W}_{\rm n}<<1$. By inserting 
this into \equs{approx_Xn_equ} we obtain to leading order in $\kappa$ 
\be 
\label{eq:marginal_growth}
\xi \simeq \pm i\left[\frac{8\Mb^{3/2}\left[\delta\left(\Mb-1\right)^2-1\right]}{\sqrt{2}\delta\left(\Mb-1\right)^2}\mathcal{W}_{\rm n}^{3/2}\right]^{1/2}\kappa^{1/2}.
\ee
This shows that the system is indeed unstable near $K_{\rm n}$, and that this 
instability has a growing mode. Using our approximation for $\mathcal{W}_{\rm n}$ 
from \equ{marginal_analytic_phi}, \equ{marginal_growth} becomes 
\be 
\label{eq:marginal_growth_2}
\xi \simeq \pm i\frac{8\sqrt{\left(\Mb-1\right)}\left[\delta\left(\Mb-1\right)^2-1\right]^{5/4}}{\sqrt{\delta}\left[n\pi\right]^{3/2}}\kappa^{1/2}.
\ee

\smallskip
The growth rate near marginal stability is 
\be 
\label{eq:marginal_growth_3}
{\rm Im}(\omega) = kv~{\rm Im}(\varpi) = k\cb |\xi|.
\ee
$\xi$ can be computed either from \equ{marginal_growth} using 
the exact (numerical) solution for $\mathcal{W}_{\rm n}$, or 
from \equ{marginal_growth_2} using our analytic approximation, 
which we saw to be a good approximation for $n>n_1$ (\equnp{marginal_n}). 

\begin{figure*}
\centering
\subfloat{\includegraphics[width =0.95 \textwidth]{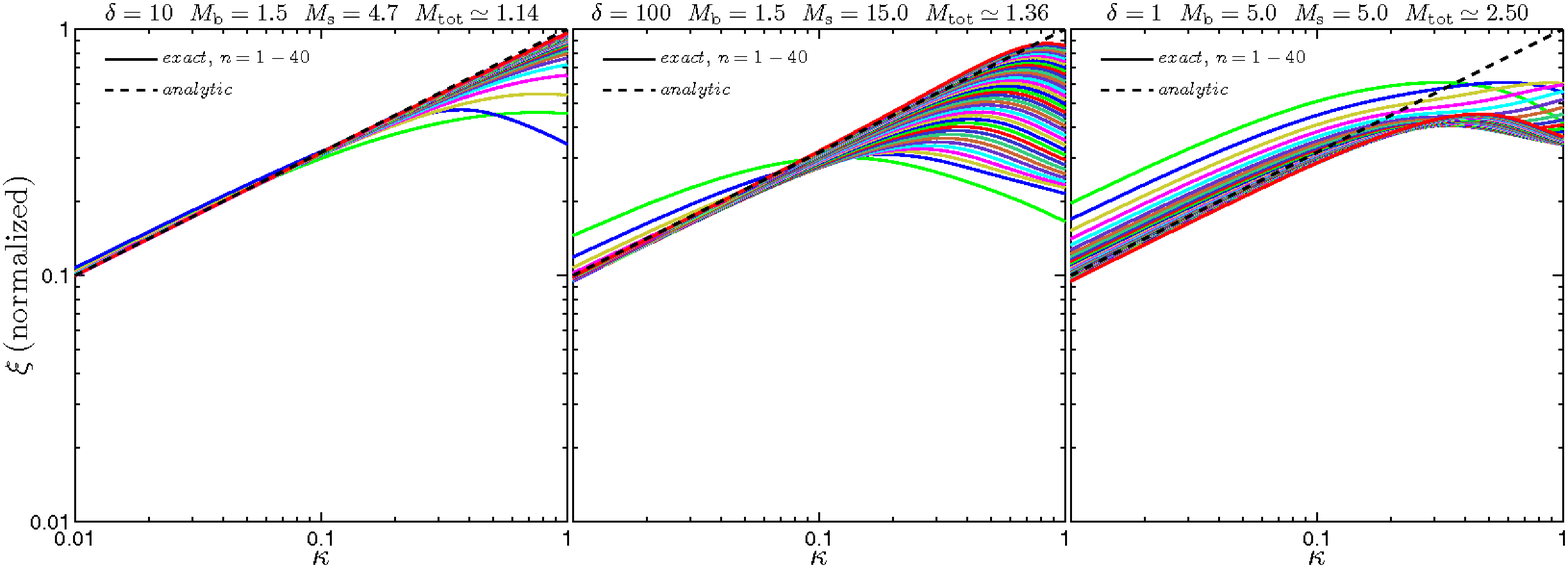}}
\caption{Comparison of our analytic estimate for the growth rate of modes with $n\ge1$ 
near marginal stability to a numerical solution of the dispersion relation. The x axis 
is $\kappa=K-K_{\rm n}$, the wavenumber relative to marginal stability. The y axis is 
$\xi={\rm Im}(\varpi)$, normalized by the $n$-dependent factor multiplying $\kappa^{1/2}$ 
in \equ{marginal_growth}. We used the exact value for $K_{\rm n}$ and $\mathcal{W}_{\rm n}$ 
when computing $\kappa$ and $\xi$ was used, obtained by numerically finding the maximum of 
$K(\varpi)$ in \equ{marginal_3}. The black dashed line shows the analytic prediction and 
the coloured solid lines show numerical calculations of the first 40 modes, $n=1-40$, 
increasing from top to bottom. For all cases examined, the slope of $\xi\propto \kappa^{1/2}$ 
is reproduced at $\kappa\lsim 0.1$. Increasing $n$ or decreasing $M_{\rm tot}$ makes the slope 
a good fit at higher $\kappa$ as well, and improves the fit to the normalization of $\xi$. However, 
even at high $M_{\rm tot}$, the normalization is well approximated by \equ{marginal_growth} by 
$n\sim 5$.
}
\label{fig:marginal_growth}
\end{figure*}

\smallskip
The analytic approximation for $\xi$ is compared to numerical solutions of the dispersion 
relation in \fig{marginal_growth}, for different values of $\delta$ and $\Mb$. The coloured 
lines show the result of a full numerical solution to the dispersion relation 
(\equnp{dispersion_slab_unitless}) for the modes $n=1-40$. The dashed black line is the 
analytic approximation. We show $\xi={\rm Im}(\varpi)$, normalized by the $n-$dependent 
prefactor of $\kappa^{1/2}$ in \equ{marginal_growth}, as a function of $\kappa$. 
When computing $\xi$ and $\kappa$ from \equ{marginal_growth}, we used the numerical 
results for $\mathcal{W}_{\rm n}$ and $K_{\rm n}$, obtained by finding the maximum 
of $K(\varpi)$ in \equ{marginal_3}. Though not fully analytic, this is still much 
easier to evaluate numerically than a full solution to the dispersion relation, and 
is thus still useful. In all cases examined, the scaling of $\xi\propto\kappa^{1/2}$ 
is captured at $\kappa\lsim 0.1$ for all $n$, even if the normalization has not yet 
converged. The normalization converges rapidly as well for low Mach numbers (left 
panel), though more slowly for high Mach numbers, qualitatively similar to the 
convergence of \equ{marginal_analytic_phi} for $\mathcal{W}_{\rm n}$ (\fig{marginal_analytic}). 

\smallskip
For high Mach numbers, $\Mb>>1$, \equ{marginal_growth_2} 
yields $\xi \propto \delta^{3/4}\Mb^3n^{-3/2}$ (for $n>n_1$), 
while $k\simeq K_{\rm n} \propto \delta^{-1/2}\Mb^{-1}n$. 
The growth rate near marginal stability thus scales as 
\be 
\label{eq:marginal_grwoth_scaling}
{\rm Im}(\omega) \propto \delta^{1/4}\Mb^{2}n^{-1/2}.
\ee
{\no}The scaling of $\Mb^{2}$ is very different than the growth rate 
near marginal stability for the fundamental modes. For the S-mode 
this is independent of $\Mb$ (\equnp{omega_fundamental_s_mode}), and for 
the P-mode it scales as $\Mb^{-1/2}$ (\equnp{omega_fundamental_p_mode}).

\section{Fastest Growing Mode in the Slab}
\label{sec:Marginal4} 

\smallskip
For each mode, the growth rate is zero at marginal stability, grows larger 
as $K$ is increased, and then goes to zero again at $K\rightarrow\infty$ 
(\se{short_comp_slab}). Each mode thus has a maximal growth rate at some 
intermediate $K$, hereafter the \textit{resonance} of the mode. It is of 
particular interest to estimate this maximal growth rate, and how it scales 
with mode number $n$, or alternatively with wavenumber $K$. This will tell 
us which mode is dominant for perturbations of a particular wavelength, and 
will give an estimate of the fastest growth rate. Below, we derive an analytic 
approximation for the maximal growth rate which is valid for very supersonic 
flows, $M_{\rm tot} = V/(\cb+\cs)>>1$ (though in practice, it is a good 
approximation for $M_{\rm tot}\gsim 1.5$).

\smallskip
We make two additional assumptions. The first is that $\varpi$ is sufficiently far 
from both 0 and 1 at resonance, so that $\delta\Mb(\varpi-1)^2>>1$ and $\Mb\varpi^2>>1$. 
The second is that the growth rate at resonance is much smaller than the angular 
frequency. In other words, if $\omega = \Or + i\Oi$, then $\Oi<<\Or$ at resonance. 
We empirically show these two assumptions to be valid by numerically solving the 
dispersion relation (see \fig{numerical_solution}). However, they can both be justified 
analytically as well. When $M_{\rm tot}>>1$, $\varpi$ grows from $\lsim\Mb^{-1}\sim 0$ 
at marginal stability to $1-\Ms^{-1}\sim 1$ as $K\rightarrow\infty$. Since in both of 
these limits the growth rate goes to zero, it is reasonable to assume that resonance 
occurs far from these points, so that $\varpi$ is sufficiently far from both 0 and 1. 
The second assumption can be justified by realizing that while both $\Pr$ and $\Pi$ 
start very small at marginal stability, $\Pr$ increases monotonically to an asymptotic 
value of $\sim 1$, while $\Pi$ flattens, reaches a maximum at resonance and then decays 
to zero. It is therefore reasonable to assume that $\Pi<<\Pr$ at resonance, and therefore 
that $\Oi<<\Or$.

\smallskip
Using these assumptions, and the fact that in our chosen branch cut 
${\rm Re}(\qbs)>0$, ${\rm Im}(\qb)<0$, ${\rm Im}(\qs)>0$ (\se{branch_cut}), 
we have $\qb=\sqrt{1-\Mb^2\varpi^2}\simeq -i\Mb\varpi$ and 
$\qs=\sqrt{1-\delta\Mb^2(1-\varpi)^2}\simeq i\sqrt{\delta}\Mb(1-\varpi)$. 
Inserting this into \equ{Z_def} yields 
\be 
\label{eq:max_growth_disp_Z}
Z \simeq \frac{\varpi}{\sqrt{\delta}(1-\varpi)}.
\ee
{\no}Writing $\varpi=\Pr+i\Pi$ and $(1+Z)/(1-Z) = A \,e^{i\theta}$, we obtain from 
\equ{inverted_dispersion_slab} 
\begin{subequations}
\label{eq:max_growth_disp_2}
\begin{equation} 
\label{eq:max_growth_disp_2_real}
\Pr \simeq 1 - \frac{\theta + n \pi }{2\sqrt{\delta}\Mb K}, 
\end{equation}
\begin{equation} 
\label{eq:max_growth_disp_2_comp}
\Pi \simeq \frac{1}{2\sqrt{\delta}\Mb K}{\rm ln}\left(A\right) \Rightarrow \Oi \simeq \frac{\cs}{2\Rs}{\rm ln}\left(A\right). 
\end{equation}
\end{subequations}
{\no}We thus conclude that resonance occurs at the maximum of $A$. 
After some algebra, it is straightforward to show from \equ{max_growth_disp_Z} 
that 
\be 
\label{eq:resonance_R}
A^2=\left| \frac{1+Z}{1-Z} \right|^2 \simeq \frac{\left[\Pr(\sqrt{\delta}-1)-\sqrt{\delta}\right]^2 + \left[\Pi(\sqrt{\delta}-1)\right]^2}{\left[\Pr(\sqrt{\delta}+1)-\sqrt{\delta}\right]^2 + \left[\Pi(\sqrt{\delta}+1)\right]^2}.
\ee

\smallskip
Neglecting terms of order $(\Pi/\Pr)^2<<1$, the maximum of $A$ occurs when 
\begin{subequations}
\label{eq:resonance_real}
\begin{equation} 
\label{eq:resonance_real_a}
\varpi_{_{\rm R,\,res}} \simeq \frac{\sqrt{\delta}}{\sqrt{\delta}+1} = \frac{\cb}{\cb+\cs}, 
\end{equation}
\begin{equation} 
\label{eq:resonance_real_R}
A_{\rm res} \simeq \frac{2\sqrt{\delta}}{\varpi_{\rm I,\,res}\left(\sqrt{\delta}+1\right)^2}.
\end{equation}
\end{subequations}
{\no}Neglecting terms of order $\Pi/\Pr$, we have 
$|\qb|\simeq |\qs|$ at resonance. This means that the perturbation penetrates the same 
depth into the slab as into the background. It also means that the angle of propagation 
relative to the normal to the slab, given by 
\be 
\label{eq:angle}
{\rm cot}\left(\Psi_{\rm b,s}\right) = \frac{\left|{\rm Im}(\qbs)\right|}{k}, 
\ee
{\no}is the same within the slab and the background. At resonance, the angle is 
\be 
\label{eq:angle2}
{\rm sin}\left(\Psi_{\rm res}\right) \simeq \frac{\sqrt{\delta}+1}{\sqrt{\delta}\Mb} = \frac{1}{M_{\rm tot}}, 
\ee
{\no}where in the last equality we have used $M_{\rm tot}^{-1} = \Mb^{-1} + \Ms^{-1}$. 
This is commonly referred to as the \textit{Mach angle}.

\smallskip
Inserting \equ{resonance_real_R} into \equ{max_growth_disp_2_comp} gives 
an estimate for the growth rate at resonance
\be 
\label{eq:growth_rate_resonance}
\tsc\omega_{_{\rm I,\,res}} + {\rm ln}\left(\tsc\omega_{_{\rm I,\,res}}\right) \simeq {\rm ln}\left(4M_{\rm tot}\frac{\sqrt{\delta}}{1+\sqrt{\delta}}K_{\rm res}\right),
\ee
{\no}where $\tsc=2\Rs/\cs$ is the sound crossing time in the slab. The solution 
to this equation can be expressed as an infinite sequence of functions 
\begin{subequations}
\label{eq:growth_rate_resonance_2}
\begin{equation}
\tsc\omega_{_{\rm I,\,1}} = {\rm ln}\left(4M_{\rm tot}\frac{\sqrt{\delta}}{1+\sqrt{\delta}}K\right),
\end{equation}
\begin{equation}
\tsc\omega_{_{\rm I,\,j}} = \tsc\omega_{_{\rm I,\,1}} - {\rm ln}\left(\tsc\omega_{_{\rm I,\,j-1}}\right).
\end{equation}
\end{subequations}
{\no}As $K\rightarrow \infty$, the sequence converges to $\omega_{_{\rm I,\,1}}$ and in 
practice, it convergences to $\omega_{_{\rm I,\,3}}$ even for small $K$. The growth rate 
at resonance diverges logarithmically with wavenumber, $\Oi \propto {\rm ln}(k)$. As shown 
in \fig{numerical_solution}, this is the effective growth rate for the slab, since at each 
wavenumber the growth rate will be dominated by the mode closest to resonance. This should 
be compared to the case of the incompressible slab and the compressible/incompressible 
sheet. In the incompressible cases, the system is always unstable and the growth rate 
diverges linearly with wavenumber, $\Oi\propto k$. In the compressible sheet, the same 
scaling applies at low Mach numbers, while at high Mach numbers the system is stable, 
$\Oi=0$. The scaling of $\Oi\propto {\rm ln}(k)$ for the effective growth rate in the 
compressible slab can be seen as a ``compromise" between these two extremes, diverging 
at short wavelengths, but only logarithmically. Also, recall that it is only the effective 
growth rate, comprised of the fastest growing modes at each wavelength, which diverges. Each 
individual mode stabilizes as $k\rightarrow \infty$, as discussed in \se{short_comp_slab}.

\smallskip
By inserting \equ{resonance_real_a} into \equ{max_growth_disp_2_real} 
we can estimate the wavenumber at resonance. Neglecting terms of order 
$\theta=O(\Pi/\Pr)$ 
\be 
\label{eq:resonant_X}
K_{\rm res} \simeq \frac{n\pi}{2M_{\rm tot}}
\ee
{\no}The resonant wavenumber increase linearly with $n$, similar to the 
marginally stable wavenumber (\equnp{marginal_analytic_x}). When 
$M_{\rm tot}>>1$, $K_{\rm res}$ is very nearly continuous.

\begin{figure*}
\centering
\subfloat{\includegraphics[width =0.95 \textwidth]{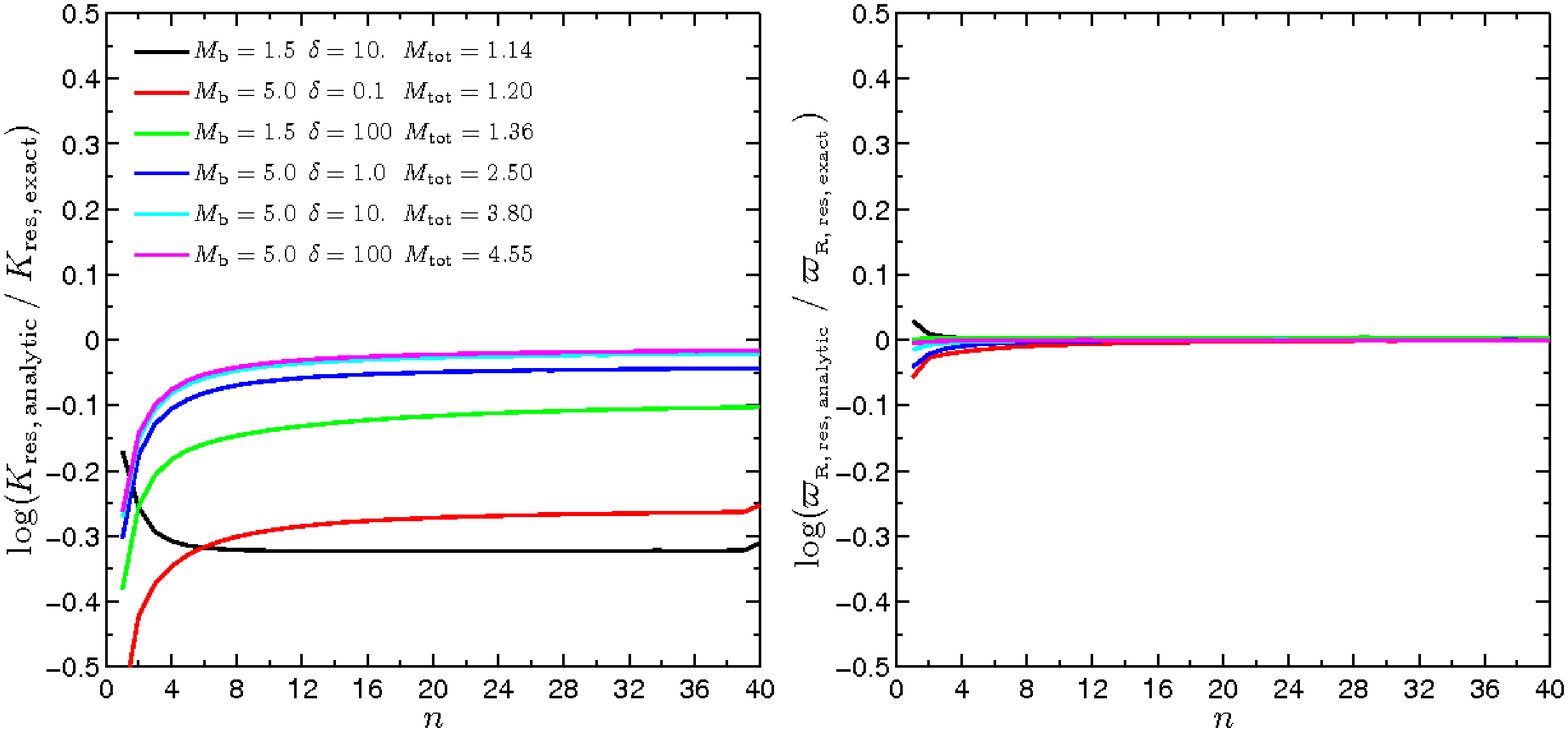}}
\caption{Comparison of our analytic estimate for the resonant wavenumber, 
$K_{\rm res}$ (\equnp{resonant_X}, left), and the real part of the resonant 
phase velocity, $\varpi_{_{\rm R,\,res}}$ (\equnp{resonance_real_a}, right), 
to a numerical solution of the dispersion relation. We compare the the first 
40 modes, $n=1-40$, for the same values of $\Mb$ and $\delta$ as in \fig{marginal_analytic}. 
The approximation for $K_{\rm res}$ is poor when $M_{\rm tot}\lsim 1.3$, but 
for $M_{\rm tot}\gsim 2$ the error is less than $\sim 10\%$ for $n>4$. On the 
other hand, the approximation for $\varpi_{_{\rm R,\,res}}$ is very good even 
at low Mach numbers.
}
\label{fig:Res}
\end{figure*}

\begin{figure*}
\centering
\subfloat{\includegraphics[width =0.95 \textwidth]{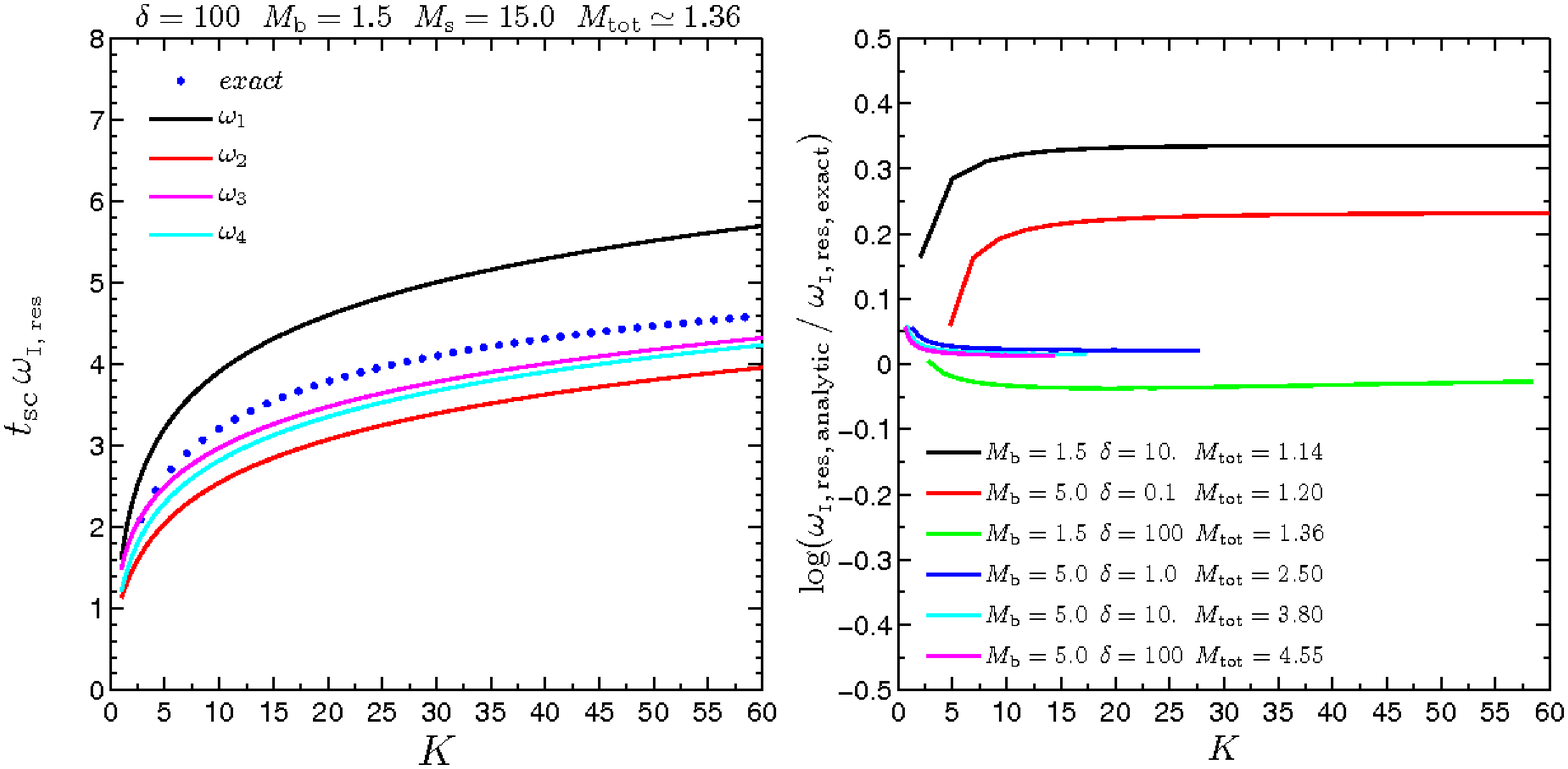}}
\caption{Analytic estimate vs. numerical solutions for the fastest growing 
mode in the slab. \textit{Left:} Comparison of the growth rate at resonance, 
$\omega_{_{\rm I,\,res}}$, computed numerically for the first 40 modes (blue 
points) and calculated analytically using the first 4 terms in the sequence in 
\equ{growth_rate_resonance_2} (solid lines), for $\Mb=1.5$ and $\delta=100$. 
The analytic series converges by $\omega_{_{\rm I,\,3}}$, though in this case 
it converges to a slightly lower value than the numerical solution. \textit{Right:} 
The ratio of $\omega_{_{\rm I,\,3}}$ from \equ{growth_rate_resonance_2} to the 
numerical solution for the same values of $\Mb$ and $\delta$ as in \fig{Res}. 
Note that the wavenumber corresponding to $n=40$ is different in each case, which 
is why different curves end at different $K$ values. The fit is quite poor for 
$M_{\rm tot}\lsim 1.3$, but for higher Mach numbers the error is less than $10\%$. 
}
\label{fig:ImPres}
\end{figure*}

\smallskip
\Fig{Res} and \fig{ImPres} compare our analytic approximations for the 
resonant frequency and wavelength to numerical solutions of the dispersion 
relation, for the first 40 modes, $n=1-40$, for different values of $\delta$ 
and $\Mb$. The left-hand panel of \fig{Res} shows the resonant wavenumber, 
$K_{\rm res}$ (\equnp{resonant_X}), the right-hand panel shows the phase 
velocity at resonance, $\varpi_{_{\rm R,\,res}}$ (\equnp{resonance_real_a}). 
and \fig{ImPres} shows the growth rate at resonance, $\omega_{_{\rm I,\,res}}$ 
(\equnp{growth_rate_resonance_2}). In the left-hand panel of \fig{ImPres}, 
we focus on one example, $\Mb=1.5$ and $\delta=100$, and examine the convergence 
of the sequence given in \equ{growth_rate_resonance_2}, which is shown to converge 
by $\omega_{_{\rm I,\,3}}$. In the right-hand panel, we compare $\omega_{_{\rm I,\,3}}$ 
to numerical solutions for the same values of $\Mb$ and $\delta$ as in \fig{Res}. 

\smallskip
The fit to $\varpi_{_{\rm R,\,res}}$ at resonance is good in all cases, with an error 
of less than $\lsim 10\%$. On the other hand, the approximations for $K_{\rm res}$ and 
$\omega_{_{\rm I,\,res}}$ are quite poor at low Mach numbers, $M_{\rm tot}\lsim 1.3$, 
but rapidly improve as $M_{\rm tot}$ is increased. The error in $\omega_{_{\rm I,\,res}}$ 
reaches $\lsim 10\%$ for $M_{\rm tot}\gsim 1.3$, and for $K_{\rm res}$ a similar error is 
achieved for $M_{\rm tot}\gsim 2$. This is expected, since the approximations we made are 
strictly valid for very high Mach numbers only. We note that in the cases where the approximation 
is particularly poor, $\Mb=1.5,\:\delta=10$ and $\Mb=5.0,\:\delta=0.1$, the sheet is still 
unstable since $\Mb<M_{\rm crit}$, so the slab instability is dominated by surface modes rather 
than body modes.

\label{lastpage} 
 
\end{document}